\documentclass[10pt,a4paper,floats,floatfix,superscriptaddress,twoside,aps,nofootinbib]{report}

\usepackage[breaklinks]{hyperref}
\usepackage{graphicx, epsfig,amssymb} %include figure files
\usepackage{epstopdf}
\usepackage{amsmath, amsfonts}
\usepackage{bm} %include bold math: \bm{} creates bold letters in math mode
\usepackage{color}
\usepackage{doi}
\usepackage[sort&compress,numbers,nonamebreak]{natbib}
\usepackage[latin1]{inputenc}
\usepackage{geometry}	
\def\be{\begin{equation}}
\def\ee{\end{equation}}

\def\FontL{% 16 pt normal
  \fontsize{16pt}{19.2pt}\selectfont}
\def\FontM{% 14 pt normal
  \fontsize{14pt}{16.8pt}\selectfont}
\def\FontS{% 12 pt normal
  \fontsize{12pt}{14.4pt}\selectfont}

\usepackage{setspace}

\begin{document}
\pagestyle{plain}
\pagenumbering{roman}

\thispagestyle {empty}

% IST Logo

%\includegraphics[bb=5.8cm 11cm 0cm 0cm,scale=0.38]{Logo.pdf}
%\vspace{-20mm}
   % Logo do IST
%   \begin{minipage}[l]{2cm}
%      \includegraphics[scale=0.29]{Logo_new}
%   \end{minipage}

\begin{center}

{\FontL \textbf{Universidade Técnica de Lisboa}} \\
\vspace{0.2mm}
{\FontL \textbf{Instituto Superior Técnico}} \\

\vspace{5cm}

% Title, author 
\vspace{0.8cm}
{\FontL \textbf{Dynamics around black holes:}} \\
\vspace{0.2cm}
{\FontM Radiation Emission and Tidal Effects}\\
\vspace{1.9cm}
{\FontM \textbf{Richard Pires Brito}} \\
\vspace{1.9cm}
{\FontM Disserta\c{c}\~{a}o para a obten\c{c}\~{a}o de Grau de Mestre em} \\
\vspace{0.3cm}
{\FontL \textbf{Engenharia Física Tecnológica}} \\
\vspace{1.9cm}
{\FontM \textbf{J\'{u}ri}} \\
\vspace{0.3cm}
{\FontS %
\begin{tabular}{ll}
Presidente: & Professora Doutora Ana Vergueiro Monteiro Cidade Mour\~{a}o \\
Orientador: & Professor Doutor Vitor Manuel dos Santos Cardoso \\
Vogais: & Professor Doutor José Pizarro de Sande e Lemos \\
        & Doutor Paolo Pani
\end{tabular} } \\

\vfill
{\FontM \textbf{Outubro 2012}} \\
\end{center}

\cleardoublepage

\vspace*{\fill} 
\begin{quote} 
\centering 
To the memory of Zulmira Matias. 
\end{quote}
\vspace*{\fill}

\cleardoublepage

\chapter*{Acknowledgements}

First of all, I would like to thank my supervisor, Vitor Cardoso, for all his guidance, support and for his patience with my mistakes throughout the development of this thesis. I have learnt a lot working with him, not only about black holes but also about physics in general, and principally about the scientific research world. 
I would also like to thank Paolo Pani for his patience in verifying the correctness of great part of my results and for very useful and critical comments. It will be a pleasure to continue to work with them in the next years.

I am very grateful to all the researchers of the Gravity Group for very interesting group meetings, where I have learnt a lot about physics.

Thanks to all my friends for these amazing five years. In particular to all the members of ``Lenços e Grunhidos'' and to the friends with whom I lived and shared almost everything at the residence. I prefer not to mention names at the risk of forgetting someone. 

Last, but not least, I would like to thank my family for all the support they gave me in the last years. A special thanks goes to my grandparents, basically for everything.  

\addcontentsline{toc}{section}{Acknowledgements}

\cleardoublepage

\chapter*{Resumo}

\addcontentsline{toc}{section}{Resumo}

Nesta tese são estudados vários processos que envolvem buracos negros em quatro e mais dimensões. Primeiro, usando técnicas de teoria de perturbações, compara-se a radiação escalar sem massa e com massa emitida por uma partícula altamente energética caindo radialmente dentro de um buraco negro de Schwarzchild. Mostra-se que em tempos muito grandes, o sinal de perturbações escalares massivas é dominado por uma cauda oscilatória universal, que aparece devido a efeitos de curvatura. Mostra-se também que o espetro de energia está de acordo com o cálculo do ZFL depois de considerar o teorema do não-cabelo. Na segunda parte, estuda-se o fenómeno de superradiância em dimensões genéricas e conjectura-se que a energia máxima extraída de um buraco negro em rotação pode ser compreendida em termos do volume próprio da ergoregião. Finalmente, estudam-se algumas consequências do efeito de superradiância na dinâmica de luas que orbitem em torno de buracos negros com rotação em cenários com mais de quatro dimensões. Em quatro dimensões, luas em torno de buracos negros geram marés de baixa amplitude, e a energia extraída da rotação do buraco negro é sempre menor do que a radiação gravitacional perdida para o infinito. Mostra-se que, em dimensões maiores do que cinco, a energia extraída do buraco negro através da superradiância é maior do que a energia emitida para o infinito. Estes resultados dão um forte apoio à conjectura de que este efeito é a regra, e não a excepção, em dimensões superiores. A superradiância domina e luas espiralam para fora; para uma determinada frequência orbital, a energia extraída no horizonte é igual à energia emitida para o infinito e `` órbitas flutuantes'' ocorrem. Dá-se uma interpretação desse fenómeno em termos do paradigma da membrana e da aceleração das marés devido à dissipação de energia no horizonte.

Parte dos resultados obtidos durante esta tese figuram na Ref.~\cite{Brito:2012prd}.

\vfill

\textbf{\Large Palavras-Chave:} Buracos negros; Campos escalares; Superradiância; Efeitos de Maré; Dimensões extra.  

\cleardoublepage

\chapter*{Abstract}

\addcontentsline{toc}{section}{Abstract}

In this thesis we study several dynamical processes involving black holes in four and higher dimensions. First, using perturbative techniques, we compare the massless and massive scalar radiation emitted by a particle radially infalling into a Schwarzchild black hole. We show that the late-time waveform of massive scalar perturbations is dominated by a universal oscillatory decaying tail, which appears due to curvature effects. We also show that the energy spectrum is in perfect agreement with a ZFL calculation once no-hair properties of black holes are taken into account. In the second part, we study the phenomenon of superradiance in higher dimensions and conjecture that the maximum energy extracted from a rotating black hole can be understood in terms of the ergoregion proper volume. We then study some consequences of superradiance in the dynamics of moons orbiting around higher-dimensional rotating black holes. In four-dimensional spacetime, moons around black holes generate low-amplitude tides, and the energy extracted from the hole's rotation is always smaller than the gravitational radiation lost to infinity. We show that in dimensions larger than five the energy extracted from the black hole through superradiance is larger than the energy carried out to infinity. Our results lend strong support to the conjecture that tidal acceleration is the rule, rather than the exception, in higher dimensions. Superradiance dominates the energy budget and moons ``outspiral''; for some particular orbital frequency, the energy extracted at the horizon equals the energy emitted to infinity and ``floating orbits'' generically occur. We give an interpretation of this phenomenon in terms of the membrane paradigm and of tidal acceleration due to energy dissipation across the horizon.  

Part of the results obtained during this thesis appeared in Ref.~\cite{Brito:2012prd}.
\vfill

\textbf{\Large Keywords:} Black holes; Scalar fields; Superradiance; Tidal effects; Extra dimensions.

\cleardoublepage

This work was supported by Fundação para a Ciência e Tecnologia,
under the grant PTDC/FIS/098025/2008.
The research included in this thesis was carried out at Centro Multidisciplinar de Astrofísica (CENTRA) in the Physics Department of Instituto Superior Técnico.

\cleardoublepage

\tableofcontents
\addcontentsline{toc}{chapter}{Contents}
\cleardoublepage

\listoftables
\addcontentsline{toc}{chapter}{\listtablename}
\cleardoublepage

\listoffigures
\addcontentsline{toc}{chapter}{\listfigurename}
\cleardoublepage 

\setcounter{page}{1}
\pagenumbering{arabic}

\chapter{Introduction}

This thesis is devoted to the study of several processes involving black holes in four and higher dimensions. In particular, we shall discuss processes that involve the emission of scalar radiation by black holes in four and higher dimensions within a perturbative approach, i.e.~by solving the linearized field equations. The first part of the thesis will be devoted to the signatures of high-energy processes around non-rotating black holes, when ordinary matter is coupled to a massive scalar field. In the second part, we shall discuss strong tidal effects which are related to the phenomenon of superradiance in higher dimensional black hole spacetimes.

Black holes are the simplest macroscopic objects in the Universe. They are easily understood using only the concept of space and time given by General Relativity. Furthermore, they can be described
by a couple of parameters, namely, their mass, angular momentum, and charge. This is the famous \emph{no-hair theorem}, which states that, in its final state, a black hole is uniquely described by these three parameters~\cite{gravitation}. In the last fifty years there have been considerable progresses in the understanding of black holes, and the mathematical tools to describe them have been highly improved~\cite{chandra}. Nowadays, it is universally recognized that black holes are not only of academic interest, but they are also of central importance in astrophysical processes and fundamental physics. For example, it is believed that most galaxies contain supermassive black holes at their centre~\cite{Melia:2007vt} and that the formation of astrophysical black holes is probably related to extreme phenomena.
In fundamental physics they play a key role since Hawking's semi-classical prediction of black hole evaporation by emission of a thermal radiation due to quantum effects~\cite{Hawking:1974sw}. Hawking's radiation was the first phenomenon to be predicted considering both gravitational and quantum effects. Black holes may therefore play a major role in the attempt to find a consistent quantum theory of gravity. 

More recently, the discovery of the gauge/gravity duality has given a whole new interest to the study of general spacetimes, and more particularly of black hole spacetimes. This duality maps the dynamics of some strongly coupled quantum field theories (QFTs) in $D$-dimensional spacetimes to the dynamics of semiclassical gravity in $D+1$-dimensional spacetimes. It is actually the only tool available to study the dynamics in strongly coupled QFT~\cite{Cardoso:2012qm}. The most widely studied gauge/gravity duality is the $AdS$/CFT correspondence, which maps the dynamics of non-Abelian conformal field theories (CFTs) in $D$-dimensional spacetimes onto semiclassical gravity in asymptotically Anti-de Sitter ($AdS$) spacetimes, i.e.~, spacetimes with a negative cosmological constant, in $D+1$ dimensions~\cite{Maldacena:1997zz,Witten:1998qj}. The field theories described by this correspondence are very different from the most important non-Abelian theory describing the physical world, Quantum Chromodynamics (QCD), the theory that describes the strong nuclear force. However, there is great hope that there could be a similar correspondence for some features of QCD (e.g.~\cite{Polchinski:2001tt,Polchinski:2002jw,Giddings:2002cd,Kang:2004jd}) and some recent results obtained at the Relativistic Heavy Ion Collider (RHIC) reinforced this hope~\cite{Luzum:2008cw}. This duality could provide tools to understand the behavior of condensed matter, quark-gluon plasmas, and other strongly coupled systems, impossible to describe using perturbative methods, in terms of $AdS$ black holes that interact semiclassically with fields.

In addition to the $AdS$/CFT duality, in the last decades, there has been a growing interest in physical phenomena in higher dimensional spacetimes, mainly motivated by higher dimensional solutions which naturally arise in the context of string theories and supergravities. In some scenarios, the extra dimensions arise naturally as an attempt to solve the hierarchy problem. Put in simple words, this problem seeks to understand why there is such a big difference between the electroweak scale $m_{EW}\sim 300$ GeV and the gravity or Planck scale $M_{4Pl}\sim 10^{19}$ GeV. In some extra-dimensional models, if the extra dimensions are highly warped and correspond to a very large volume, the Planck scale can be as low as $M_{dPl}\sim \mathcal{O}(1)$ TeV, eliminating the large difference between this scale and the electroweak scale. These are the so-called \emph{TeV-gravity} scenarios~\cite{ArkaniHamed:1998rs,ArkaniHamed:1998nn,Antoniadis:1998ig,Randall:1999vf,Cardoso:2012qm}. The idea behind that comes from string theory and the concept of $p$-branes. A $p$-brane is a hypersurface with $p$-spatial dimensions. Gravity would live in a $D$-dimensional spacetime, and the gauge interactions, i.e.~, the electromagnetic and nuclear forces, would be constrained to a $3+1$-dimensional brane. The gravitational attractive power would then be ``diluted'' in the extra dimensions, appearing to be much smaller in the brane~\cite{Cardoso:2012qm,DeWolfe:2002nn}.     
Presently the LHC has begun to search for evidences of Tev-gravity models, looking for trans-Planckian signatures, i.e.~, processes where the energy involved exceed the Planck energy. At the Planck scale, non-linear quantum gravity effects dominate and we would need a complete quantum gravity theory to describe the physics near this domain. However, in some conditions gravity becomes highly non-linear, and black hole formation is expected~\cite{Banks:1999gd,Giddings:2001bu,Dimopoulos:2001hw,Giddings:2000ay}. This happens when the impact parameter $b=J/\sqrt{s}$, with $J$ being the angular momentum and $\sqrt{s}$ the center-of-mass energy, is of the order of the Schwarzschild-Tangherlini radius $R_s$, which is a generalization of the $4$-dimensional Schwarzschild radius $R_s=2M$ in higher-dimensions. The formation of these higher-dimensional black holes would carry a clear signature through the decay by Hawking radiation, emission of gravitational waves, and in some theories, other types of radiation, such as scalar radiation. Although at the time of writing, none event of black hole formation has been recorded at the LHC~\cite{Khachatryan:2010wx,Chatrchyan:2012taa}, further study is necessary to fully understand  the trans-Planckian regime and the signatures that higher-dimensional black hole formation could carry.

%Generally an analytic description of processes involving black holes are extremely difficult to obtain. A full understanding of these processes involve using approximate methods, and in some cases only a numerical solution is possible. Doing numerical simulations requires a good understanding of the techniques used and, in many situations, interpreting the physical results is not an easy task. Therefore, a wider knowledge about possible effects arising in four, and principally in higher dimensions, is mandatory. 
Today it is widely accepted that there are at least three complementary approaches to solve the Einstein's equations for a given problem: (i) analytical computations, (ii) semianalytical methods and soft numerics, (iii) fully numerical simulations. Purely analytical computations are normally only possible at a linear level and, in most cases, it is impossible to obtain a full analytical solution. In fact, solving the Einstein's equations exactly, i.e.~, with no approximations, is a formidable task which requires supercomputers and sophisticated numerical methods. Complementary to these, semianalytical methods are instrumental for a better understanding of dynamical processes in General Relativity and sometimes for interpreting the results of the simulations. There is a need for the development of new approximation schemes to accompany large scale simulations. Fully numerical simulations give only numerical answers to problems. Often these answers do not provide physical understanding, or even what principles are behind the process. Furthermore, in some cases, simulation results can be simply incorrect or misleading. By closely coupling various perturbation schemes it is possible to interpret and confirm simulation results. On the other hand, even using a perturbative approach usually requires numerical solutions. The development of these complementary approaches cannot be done independently. 
This is the main motivation of this thesis, to study and understand processes that could be seen in future numerical simulations and could also be of great importance, both in the astrophysical context and in more theoretical applications, like the gauge/gravity duality and TeV-gravity scenarios.

\section{Outline of the thesis}
This thesis is organized as follows. 
In chapter~\ref{chap:myers} we compute the fundamental equations of scalar radiation in terms of black hole perturbations sourced by a test-particle in geodesic motion around a spinning black hole. We derive the Teukolsky equations and solve the wave equation using the Green's functions approach. 
In chapter~\ref{chap:radial} we specialize the problem to the case of a particle falling radially into a Schwarzschild black hole. We compute the massive scalar radiation waveforms emited by the particle and the correspondent energy spectra.  
In chapter~\ref{chap:super} we study the superradiant scattering of a massless scalar field with a singly spinning black hole in $D=4+n$ dimensions.
In chapter~\ref{chap:tides}, extending the discussion of Ref.~\cite{Cardoso:2012zn}, we study the effect of an electrically charged particle orbiting a neutral central object in $4+n$ dimensions. By applying the membrane paradigm~\cite{Thorne:1986iy}, we derive a simple formula for the ratio between the energy flux at infinity and at the horizon in $4+n$ dimensions. Furthermore, we confirm the results obtained at the Newtonian level by solving the wave equations using black hole perturbation theory both analytically and numerically.
We conclude in chapter~\ref{chap:conclusion}. 

%The results presented in chapter~\ref{chap:tides} have been published in Ref.~\cite{Brito:2012prd}.
Throughout the thesis we use $G=c=\hbar=1$ units, except in the first section of chapter~\ref{chap:tides} where, for clarity, we show $G$ and $c$ explicitly.

\cleardoublepage

%%%%%%%%%%%%%%%%%%%%%%%%%%%%%%%%%%%%%%%%%%%%%%%%%%%%%%%%%%%%%%%%%%%%%%%%%%%%%%%%%%%%%%%%%%%%%%%
\chapter{Scalar perturbations of singly spinning Myers--Perry black holes}\label{chap:myers}

The way a black hole reacts to external perturbations provides us with a deep understanding of the space-time around it. In this thesis we will be interested, more particularly, in scalar perturbations of singly-spinning Myers--Perry black holes due to the presence of a test particle coupled to the scalar field.

Scalar fields are predicted by several theories aiming to unify Einstein's theory with the Standard Model of particle physics~\cite{Arvanitaki:2010prd,Arvanitaki:2011prd,Fujii,Sotiriou:2008rp}. 
%They also naturally arise in purely classical modified gravity theories, such as scalar-tensor theories and $f(R)$ theories~\cite{Fujii,Sotiriou:2008rp}. 
Their existence is being widely looked, and therefore, the study of such scalar fields is very relevant and timely.     

 We will not go into details about the mathematical formalism of the problem, since it as been deeply studied over the years and it is not the purpose of this thesis (for further detail about black hole perturbation theory see~\cite{chandra}). The starting point of a relativistic perturbation theory is to consider two different spacetimes, the \emph{physical} one, which carries the information about the actual physical system, and the \emph{background} one, which corresponds to a simpler idealized spacetime, solution of Einstein's equations. The background and physical spacetimes can be related using different maps, each of them corresponding to different a \emph{gauge} choice. The transformation between maps is then called \emph{gauge} transformation. 
For a particular map, we can write the spacetime metric as 
\be  
g_{ab}(x^{\mu})=g_{ab}^{(0)}(x^{\mu})+h_{ab}(x^{\mu})\,,
\ee
where $g_{ab}^{(0)}(x^{\mu})$ is the background metric, and $h_{ab}(x^{\mu})$ is a small perturbation that can be obtained solving Einstein's equations order by order in a small curvature and small velocity expansion. To avoid possible physical interpretation problems due to the choice of the gauge, one usually looks for gauge-invariant quantities, i.e.~, quantities that have the same value independently of the chosen gauge.
%Varying Einstein's equations $G_{ab}=8\pi T_{ab}$ with respect to the perturbation, we then arrive at a couple of gauge-invariant differential equations describing the perturbation.
 
First-order perturbations of black holes began with the pioneering work of Regge and Wheeler~\cite{Regge:1957td}, who studied the stability of Schwarzschild black holes, followed by the work of Zerilli, who first gave a fully relativistic treatment of the gravitational radiation emitted by a particle falling radially into a Schwarzschild black hole~\cite{Zerilli:1971wd}.   
Finding a couple of decoupled master equations describing generic perturbations in the case of a rotating black hole was much more difficult to solve, but scalar perturbations were shown to be easily treated even in this case~\cite{teunature,Brill:1972prd}. The separability of the scalar wave equation in the Kerr family of geometries, as demonstrated by Carter in 1968~\cite{Carter:1968pr}, and explicitly exposed by Brill \emph{et al} in 1971~\cite{Brill:1972prd}, opened a new door for the study of rotating black hole perturbations. Among many others, one of the major achievements of this era, the so-called "golden age" of General Relativity, was the discovery, by Teukolsky in 1973~\cite{teu}, of fundamental equations governing, not only scalar perturbations, but also, gravitational, electromagnetic, and massless fermionic perturbations.    

In this chapter we shall generalize some of these results to higher-dimensional backgrounds, considering a test-particle coupled to a massive scalar field around the spinning black hole. The results here derived will serve as a framework for the subsequent chapters.   

%%%%%%%%%%%%%%%%%%%%%%%%%%%%%%%%%%%%%%%%%%%%%%%%%%%%%%%%%%%%%%%%%%%%%%%%%%%%%%%%%%%%%%%%%%%%%%%
\section{The background metric}
%%%%%%%%%%%%%%%%%%%%%%%%%%%%%%%%%%%%%%%%%%%%%%%%%%%%%%%%%%%%%%%%%%%%%%%%%%%%%%%%%%%%%%%%%%%%%%%
In four dimensions, there is only one possible angular momentum parameter for an axisymmetric spacetime, and rotating black hole solutions are uniquely described by the Kerr family. In higher dimensions there are several choices of rotation axis, which correspond to a multitude of angular momentum parameters~\cite{myers}. Here we shall focus on the simplest case, where there is only a single axis of rotation. In the following we shall adopt the notation used in Refs.~\cite{Ida,Cardoso:2004cj,Cardoso:2005vk}, to which we refer for details.

 The metric of a ($4 + n$)-dimensional Kerr--Myers--Perry black hole with only one nonzero angular momentum parameter is given in Boyer--Lindquist coordinates by~\cite{myers}
\begin{align} 
\label{Myers-Perry}
ds^2=&-\frac{\Delta-a^2\sin^2\vartheta}{\Sigma}dt^2
-\frac{2a(r^2+a^2-\Delta)\sin^2\vartheta}{\Sigma}dtd\phi
+\frac{(r^2+a^2)^2-\Delta a^2\sin^2\vartheta}{\Sigma}\sin^2\vartheta d\phi^2\nonumber\\
&+\frac{\Sigma}{\Delta}dr^2
+\Sigma d\vartheta^2+r^2\cos^2\vartheta d\Omega_n^2\,,
\end{align}
where
\begin{equation}
\Sigma= r^2+a^2\cos^2\vartheta\,,\qquad \Delta= r^2+a^2-2M r^{1-n}\,,
\end{equation}
and $d\Omega_n^2$ denotes the standard line element of the unit $n$-sphere. 
This metric describes a rotating black hole in asymptotically flat, vacuum spacetime, whose physical mass ${\cal M}$ and angular momentum ${\cal J}$ (transverse to the $r\phi$ plane), respectively read
%%%
\begin{equation}
 {\cal M}=\frac{(n+2) A_{n+2}}{8\pi}M\,,\qquad {\cal J}=\frac{2}{n+2} {\cal M} a\,,
\end{equation}
%%%
where $A_{n+2}=2\pi^{{n+3}/2}/\Gamma[(n+3)/2]$.% and we assume $\mu, a>0$.

The event horizon is located at $r=r_H$, defined as the largest real root of $\Delta$. In four dimensions, an event horizon exists only for $a\leq M$. In five dimensions, an event horizon exists only for $a\leq\sqrt{2M}$, and the black hole area shrinks to zero in the extremal limit $a\rightarrow \sqrt{2M}$. On the other hand, when $D>5$, there is no upper bound on the black hole spin and a horizon exists for any $a$.
%%%%%%%%%%%%%%%%%%%%%%%%%%%%%%%%%%%%%%%%%%%%%%%%%%%%%%%%
\section{The wave equation}
%\section{Setup}
%%%%%%%%%%%%%%%%%%%%%%%%%%%%%%%%%%%%%%%%%%%%%%%%%%%%%%%%
We consider a small object in a geodesical curve around a spinning black hole and a scalar field of mass $m_s=\mu\hbar$ coupled to matter (from now on we set $\hbar=1$. In these units $\mu$ has the dimensions of 1/length). At first order in perturbation theory, the scalar field equation in the background~(\ref{Myers-Perry}) reads
\be\label{fieldeq}
\square\varphi-\mu^2\varphi\equiv\frac{1}{\sqrt{-g}}\frac{\partial}{\partial x^{\mu}}\left(\sqrt{-g}g^{\mu\nu}\frac{\partial}{\partial x^{\nu}}\varphi\right)-\mu^2\varphi
=\alpha {\cal T}\,,
\ee
where $\alpha$ is some coupling constant. For simplicity we focus on source terms of the form
\be
{\cal T}=-\int \frac{d\tau}{\sqrt{-g}} q_p\delta^{(4+n)}(x-X(\tau))\,,
\ee
which corresponds to the trace of the stress-energy tensor of a point particle with scalar charge $q_p$.
Because of the coupling to matter, the object emits scalar radiation, which is governed by Eq.~(\ref{fieldeq}). 
To separate Eq.~\eqref{fieldeq}, we consider the ansatz
\be\label{eq:ansatz}
\varphi(t,r,\vartheta,\phi)=\sum_{l,m,j}\int d\omega e^{im\phi-i\omega t}R(r)S_{lmj}(\vartheta)Y_j\,,
\ee
where $Y_j$ are hyperspherical harmonics~\cite{Cardoso:2004cj,Berti:2005gp} on the $n$-sphere with eigenvalues given by $-j(j+n-1)$ and $j$ being a non-negative integer. The radial and angular equations read
\begin{equation}\label{teuradial}
r^{-n}\frac{d}{dr}\left(r^n\Delta\frac{dR}{dr}\right)
+\left\{\frac{\left[\omega(r^2+a^2)-ma\right]^2}{\Delta}\right.
\left.-\frac{j(j+n-1)a^2}{r^2}-\lambda-\mu^2r^2\right\}R=T_{{lmj}}\,,
\end{equation}
and
\begin{equation}\label{ang}
\frac{1}{\sin\vartheta\cos^n\vartheta}\frac{d}{d\vartheta}\left(\sin\vartheta\cos^n\vartheta\frac{dS_{lmj}}{d\vartheta}\right)
+\left[(\omega^2-\mu^2)a^2\cos^2\vartheta
-\frac{m^2}{\sin^2\vartheta}-
\frac{j(j+n-1)}{\cos^2\vartheta}+A_{lmj}\right]S_{lmj}=0\,,
\end{equation}
where $\lambda=A_{lmj}-2m\omega a+\omega^2a^2$, $A_{lmj}$ are the eigenvalues of the angular equation, and we have defined
\be
\alpha \Sigma \mathcal{T}=\sum_{l,m,j}\int d\omega e^{im\phi-i\omega t}T_{lmj}S_{lmj}(\vartheta)Y_j\,.
\ee
%
%\begin{equation}
%T_{{lmj}}=-\frac{m_p\alpha}{U^tr^n}S^*_{lmj}(\pi/2)Y_{j}^{*}(\pi/2,\pi/2,\ldots)
%\delta(r-r_0)\delta(m\Omega_p-\omega)\,,
%\label{tlmw}
%\end{equation}
%
%which has been derived from the stress-energy tensor of the point particle (see Appendix~\ref{chapter:appendix2}).
%

\subsection*{Radial equation}
Defining a new radial function $X_{lmj}(r)$ 
\be
X_{lmj}=r^{n/2}(r^2+a^2)^{1/2}R\,,
\ee
we get the nonhomogeneous equation for the scalar field
\be\label{radial1}
\left[\frac{d^2}{dr_*^2}+V\right]X_{{lmj}}(r^*)=\frac{\Delta}{(r^2+a^2)^{3/2}}r^{n/2}T_{{lmj}}\,,
\ee
where $dr/dr_*=\Delta/(r^2+a^2)$ defines the standard tortoise coordinates and the effective potential $V$ reads

\begin{align}
V=&\omega^2-\frac{\Delta\mu^2}{r^2+a^2}+\frac{3r^2\Delta^2}{(r^2+a^2)^4}-\frac{\Delta\left[3r^2+a^2-2M r^{1-n}(2-n)\right]}{(r^2+a^2)^3}
+\frac{1}{(r^2+a^2)^{2}}\left\{a^2m^2-4M\right.\nonumber\\
+&\frac{am\omega}{r^{n-1}}-\Delta\left(\omega^2a^2+A_{lmj}\right)
+\left.
\Delta\left[\frac{n(2-n)\Delta}{4r^2}-n+\frac{2n\left(1-n\right)M}{2r^{n+1}}\right.\right.
\left.\left.-\frac{j(j+n-1)a^2}{r^{2}}\right]\right\}.
\end{align}

\subsection*{Angular equation}
In the low-frequency limit the angular Eq.~(\ref{ang}) can be solved exactly. For the massless case $\mu=0$ and at first order in $a\omega$, the eigenvalues can be computed analytically~\cite{Berti:2005gp}
\be 
A_{kjm}=(2k+j+|m|)(2k+j+|m|+n+1)+{\cal O}(a\omega)\,.\label{Akjm}
\ee
By setting $2k=l-(j+|m|)$, the eigenvalues above take the form $A_{ljm}=l(l+n+1)$ and $l$ is such that $l\geq(j+|m|)$, which generalizes the four-dimensional case. An important difference from the four-dimensional case is that regularity of the angular eigenfunctions requires $k$ to be a non-negative integer; i.e.~for given $j$ and $m$ only specific values of $l$ are admissible. In fact, it is convenient to label the eigenfunctions and the eigenvalues with the ``quantum numbers'' $(k,j,m)$ rather than with $(l,j,m)$ as in the four-dimensional case.
Note that when $a\omega\gtrsim1$, these eigenvalues might not be accurate. Therefore, in some cases, we will use the exact numerical eigenvalues. These can be obtained assuming an expansion for the eigenfunctions of the form~\cite{Berti:2005gp},
\be\label{eigenseries}  
S_{kjm}=\sin(\vartheta)^{|m|}\cos(\vartheta)^j \sum_{p=0}^{\infty}\tilde{a}_p(\cos^2\vartheta)^p\,.
\ee
This series (if convergent) automatically satisfies the regularity conditions at $\theta=0, \pi/2, \pi$. Upon substitution of \eqref{eigenseries} into the angular equation~(\ref{ang}), we obtain a three-term recursion relation~\cite{Cardoso:2004cj}
\begin{align}
&\tilde{\alpha}_0 \tilde{a}_1+\tilde{\beta}_0 \tilde{a}_0=0\,,\nonumber\\
&\tilde{\alpha}_p \tilde{a}_{p+1}+\tilde{\beta}_p \tilde{a}_p+\tilde{\gamma}_p \tilde{a}_{p-1}=0\,,\qquad (p=1,2,\ldots) 
\end{align}
where
\begin{align}
\tilde{\alpha}_p&=-2(p+1)(2j+n+2p+1)\,,\nonumber\\
\tilde{\beta}_p&=(j+|m|+2p)(j+n+|m|+2p+1)-A_{kjm}\,,\nonumber\\
\tilde{\gamma}_{p}&=-(a\omega)^2\,.
\end{align}
Then, given a value for $a\omega$, the eigenvalues $A_{kjm}$ can be obtained solving numerically the continued fraction equation~\cite{Leaver:1985qnm}
\be 
\tilde{\beta}_0-\frac{\tilde{\alpha}_0 \tilde{\gamma}_1}{\tilde{\beta}_1-\frac{\tilde{\alpha}_1 \tilde{\gamma}_2}{\tilde{\beta}_2-\frac{\tilde{\alpha}_2 \tilde{\gamma}_3}{\tilde{\beta}_3-\ldots}}}
=0\,.
\ee

The (non-normalized) zeroth-order eigenfunctions are given in terms of hypergeometric functions~\cite{Cardoso:2004cj,Berti:2005gp}
\begin{equation}\label{zerothorder} 
S_{kjm}\propto\sin(\vartheta)^{|m|}x^{j}F\Big[-k,k+j+|m|+\frac{n+1}{2},j+\frac{n+1}{2};x^2\Big]\,,
\end{equation}
%%%
where $x=\cos(\vartheta)$.
We adopt the following normalization condition:
\be
\int_0^{\pi/2} d\vartheta\sin\vartheta\cos^n\vartheta S_{kjm}S^*_{kjm}=1\,,
\ee
where the integration domain has been chosen in order to have a nonvanishing measure also in the case of odd dimensions. Note that this normalization differs from that adopted in Ref.~\cite{Berti:2005gp}.

%We note that at $\vartheta=\pi/2$ only hyperspherical harmonics with $j=0$ are nonvanishing. Thus, in order to calculate the fluxes on circular orbits, one only needs to consider terms with $j=0$. In this case, the hyperspherical harmonics $Y_0$ are constant. 

%
%%%%%%%%%%%%%%%%%%%%%%%%%%%%%%%%%%%%%%%%%%%%%%%%%%%%%%%
\section{Green function approach}
%%%%%%%%%%%%%%%%%%%%%%%%%%%%%%%%%%%%%%%%%%%%%%%%%%%%%%%
To solve the wave equation, let us choose two independent solutions $X_{{kjm}}^{r_H}$ and $X_{{kjm}}^{\infty}$ of the homogeneous equation, which satisfy the following boundary conditions:
\begin{equation} \label{boundinf}
\left\{
 \begin{array}{l}
 X_{{kjm}}^{\infty}\sim e^{ik_{\infty} r_*}\,,\\
X_{{kjm}}^{r_H}\sim A_{\rm{out}}e^{ik_{\infty} r_*}+A_{\rm{in}}e^{-ik_{\infty} r_*}\,,  
\end{array}\right.
 \qquad r\to \infty
\end{equation}
\begin{equation}\label{boundhor}
\left\{
 \begin{array}{l}
X_{{kjm}}^{\infty}\sim B_{\rm{out}}e^{ik_H r_*}+B_{\rm{in}} e^{-ik_H r_*}\,,\\
X_{{kjm}}^{r_H}\sim e^{-ik_H r_*}\,,
\end{array}\right.
 \qquad r\to r_H\,.
\end{equation}
Here $k_H=\omega-m\Omega_H, k_{\infty}=\sqrt{\omega^2-\mu^2}$, and $\Omega_H\equiv-
\lim_{r\to r_H}g_{t\phi}/g_{\phi\phi}={a}/({r_H^2+a^2})$ is the angular velocity at the horizon of locally nonrotating observers.
The Wronskian of the two linearly independent solutions reads
\begin{equation} \label{wronskian}
W=X_{{kjm}}^{r_H}\frac{dX_{{kjm}}^{\infty}}{dr_*}-X_{{kjm}}^{\infty}\frac{dX_{{kjm}}^{r_H}}{dr_*}=2ik_{\infty} A_{\rm in}\,,
\end{equation}
and it is constant by virtue of the field equations.

Imposing the usual boundary conditions (see Appendix \ref{chapter:appendix1}), i.e.~, only ingoing waves at the horizon and outgoing waves at infinity, Eq.~(\ref{radial1}) can be solved in terms of the Green function~\cite{Detweiler:1978ge}
\begin{equation}\label{greensol}
X_{{kjm}}(r_*)=\frac{X_{{kjm}}^{\infty}}{W}\int_{-\infty}^{r_*}{{T}}_{{kjm}}(r^{\prime})\frac{\Delta r^{\prime n/2}}{(r^{\prime 2}+a^2)^{3/2}}X_{{kjm}}^{r_H}dr_*^{\prime}
+\frac{X_{{kjm}}^{r_H}}{W}\int_{r_*}^{\infty}{{T}}_{{kjm}}(r^{\prime})\frac{\Delta r^{\prime n/2}}{(r^{\prime 2}+a^2)^{3/2}}X_{{kjm}}^{\infty}dr_*^{\prime}\,.
\end{equation}

In the next chapters we shall use these results in two different scenarios: a particle radially infalling into a Schwarzschild black hole in four dimensions, where the particle is coupled to a massive scalar field; and a particle coupled to a massless scalar field in an equatorial circular geodesic around a singly spinning Myers--Perry black hole. 

\cleardoublepage

\chapter{Scalar radiation from an infall of a particle into a Schwarzschild black hole}\label{chap:radial}

In this chapter we will use the results presented in the last chapter in a particular scenario, a particle radially infalling into a Schwarzschild black hole coupled to a massive scalar field. All the formulae derived in the last chapter apply, as long as we use $a=0$ and $n=0$, i.e.~, a non-rotating black hole in the usual four dimensions.

In the context of TeV-scale gravity and gauge/gravity duality scenarios, the signature of these kind of collisions could be extremely relevant. Although we restrict to the non-rotating case and to four dimensions, it would be very interesting to generalize the results derived below to rotating black holes in higher dimensions, and even to non-asymptotically flat geometries.

The use of perturbative techniques to study the energy radiated by an infalling particle into a Schwarzschild black hole goes back to 1970 with the work of Zerilli~\cite{Zerilli:1971wd} and Davis \emph{et al}~\cite{Davis:1971gg}, who first computed the gravitational energy radiated away by a small test particle of mass $m_p$, falling radially, from rest at infinity, into a Schwarzschild black hole of mass $M$. A recent work by Mitsou~\cite{Mitsou:2010jv} has confirmed and improved significantly the numerical accuracy of the results obtained by Davis \emph{et al}. Later, Ruffini generalized these results, allowing the particle to fall with an initial velocity at infinity~\cite{Ruffini:1973ky}. The importance of the infalling particle model lies in the fact that it sometimes appears as a limit case of more general scenarios such as the coalescence of black hole binaries in the extreme mass-ratio limit~\cite{Berti:2010ce}. 
Furthermore, the limit $m_p\to M$ describing the collision of two black holes, do predict reasonable results still within perturbation theory, making perturbation theory a fundamental tool to study important phenomena~\cite{Anninos:1994gp,Berti:2010ce}. 
Recently, all these results where extended to the case of large boost factors by Cardoso and Lemos~\cite{Cardoso:2002ay}, where they considered a massless particle falling radially from infinity.  

One interesting common behavior in this kind of processes, in the case of massless fields, is that the black hole fundamental quasinormal frequency acts as a cutoff in the energy spectra. Furthermore, the signal is always dominated by the quasinormal ringing at intermediate times. Quasinormal modes are the characteristic oscillations of a black hole. They are completely independent of the initial configuration that caused such vibrations. Therefore, they are a characteristic of black holes. The name derive from the similarity between these excitations and normal mode systems, such as the normal modes of a guitar string. However, they are called quasinormal for two reasons: first, they are not stationary modes, since they are exponentially damped; secondly, unlike a normal mode system, quasinormal modes seem to appear only over a limited time interval, at very late times the quasinormal ringing gives way to a power-law falloff. Mathematically, this is related to their incompleteness. For full reviews about quasinormal modes in different black hole spacetimes see for example Refs.~\cite{Kokkotas:1999bd,Berti:2009kk,Konoplya:2011qq}.

Here we will be interested in the energy radiated due to the presence of a massive scalar field. Massive scalar fields present a number of interesting behavior in the presence of black holes. For instance, it is known that the quasinormal decay of massive scalar fields is slower than massless scalar fields, and the greater the mass of the field the slower it decays~\cite{Konoplya:2004wg,Konoplya:2006br}. At a linear level, purely real modes which corresponds to non-damping oscillations can appear, leading to the appearance of infinitely long living modes, the so-called \emph{quasi-resonance} modes~\cite{Ohashi:2004wr,Konoplya:2004wg,Konoplya:2006br}.   
 
Besides that, massive scalar fields seem to behave very differently at very late times. At late times, massless scalar and gravitational fields perturbations of Schwarzschild black holes were shown by Price~\cite{Price:1971fb,Price:1972pw}, to decay according to a power law of the form 
\be   
|\Psi|\sim t^{-(2l+3)}\,,
\ee
where $l$ is the multipole number. Instead, massive fields have oscillatory late-time tails. One of the reasons, is that massive field tails appear already in Minkowsky spacetime. This is related to the fact that different frequencies forming a massive wave packet have different phase velocities. In flat spacetime the late-time tails of the scalar field are given by~\cite{Konoplya:2011qq,Hod:1998ra}
\be   
|\Psi|\sim t^{-(l+3/2)}\sin(\mu t)\,,
\ee
where $\mu$ is the mass of the field. Normally, the Minkowsky spacetime tail shows itself in the black hole tails at intermediate late times. In the presence of a black hole, these intermediate tails are not the final asymptotic behavior. In fact, at very late times, the massive scalar field decays as~\cite{Koyama:2001ee,Koyama:2001qw} 
\be   
|\Psi|\sim t^{-5/6}\sin(\mu t)\,,
\ee
independently of the number $l$. These asymptotic tails are believed to be a resonance backscattering due to the curvature-induced potential. 

Here we show that in a collision of a point particle with a Schwarzschild black hole, the late-time tails due to the curved background, for the lowest radiatable multipoles of the massive scalar field, are dominant even at intermediate late times over all other contributions, namely, the quasinormal ringing and the Minkowsky spacetime tail.

%%%%%%%%%%%%%%%%%%%%%%%%%%%%%%
\section{Zero-frequency limit}
%%%%%%%%%%%%%%%%%%%%%%%%%%%%% 
In the massless scalar radiation case some characteristic features of energetic collisions of point particles with black holes are expected, namely: the spectrum and the waveform largely depend on the lowest quasinormal frequency of the spacetime under consideration which works as a cutoff for the energy spectra; there is a non-vanishing zero-frequency limit (ZFL) for the spectra, whereas for low-energy collisions the ZFL is zero~\cite{Adler:1975dj,Smarr:1977fy,Cardoso:2002ay}.

To understand this, let us do the classic ZFL calculation for head-on collisions~\cite{Adler:1975dj,Smarr:1977fy}, but considering the emission of scalar radiation instead of gravitational radiation. 

The initial configuration consist of one point particle with scalar charge $q_p$ freely moving toward a chargeless particle, with four velocity $u^{\mu}$ and constant positive velocity $v$ corresponding to a boost factor of $\gamma=\frac{1}{(1-v^2)^{1/2}}$. At $t=0$ the two particles collide instantaneously, forming one chargeless particle and emitting scalar radiation. Note that the non conservation of the scalar charge is motivated by the no-hair theorem. Since the initial chargeless particle and the final particle do not contribute to the emitted scalar radiation, the stress-energy tensor for this system is simply
\be\label{zfltensor}
T^{\mu\nu}(\textbf{x},t)=\frac{q_p}{\gamma}\,u^{\mu}u^{\nu}\delta^3(\textbf{x}-\textbf{v} t)\Theta(-t)\,,
\ee
where the boldface denotes a three-vector.

Using eq.~\eqref{fieldeq}, the energy per solid angle and per unit frequency emitted in the wave direction $\hat{\textbf{k}}=\textbf{k}/\omega$ is
\be\label{spectraflat}
\frac{d^2 E}{d\omega d\Omega}=\frac{\omega^2}{4}|T^{\,\lambda}_{\lambda}(\textbf{k},\omega)|^2 \,,
\ee
where the Fourier transform of the stress-energy tensor~\eqref{zfltensor} is given by
\be\label{zflfourier}
T^{\mu\nu}(\textbf{k},\omega)=\frac{q_p\,u^{\mu}u^{\nu}}{2\pi\,i\,\gamma\,
(\omega-\textbf{v}\cdot \textbf{k})}\,.
\ee

For convenience the axes are oriented such that the motion of the initial particles is in the $z$ axis. We then trivially find
\be\label{spectra2}
\frac{d^2 E}{d\omega d\Omega}=\frac{q_p^2}{16\pi^2\,\gamma^2
(1-v\cos\theta)^2}\,.
\ee
where $\theta$ is the angle between the particles motion direction and the wave direction. 
The total energy diverges unless a cutoff frequency $\omega_c$ is introduced, being the total radiated energy given by $\Delta E=\frac{dE}{d\omega}\omega_c$. Numerically we will see that this is indeed the case: the energy spectra for high velocities is approximately flat until a cutoff, which corresponds to the fundamental quasinormal frequency of the black hole. 

In order to compare the ZFL calculations against numerical results we perform a multipolar decomposition of the radiated energy using~\cite{Berti:2010ce} 
\be\label{decom}
\frac{d^2 E}{d\omega d\Omega}=\left(\sum_{lm}\sqrt{\frac{dE_{lm}}{d\omega}}Y_{lm}\right)^2\,,
\ee
where $Y_{lm}$ are spherical harmonics and $\sqrt{\frac{dE_{lm}}{d\omega}}$ are yet undetermined functions of $\omega$. From the orthonormality of the spherical harmonics it follows that
\be\label{spectramodes}
\sqrt{\frac{dE_{lm}}{d\omega}}=\int\,d\Omega\sqrt{\frac{d^2 E}{d\omega d\Omega}}Y_{lm}\,.
\ee
Since we considered a collision along the $z$ axis there is no dependence on the azimuthal angle $\phi$, and thus only the $m=0$ modes contribute. The monopole ($l=0$) contribution then reads
\be\label{specmono}
\frac{dE_{00}}{d\omega}=\frac{q_p^2}{8\pi^2\gamma^2\,v^2
}[\arctan(v)]^2\,.
\ee

In Table~\ref{tab:ZFl} we compare the ZFL for the $l=0$ mode evaluated using Eq.~\eqref{specmono} against $\frac{dE_{0}}{d\omega}|_{\omega=0}$ obtained numerically using the point-particle method discussed in the next section. Our numerical results are in very good agreement with the ZFL predictions.
 
\begin{table}[th!]
%%%%%%%%%%%%%%%%%%%%%%%
\begin{center}
%%%%%%%%%%%%%%%%%%%%%%
\begin{tabular}{c|c|c}
\hline
\hline
$E$ & ${\rm ZFL_{ana}}(\times q_p^{-2})$ & ${\rm ZFL_{num}}(\times q_p^{-2})$\\
\hline 
$1.5$	&	$0.00938$ & $0.00939$\\
$5$	&	$0.00277$ & $0.00276$\\
$10$	&	$0.00115$ & $0.00113$\\
$15$	&	$0.000653$ & $0.000663$\\
\hline
\end{tabular}
\caption[Radial Infall: Comparison between the ZFL of the $l=0$ mode evaluated analytically and numerically, for $E=1.5,5,10,15$.]{\label{tab:ZFl} Comparison between the ZFL of the $l=0$ mode evaluated analytically and numerically, for $E=1.5,5,10,15$. The agreement is remarkable.}
\end{center}
\end{table}

\section{Numerical Setup}
The formalism discussed in the last section is a flat-space approximation valid for the low-frequency part of the energy spectrum. In this section we consider a test particle with scalar charge $q_p$ and gravitational mass $m_p$, and a massive scalar field coupled to matter, falling into a Schwarzschild black hole along a radial timelike geodesic. This is an accurate description at all frequencies in the limit where one of the binary components is much more massive than the other.

%In this section we consider a test particle with mass $m_p$ and a massive scalar field coupled to matter, falling into a Schwarzschild black hole along a radial timelike geodesic. This is an accurate description at all frequencies in the limit where one of the binary components is much more massive than the other.

In this particular case the background metric is the Schwarzschild metric which is given by
\begin{equation}
ds^2=-f(r)dt^2+\frac{dr^2}{f(r)}+r^2(d\theta^2+\sin^2\theta d\phi^2)\,,
\end{equation}
where $f(r)=1-\frac{2M}{r}$. The test particle is described by the stress-energy tensor
\be
T^{\mu\nu}=\int \frac{d\tau}{\sqrt{-g}} q_p\delta^{4}(x-X(\tau))\dot{x}^{\mu}\dot{x}^{\nu}\,,
\ee
where $x^{\mu}$ is the trajectory of the particle along the word-line parameterized by his proper time. 

Using the framework introduced in chapter~\ref{chap:myers}, and specializing for the case of a test-particle radially infalling into a Schwarzschild black hole in four dimensions, we arrive at a wavefunction for the scalar field whose evolution is given by the wave equation
\be\label{radialschwar}
\left[\frac{d^2}{dr_*^2}+(\omega^2-V(r))\right]\tilde{X}(\omega,r)=f(r)S\,,
\ee
where the potential $V$ is given by,
\be
V=f(r)\left(\mu^2+\frac{l(l+1)}{r^2}+\frac{2M}{r^3}\right)\,.
\ee
Note that this equation corresponds to the wave equation~\eqref{radial1}, setting $a=0$, $n=0$ and redefining the source term in a way that it coincides with the literature~\cite{Cardoso:2002ay}.
The source term $S$ depends uniquely on the stress-energy tensor and on the geodesic the particle follows. For massive particles, the radial timelike geodesics can be written as,
\be
\frac{dT}{dr}=-\frac{E}{f(r)\sqrt{E^2-1+2M/r}}\,, \quad \frac{dT}{d\tau}=-\frac{E}{f(r)} \,,
\ee
where $E$ is a conserved energy parameter. If we consider a particle with velocity $v_{\infty}$ at infinity, then, $E=\frac{1}{(1-v_{\infty}^2)^{1/2}}\equiv \gamma$.

For a massive point particle the source $S$ is given by (see Appendix~\ref{chapter:appendix2}),
\be
S=-\frac{q_p}{\sqrt{2\pi}r}Y^*_{lm}(0,0)e^{i\omega T(r)}(\frac{dr}{d\tau})^{-1}\,.
\ee
Here, $Y_{lm}(\theta,\phi)$ are the spherical harmonics and the particle velocity is given by $\frac{dr}{d\tau}=-\sqrt{E^2-1+2M/r}$. 

The energy spectra is given by (see Chapter~\ref{chap:tides}, Eq.~\eqref{flux})
\be\label{spectra}
\frac{dE}{d\omega}=\omega\sqrt{\omega^2-\mu^2}|\tilde{X}(\omega,r)|^2 \,,
\ee
and to reconstruct the wavefunction as a function of the time $t$ we use the inverse Fourier transform
\be
X(t,r)=\frac{1}{\sqrt{2\pi}}\int_{-\infty}^{+\infty}\,e^{-i\omega t}\tilde{X}(\omega,r)d\omega\,.
\ee

To find $\tilde{X}(\omega,r)$ we use the Green's function technique described in chapter~\ref{chap:myers}. Using Eq.~\eqref{greensol} we get, at infinity,
\be\label{waveform}
\tilde{X}(\omega,r\to \infty)=\frac{e^{i\sqrt{\omega^2-\mu^2}\,r_*}}{W}\int_{r_H}^{\infty}X^{r_H}\,S\,dr\,,
\ee
where $X^{r_H}$ is the solution of the homogeneous wave equation with the correct boundary condition at the horizon and $W$ is the wronskian of the homogeneous solutions of Eq.~\eqref{radialschwar} given by~\eqref{wronskian}. We find $A_{\rm in}$ by solving Eq.~\eqref{radialschwar} with the right hand side set to zero, and using the boundary condition given in Eq.~\eqref{boundinf}. Matching the solution to the asymptotic solution at infinity we then find $A_{\rm in}$. For computational purposes, good accuracy is hard to achieve, so we keep higher-order terms of the condition at infinity, using an expansion of the form:
\begin{equation}\label{eq:improvecon}
X^{r_H}\sim e^{i\sqrt{\omega^2-\mu^2} r}r^{\beta}\left(1+\frac{C_1}{r}+\frac{C_2}{r^2}+\dots\right),
\end{equation}
with,
\begin{align}
\beta&=\frac{i M \left(2 \omega ^2-\mu ^2\right)}{\sqrt{\omega ^2-\mu ^2}}\,,\nonumber\\
C_1&=\frac{-i A+i \beta ^2+\beta  \left(8 M \sqrt{\omega ^2-\mu ^2}-i\right)+2 M \left[-\sqrt{\omega ^2-\mu ^2}-2 i M \left(\omega ^2-\mu ^2\right)\right]}{2 (\beta -1)
   \sqrt{\omega ^2-\mu ^2}-2 i M \left(2 \omega ^2-\mu ^2\right)}\,,\nonumber\\
C_2&=\frac{2 M \left[i A-2 i \beta ^2+\beta  \left(4 C_1 \sqrt{\omega ^2-\mu ^2}+3 i\right)-5 C_1 \sqrt{\omega ^2-\mu ^2}-i\right]-i C_1 \left(A-\beta
   ^2+3 \beta -2\right)}{2 (\beta -2) \sqrt{\omega ^2-\mu ^2}-2
   i M \left(2 \omega ^2-\mu ^2\right)}+\nonumber\\
&\frac{M^2 \left[4 (1-2 \beta ) \sqrt{\omega ^2-\mu ^2}-4 i \omega ^2 C_1+4 i \mu ^2 C_1\right]}{2 (\beta -2) \sqrt{\omega ^2-\mu ^2}-2
   i M \left(2 \omega ^2-\mu ^2\right)}\,,
\end{align}
where $A=l(l+1)$.

When doing the numerical integration of~\eqref{waveform}, convergence is hard to achieve. To assure convergence we integrate outward  until a large value of $r$, typically $r=5000/\omega$.

%%%%%%%%%%%%%%%%%%%%%%%%%%%%%%
\section{Numerical Results}
%%%%%%%%%%%%%%%%%%%%%%%%%%%%% 
%In the massless scalar radiation case some characteristic features of energetic collisions of point particles with black holes are expected, namely: the spectrum and the waveform largely depend on the lowest quasinormal frequency of the spacetime under consideration which works as a cutoff for the energy spectra; there is a non-vanishing zero-frequency limit (ZFL) for the spectra, whereas for low-energy collisions the ZFL is zero~\cite{Adler:1975dj,Smarr:1977fy,Cardoso:2002ay}.

\begin{figure*}[p]
\begin{center}
\epsfig{file=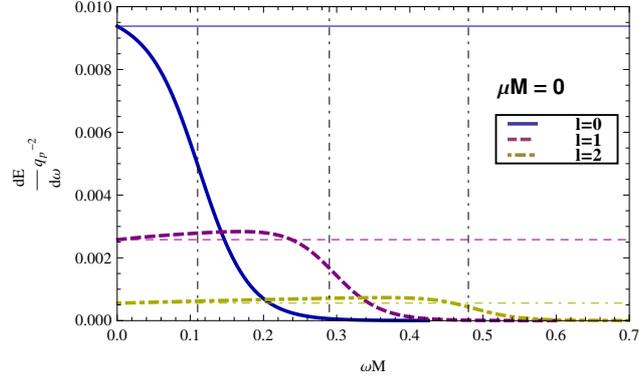,width=8.5cm,angle=0,clip=true}
\caption[Radial infall: Energy spectra of the massless scalar radiation.]{Energy spectra of the massless scalar radiation for the three lowest multipoles, for a massive particle falling from infinity into a Schwarzschild black hole with $E=1.5$. The vertical lines correspond to the real part of the fundamental quasinormal mode for $l=0$, $l=1$ and $l=2$ given respectively by, $M\omega_R=0.11$, $M\omega_R=0.29$, and $M\omega_R=0.48$. The horizontal lines correspond to the ZFL predictions.\label{fig:spectra_mu0}}
\end{center}
\end{figure*}

\begin{figure*}[p]
\begin{center}
\epsfig{file=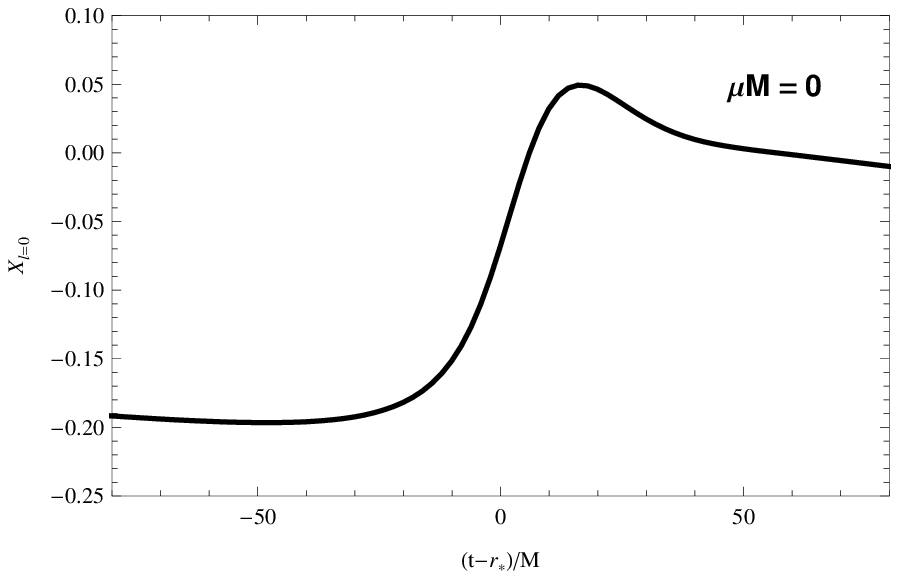,width=7.5cm,angle=0,clip=true}
\epsfig{file=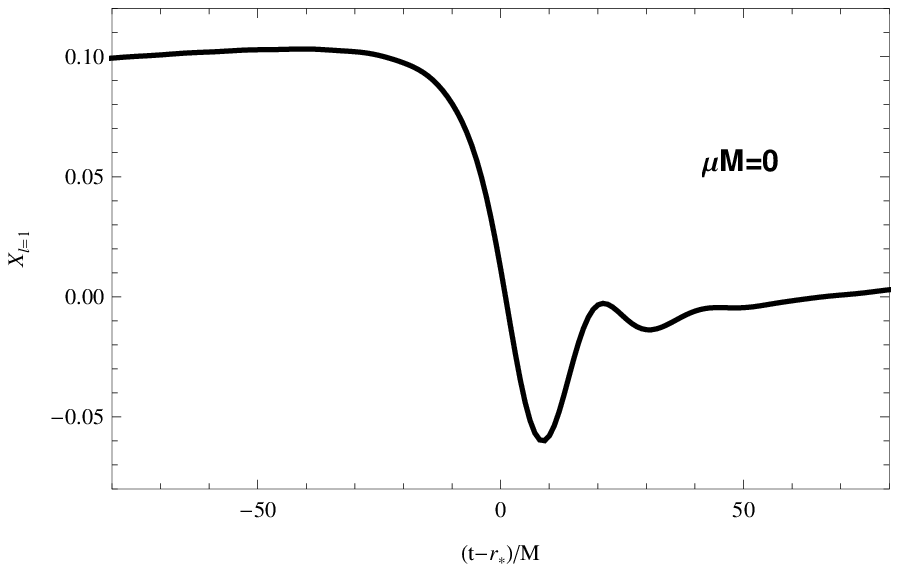,width=7.5cm,angle=0,clip=true}
\epsfig{file=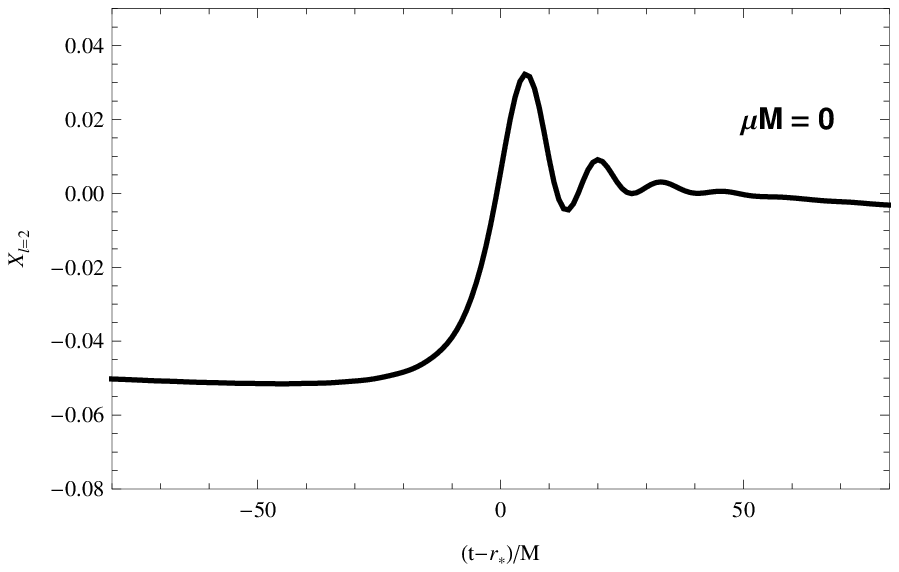,width=7.5cm,angle=0,clip=true}
\caption[Radial infall: Waveforms of the massless scalar radiation.]{Waveforms of the massless scalar radiation for the three lowest multipoles, for a massive particle falling from infinity into a Schwarzschild black hole with $E=1.5$. Here, the wavefunction $X$ is measured in units of $q_p$.\label{fig:waveform_mu0}}
\end{center}
\end{figure*}

In Fig.~\ref{fig:spectra_mu0} we show the energy spectra with an energy parameter of $E=1.5$, for the three lowest values of $l$. As expected the ZFL predictions are in very good agreement with our numerical results and the fundamental quasinormal frequency for each mode act as a cutoff.

The waveform as a function of the retarded time $u\equiv t-r_*$ for the massless case is shown in Fig.~\ref{fig:waveform_mu0}. At early times the wavefunction is not zero, reflecting the fact that the particle begins to fall with a non zero velocity. At late times the signal is dominated by the quasinormal ringing. As expected, our results are very similar to the ones obtained in the work of Ruffini~\cite{Ruffini:1973ky} and Cardoso and Lemos~\cite{Cardoso:2002ay}, where they computed the gravitational radiation emitted in the same process, but with $E\to \infty$.

\begin{figure*}[htb]
\begin{center}
\epsfig{file=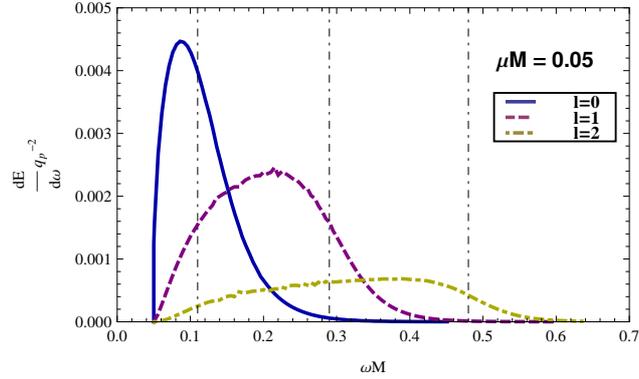,width=8.5cm,angle=0,clip=true}
\caption[Radial infall: Energy spectra of the massive scalar field.]{Energy spectra of the massive scalar field of mass $\mu M=0.05$ for the three lowest multipoles, for a massive particle falling from infinity into a Schwarzschild black hole with $E=1.5$. The vertical lines correspond to the real part of the fundamental quasinormal mode for $l=0$, $l=1$ and $l=2$ given respectively by, $M\omega_R=0.11$, $M\omega_R=0.29$, and $M\omega_R=0.48$.\label{fig:spectra_mu005}}
\end{center}
\end{figure*}

The energy spectra for the massive scalar field ($\mu M=0.05$) case is shown in Fig.~\ref{fig:spectra_mu005}. In this case the field mass acts as lower cutoff since no energy can be radiated for frequencies below the field mass. We can also see that the quasinormal frequency of the black hole acts as a upper cutoff, as in the massless case. Comparing these spectra with the ones obtained for the massless case, we can see that the real part of the quasinormal frequencies of a Schwarzschild black hole for massless and massive scalar perturbations, are very similar, as pointed out in Ref.~\cite{Konoplya:2006br}.

\begin{figure*}[htb]
\begin{center}
\epsfig{file=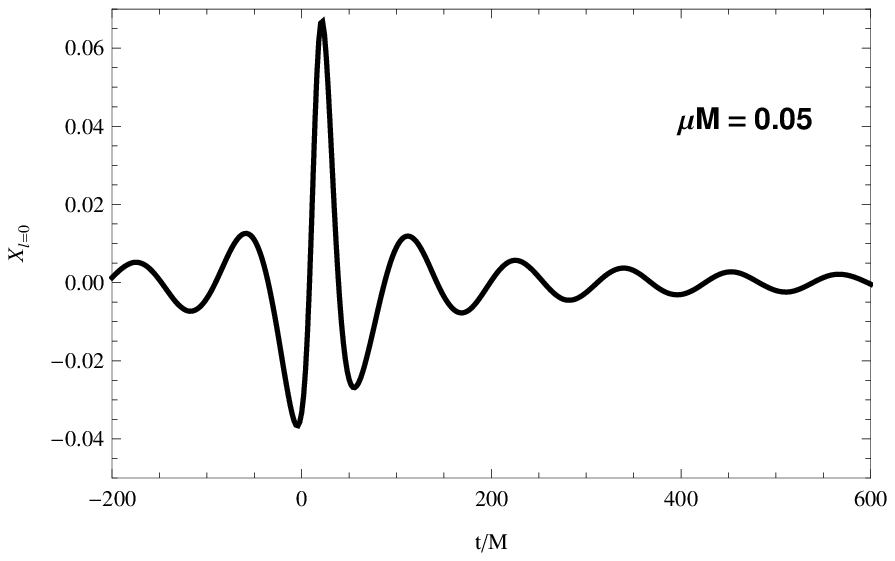,width=7.5cm,angle=0,clip=true}
\epsfig{file=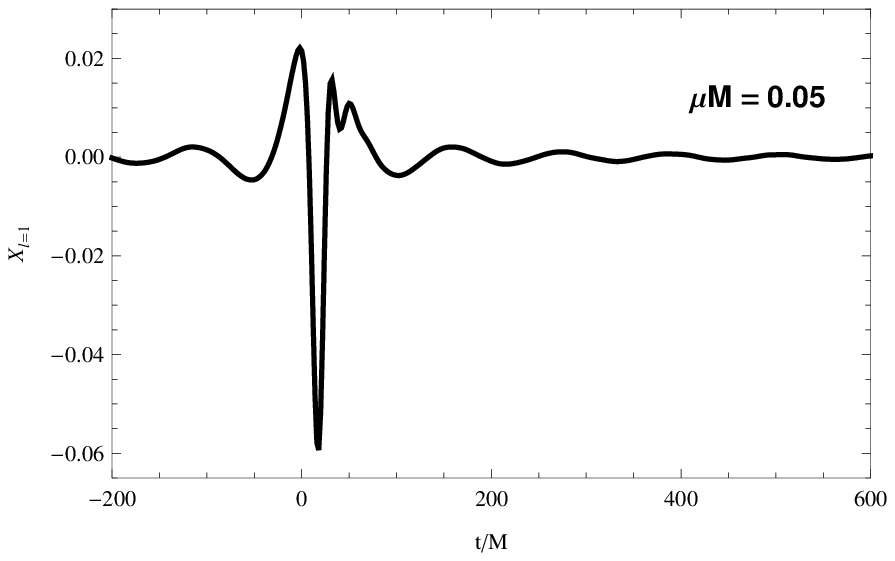,width=7.5cm,angle=0,clip=true}
\epsfig{file=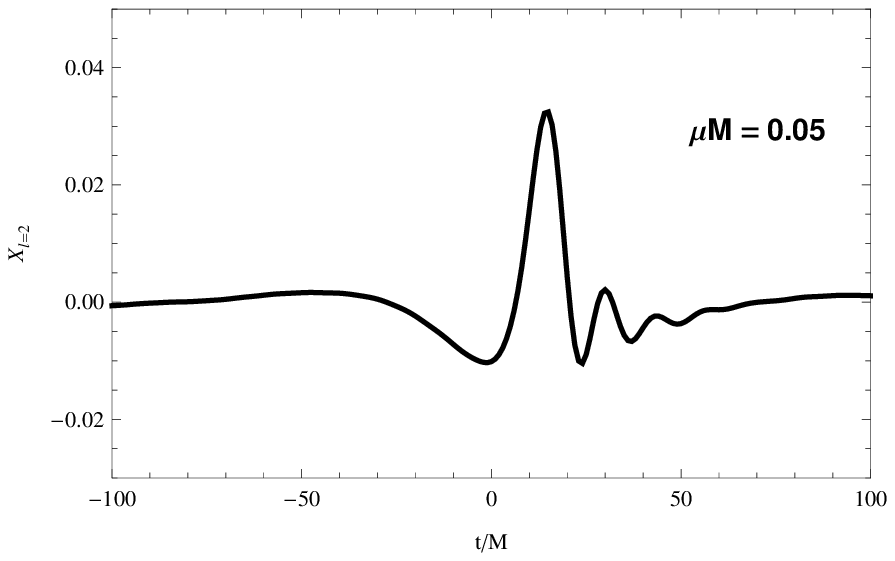,width=7.5cm,angle=0,clip=true}
\caption[Radial infall: Waveforms of the massive scalar radiation.]{Waveforms of the massive scalar radiation of mass $\mu M=0.05$ for the three lowest multipoles at $r_*=10M$, for a massive particle falling from infinity into a Schwarzschild black hole with $E=1.5$. Here, the wavefunction $X$ is measured in units of $q_p$.\label{fig:waveform_mu005}}
\end{center}
\end{figure*}

\begin{figure*}[htb]
\begin{center}
\epsfig{file=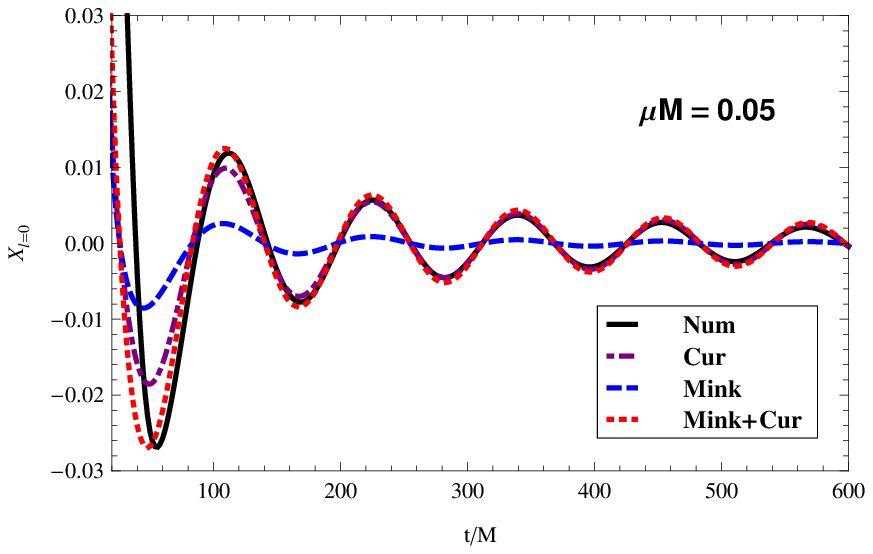,width=7.5cm,angle=0,clip=true}
\epsfig{file=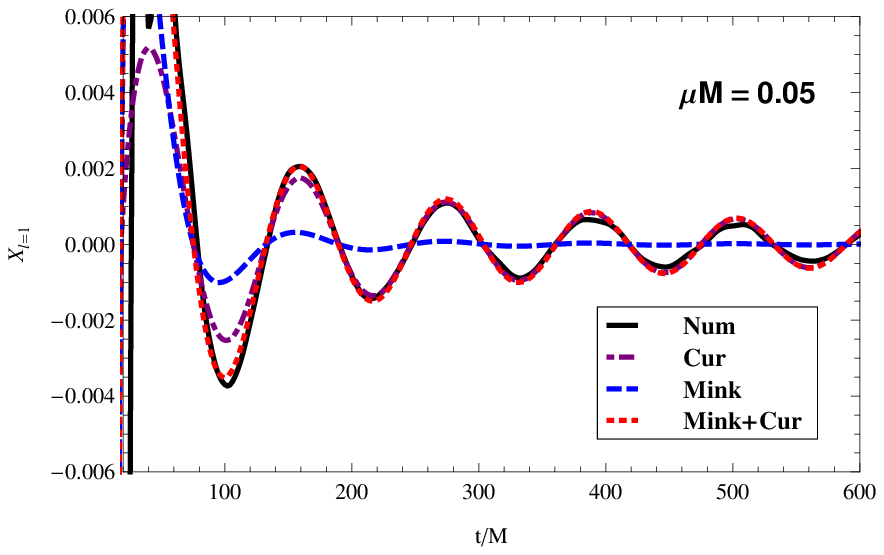,width=7.5cm,angle=0,clip=true}
\caption[Radial infall: Late-time tails of the massive scalar radiation.]{Late-time tails of the massive scalar radiation ($\mu M=0.05$) for the two lowest multipoles. The blue line corresponds to the theoretical contribution of the Minkowsky tail $X\sim t^{-(l+3/2)}\sin(\mu t)$, the purple line is the contribution of the curvature tail $X\sim t^{-5/6}\sin(\mu t)$ and the red line is the sum of the two tails.\label{fig:tail_mu005}}
\end{center}
\end{figure*}

The waveforms for a scalar field of mass $\mu M=0.05$ at a fixed radius $r_*=10M$, are given in Fig.~\ref{fig:waveform_mu005}. For the quadrupolar mode $l=2$, at late times, the signal is clearly dominated by the quasinormal ringing with frequency $\omega\sim 0.48/M$. However in the lowest radiatable modes, $l=0$ and $l=1$, the quasinormal ringing does not seem to appear even at intermediate late times. Instead the signal is dominated by a tail of the form $X\sim t^{-5/6}\sin(\mu t)$, independent of the angular number $l$. This is shown in Fig.~\ref{fig:tail_mu005} where we can see that this tail fits very well the numerical curve at late times. The contribution from the $l$ dependent Minkowsky tail $X\sim t^{-(l+3/2)}\sin(\mu t)$ is also shown, and it is clear that at intermediate late-times this contribution is not negligible. 

It is important to point that the curvature dependent tail $X\sim t^{-5/6}\sin(\mu t)$ is quite universal not only because it does not depend on $l$, but also because it appears also in Kerr black holes~\cite{He:2006jv}, and for other massive fields (Dirac~\cite{Jing:2004zb,He:2006jv} and Proca~\cite{Konoplya:2006gq}). It is thus expected that this behavior is universal for massive fields and does not depend on the details of the black hole horizon geometry~\cite{Konoplya:2006gq}. Therefore, the signal emitted by the lowest multipoles of the massive scalar radiation does not give us much information about the black hole parameters, but can give us information about the field mass. This behavior was also found for other values of the mass $\mu$, thus confirming our results.  

\cleardoublepage

\chapter{Superradiance in $4+n$ dimensions}\label{chap:super}

The phenomenon known by the name of superradiance was first proposed by Zel'dovich in 1971\cite{zeldo1,zeldo2}. He showed that a cylinder made of absorbing material, rotating about its longitudinal axis with frequency $\Omega$, can amplify modes of scalar radiation of frequency $\omega$, when the condition $\omega<m\Omega$ (where $m$ is the azimuthal quantum number) is satisfied. He then pointed out that in a Kerr black hole a similar situation should arise, noting that the gravitational capture of particles and waves at the event horizon replace the absorption occurring in the cylinder. Superradiance is thus a process where certain wave-modes can be amplified by scattering off the rotating black hole at the expense of the hole energy and angular momentum. Press and Teukolsky showed later that the maximum amplification factor, for a Kerr black hole, is finite for scalar, electromagnetic and gravitational radiation and it depends on the spin of the perturbing wave: $0.3$ percent for scalar fields, $4.4$ percent for electromagnetic waves and $138$ percent for gravitational radiation~\cite{teunature,Teukolsky:1974yv}. Furthermore, they showed that there is no classical superradiance for fermions~\cite{Teukolsky:1974yv}. This is to be contrasted with the well known Penrose process~\cite{Penrose:1971nature,Penrose:1969rnc}. In fact, to extract rotational energy from the black hole, the Penrose process only needs some decay within the ergoregion, regardless of the spin of the particles, thus occurring also for fermions.

Superradiance is responsible for many interesting effects (e.g.~\cite{misner,teunature,Cardoso:2004nk,Cardoso:2007az,Cardoso:2011xi,Cardoso:2012zn,Yoshino:2012kn}).
For example, we can state the strong instability of compact rotating objects  without event horizons but where an ergoregion is present~\cite{Cardoso:2007az}. Instabilities develop when the process of superradiance repeats itself \emph{ad infinitum}. For example, a black hole can be made unstable by placing a ``mirror'' around it. The wave will bounce back and fourth between the mirror and the black hole, amplifying at each scattering. The energy extracted will grow exponentially until the radiation pressure destroys the mirror. This process was named \emph{black hole bomb} by Press and Teukolsky~\cite{Cardoso:2004nk,teunature}. One particularly interesting example of this process is the case where we consider a massive scalar field, with mass $\mu$, scattering off a Kerr black hole. In this case, for $\omega\lesssim\mu$, the mass $\mu$ will play the role of the mirror~\cite{Cardoso:2004nk,Cardoso:2005vk,Cardoso:2011xi,Yunes:2011aa,dolan}. Another example where a natural mirror is present is the case of asymptotically AdS spacetimes. In this case, the boundary of the AdS behaves as a mirror and one would expect that in some cases instabilities develop. In fact, it has been shown that small Kerr--AdS black holes are unstable~\cite{Cardoso:2004hs}, but not large Kerr--AdS black holes~\cite{Hawking:1999dp}.
  
We will see in the next chapter that superradiance is also responsible for strong tidal effects around higher-dimensional rotating black holes. Therefore, in the context of this thesis, studying the phenomenon of superradiance in higher-dimensional rotating black holes is an interesting topic. In this chapter we shall use the framework introduced in chapter \ref{chap:myers} to compute the energy amplification of a massless scalar field scattering off a singly spinning Myers--Perry black hole due to superradiance. 
 
\section{Superradiance}

As we have shown in chapter~\ref{chap:myers}, the wave equation describing a massless scalar perturbation of a singly spinning Myers--Perry black hole can be decoupled into a radial, Eq.~(\ref{teuradial}), and an angular equation, Eq.~(\ref{ang}). 

Imagine now a massless scalar wave scattering off the black hole. To see what happens to the wave we must impose a suitable boundary condition close to the event horizon (see Appendix \ref{chapter:appendix1} for a discussion about the correct boundary condition at the horizon). Since classically nothing can escape from the black hole we consider only in-going waves at the horizon,
\begin{equation}\label{eq:horizonwave}
X_{{kjm}}\sim \mathcal{T}e^{-ik_H r_*},\qquad r\to r_H\,,
\end{equation}
where $\mathcal{T}$ is the transmission coefficient. 

At infinity we consider that the solution is given by an incident wave and a reflected out-going wave,
\begin{equation}\label{eq:inftywave}
X_{{kjm}}\sim e^{-i k_{\infty} r_*}+\mathcal{R}e^{i k_{\infty} r_*},\qquad r\to \infty\,,
\end{equation}
where $\mathcal{R}$ is the reflection coefficient.

It can be easily shown that the Wronskian of the solution $X_{{kjm}}$, with the asymptotic behavior described above, and its complex conjugate, which is linearly independent of $X_{{kjm}}$, is constant everywhere by virtue of the field equations.
Thus, using the boundary conditions (\ref{eq:horizonwave}) and (\ref{eq:inftywave}), we obtain after some algebra,
\begin{equation}\label{eq:reflection_coeff}
W(r\to r_H)=W(r\to \infty)\Rightarrow \left|\mathcal{R}\right|^2=1-(1-\frac{m\Omega_H}{\omega})\left|\mathcal{T}\right|^2\,.
\end{equation}
It is evident that superradiance ($\left|\mathcal{R}\right|^2>1$) will occur if,
\begin{equation}\label{eq:consuper}
0<\omega< m\Omega_H\,.
\end{equation}
%Note that this condition is exactly the same as the one obtained from the energy flux at the horizon, Eq.~(\ref{flux}).

\section{Numerical Results}
Here we want to compute the superradiant amplification for various dimensions. To do so, we numerically integrate the homogeneous radial equation~(\ref{teuradial}), with the appropriate boundary conditions at the horizon, Eq.~(\ref{eq:horizonwave}). We adopt $r$ as being the independent variable, avoiding the numerical inversion of $r_*(r)$. We start the integration near the horizon, $r_i=r_H+\epsilon r_H$, where $\epsilon$ is typically $10^{-7}$. Then we integrate outward  until a large value of $r=r_f$, typically $r_f=1000/\omega$. We finally find the reflection coefficient, Eq.~(\ref{eq:reflection_coeff}), matching the numerical solution at $r_f$, to the boundary condition at infinity, Eq.~(\ref{eq:inftywave}).
  
\begin{figure*}[htb]
\begin{center}
\begin{tabular}{cccc}
\epsfig{file=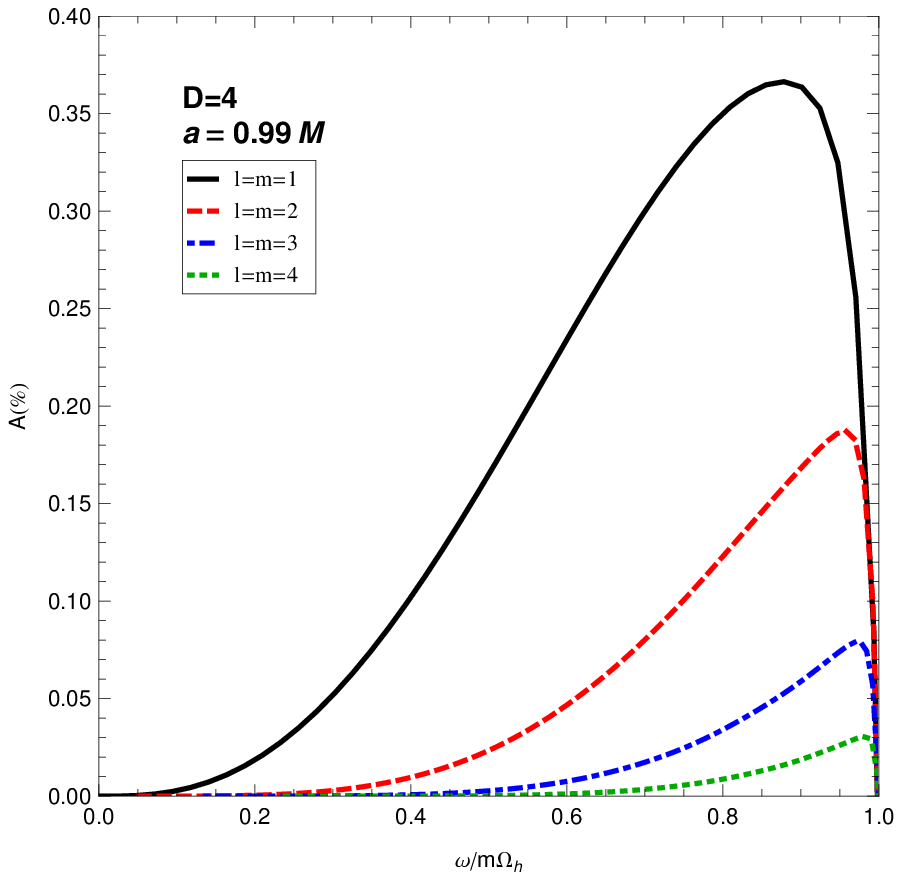,width=7.5cm,angle=0,clip=true}&
\epsfig{file=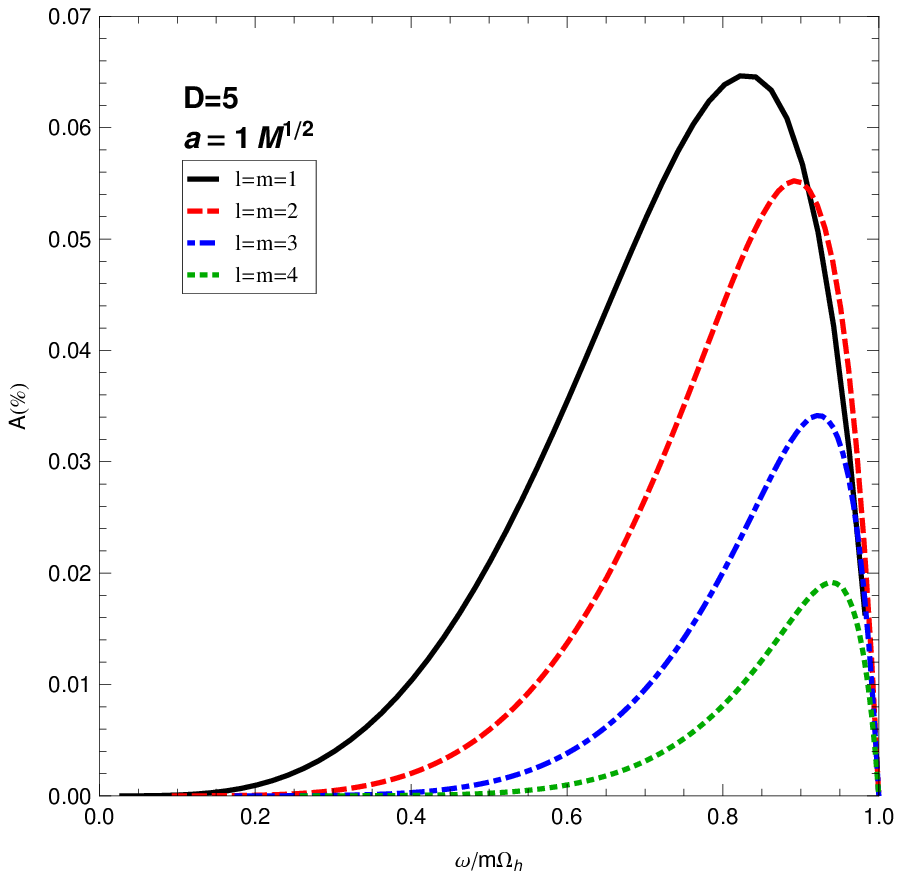,width=7.5cm,angle=0,clip=true}
\\
\epsfig{file=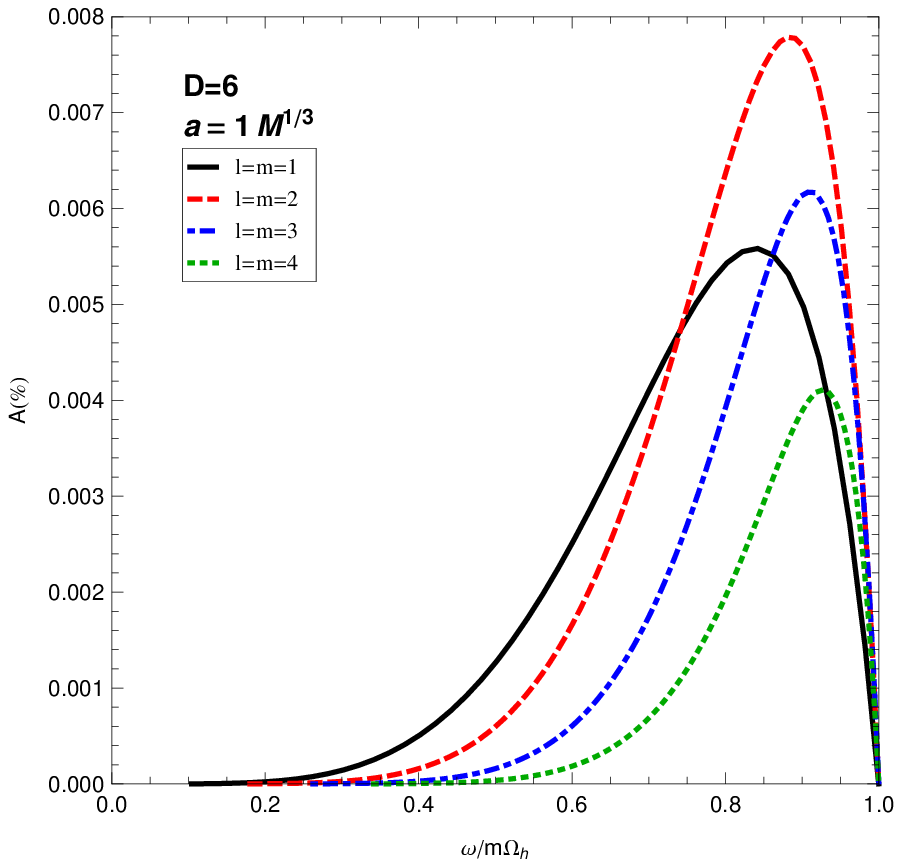,width=7.5cm,angle=0,clip=true}&
\epsfig{file=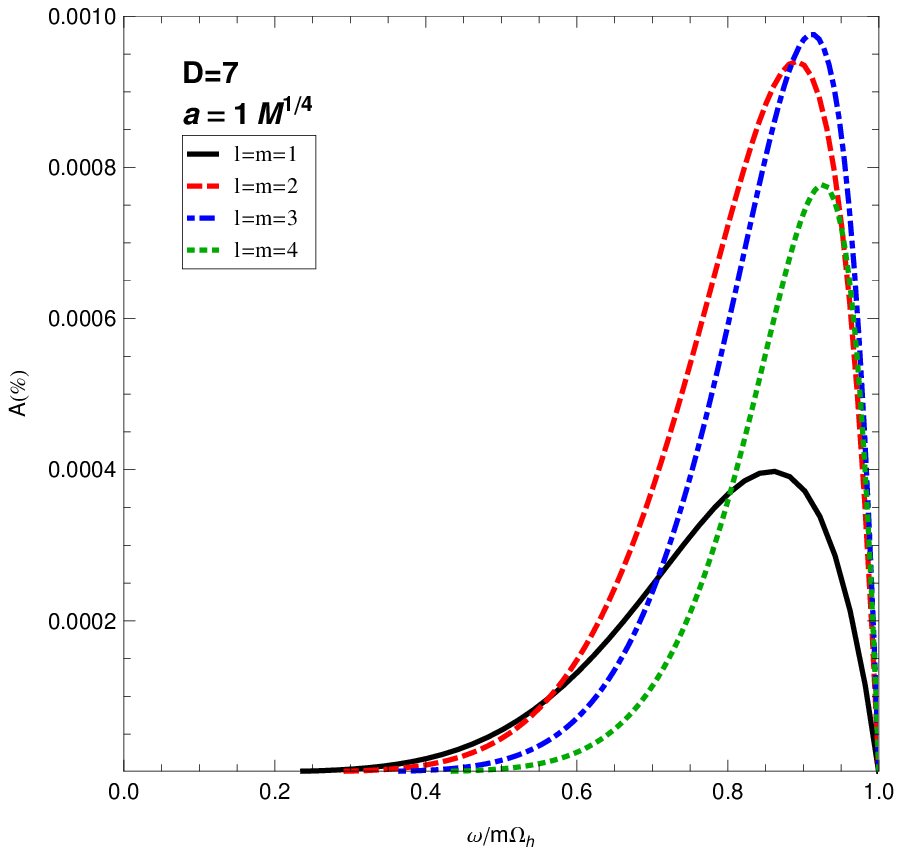,width=7.5cm,angle=0,clip=true}
\end{tabular}
\caption[Superradiance: Amplification factor $A=(1-|\mathcal{R}|^2)\times 100 \%$ as a function of the frequency.]{Amplification factor $A=(1-|\mathcal{R}|^2)\times 100 \%$ as a function of the frequency for several values of the modes $l=m$, and $j=0$.\label{fig:super}}
\end{center}
\end{figure*}

Fig.~\ref{fig:super} shows the amplification factor for several values of the modes $l=m$ and $j=0$, which are the modes where the superradiant amplification is most significant, in different dimensions. Unlike the $D=4$ case, in higher dimensions, the most amplified mode is not necessarily the $l=m=1$ mode. For example, for $a=M^{1/1+n}$, in $D=6$, it is the $l=m=2$ mode, and in $D=7$, the $l=m=3$ mode. This behavior is also seen in dimensions greater than seven and is confirmed by the results obtained in Ref.~\cite{Casals:2008jhep}.  

\begin{figure*}[p]
\begin{center}
\begin{tabular}{cccc}
\epsfig{file=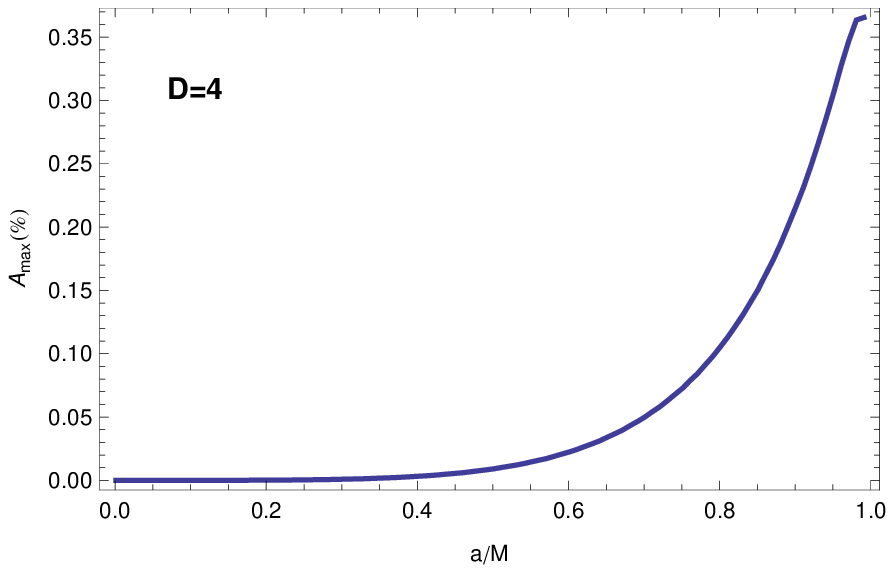,width=7.5cm,angle=0,clip=true}&
\epsfig{file=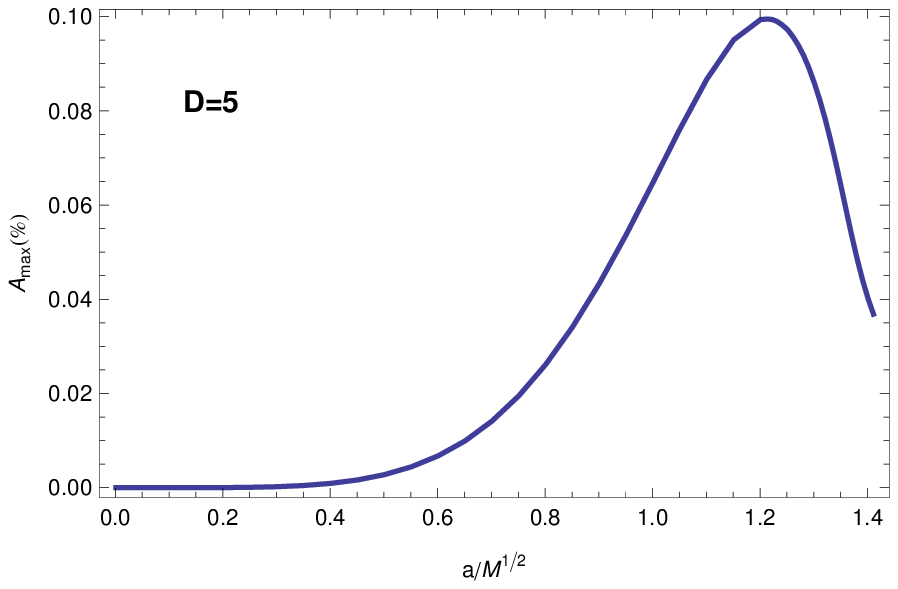,width=7.5cm,angle=0,clip=true}
\\
\epsfig{file=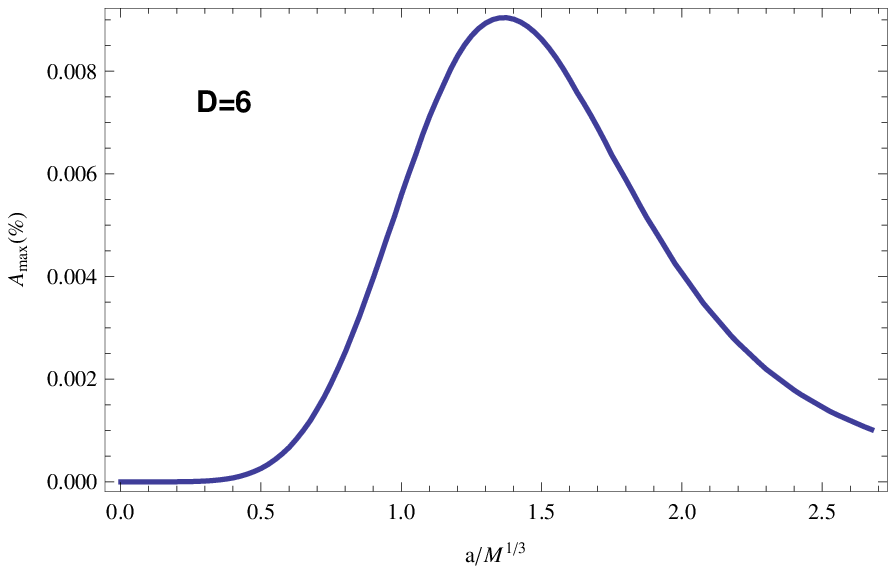,width=7.5cm,angle=0,clip=true}&
\epsfig{file=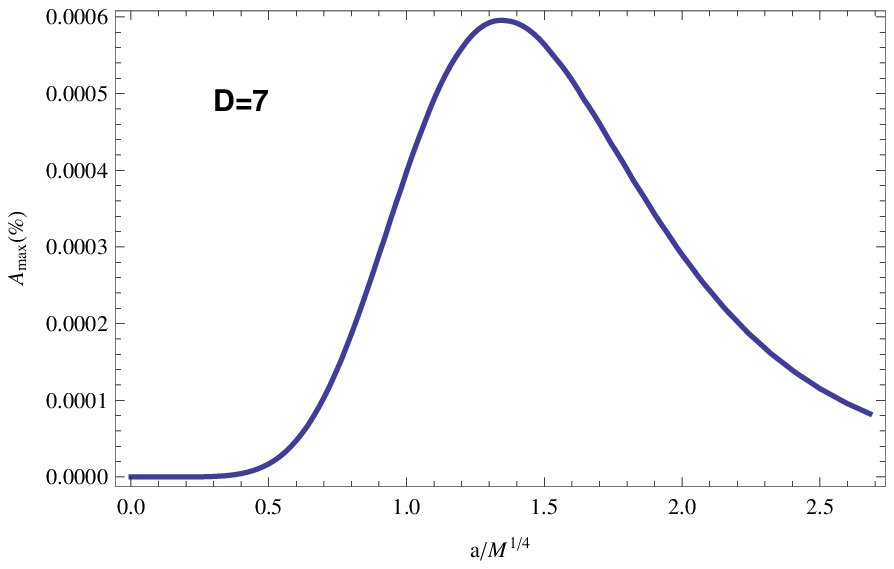,width=7.5cm,angle=0,clip=true}
\end{tabular}
\caption[Superradiance: Maximum amplification factor $A_{\rm max}$ as a function of the spin parameter $a/M^{1/(1+n)}$.]{Maximum amplification factor $A_{\rm max}$ as a function of the spin parameter $a$ for the $l=m=1,j=0$ mode. Top: when D=4 (left), the maximum amplification grows with the spin parameter and approaches $A_{\rm max}\sim 0.366\%$ at $a=0.99M$. When D=5 (right) the maximum amplification grows with the spin parameter until $a\sim 1.2 M^{1/2}$ and then decreases monotonically. Bottom: in $D=6$ (left), and $D=7$ (right) the maximum amplification factor increases monotonically until $a\sim 1.4 M^{1/(n+1)}$ and then decreases for large spins. \label{fig:Amax}}
\begin{tabular}{cccc}
\\
\epsfig{file=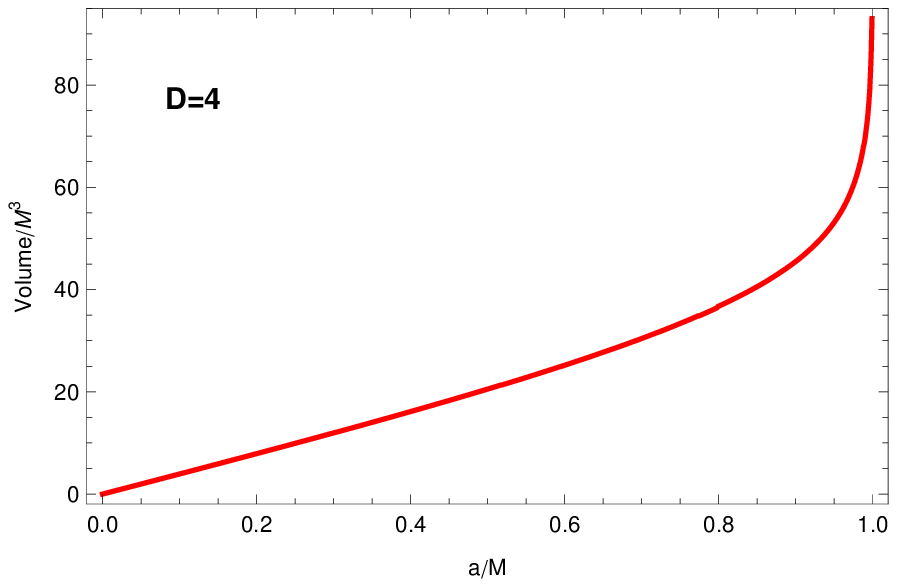,width=7.5cm,angle=0,clip=true}&
\epsfig{file=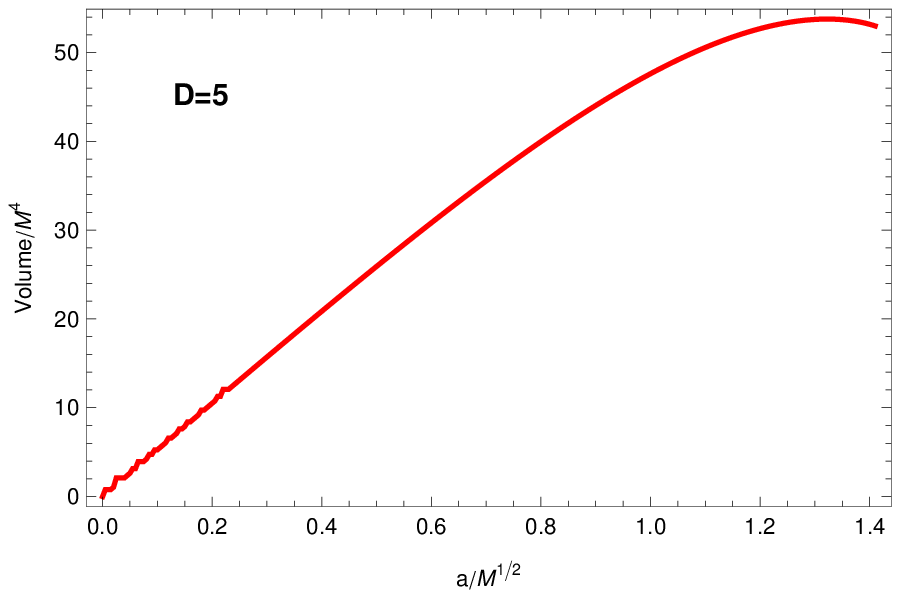,width=7.5cm,angle=0,clip=true}
\\
\epsfig{file=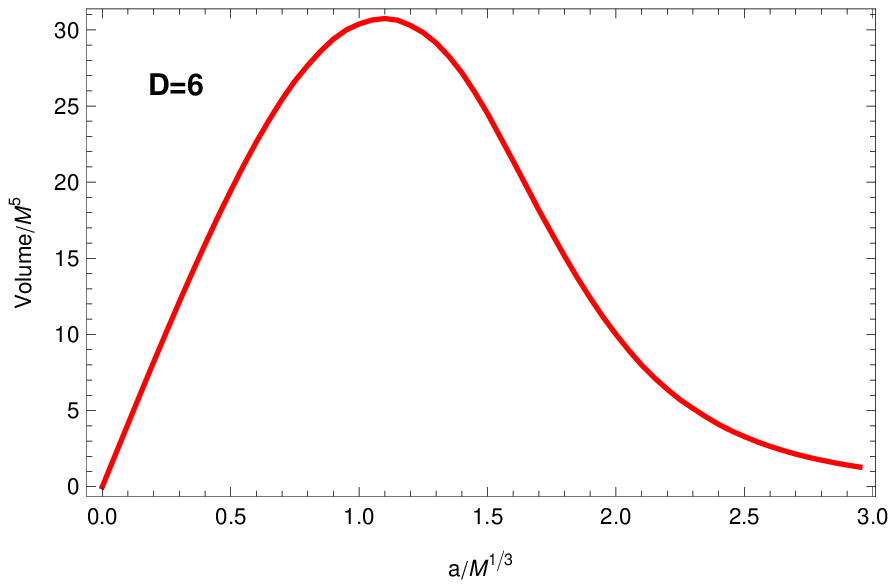,width=7.5cm,angle=0,clip=true}&
\epsfig{file=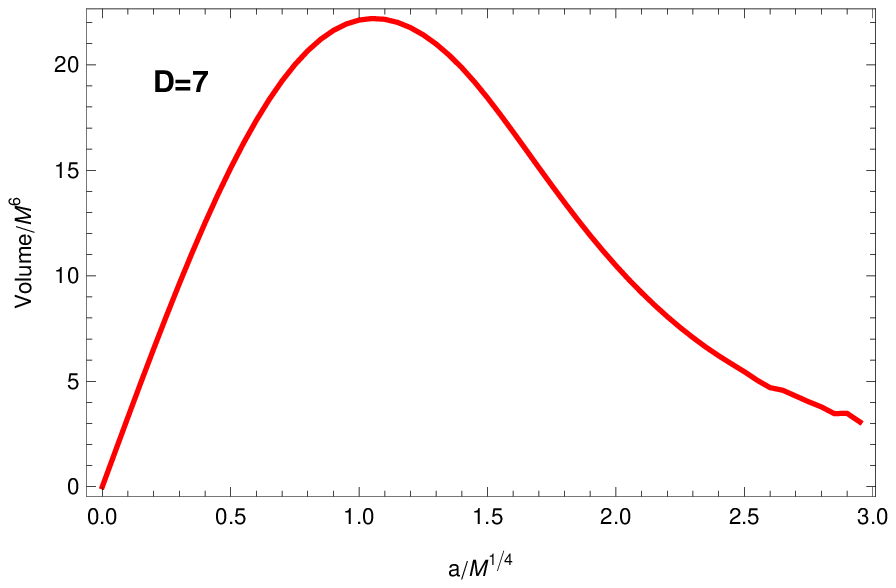,width=7.5cm,angle=0,clip=true}
\end{tabular}
\caption[Superradiance: Proper volume of the ergoregion as a function of the spin parameter $a/M^{1/(1+n)}$]{Proper volume of the ergoregion as a function of the spin parameter $a/M^{1/(1+n)}$.\label{fig:volume}}
\end{center}
\end{figure*}

In Table~\ref{tab:super} we show the maximum amplification factor for the mode $l=m=1$, as well as the frequency and spin at which it occurs for $D=4,5,6,7$. This table shows that the peak of the maximum amplification factor occurs at a finite spin parameter in all dimensions.
%%%%%%%%%%%%%%%%%%%%%%%
 \begin{table}[th!]
%%%%%%%%%%%%%%%%%%%%%%%
\begin{center}
%%%%%%%%%%%%%%%%%%%%%%
\begin{tabular}{c|c|c|c}
\hline
\hline
$n$ & $A_{\rm max}$ & $\omega M^{1/(1+n)}$ & $a/M^{1/(1+n)}$\\
\hline 
$0$	&	$0.366$	& $0.380$ &	$0.99$\\
$1$	&	$0.0995$	& $0.505$ &	$1.21$\\
$2$	&	$0.00907$	& $0.455$ &	$1.37$\\
$3$	&	$0.000596$	& $0.445$ &	$1.37$\\
\hline
\end{tabular}
\caption[Superradiance: Maximum amplification factor of the $l=m=1$ mode, for $D=4,5,6,7$, as well as the correspondent frequency and spin parameter.]{\label{tab:super} Maximum amplification factor of the $l=m=1$ mode, for $D=4,5,6,7$, as well as the correspondent frequency and spin parameter.}
\end{center}
\end{table}
%%%%%%%%%%%%%%%%%%%%%%%%%%%%%%%%%%%
This can be seen in Fig.~\ref{fig:Amax} where we show the maximum amplification as a function of the spin parameter in different dimensions. To evaluate $A_{\rm max}$ for a given $a$, we varied the frequency in order to find a maximum for the amplification, and then repeated this for each value of $a$. The superradiant amplification depends strongly on the dimension. The higher the dimension, the less the wave is amplified. In $D=4$ (upper-left panel), the amplification increases with the rotation of the black hole and approaches the maximum value, $A_{\rm max}\sim 0.366\%$, at $a=0.99M$. In $D=5$ (upper-right panel) the amplification doesn't always grow as it would be naively expected. In fact, it grows until $a\sim 1.2M^{1/2}$, and then decreases until the extremal limit $a=(2M)^{1/2}$. This behavior is also seen in higher dimensions. In $D=6$ and $D=7$ (lower panels), where there is no upper bound on the black hole spin, the amplification factor doesn't increase without limit as we go to large spins. Instead, for large spins the superradiant amplification decreases and eventually becomes negligible.
 
\begin{figure*}[htb]
\begin{center}
\begin{tabular}{cccc}
\\
\epsfig{file=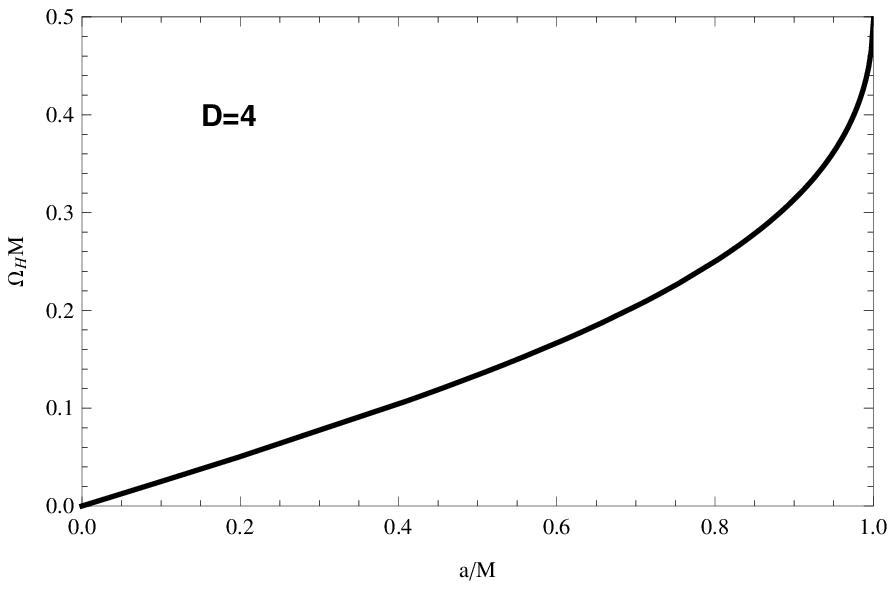,width=7.5cm,angle=0,clip=true}&
\epsfig{file=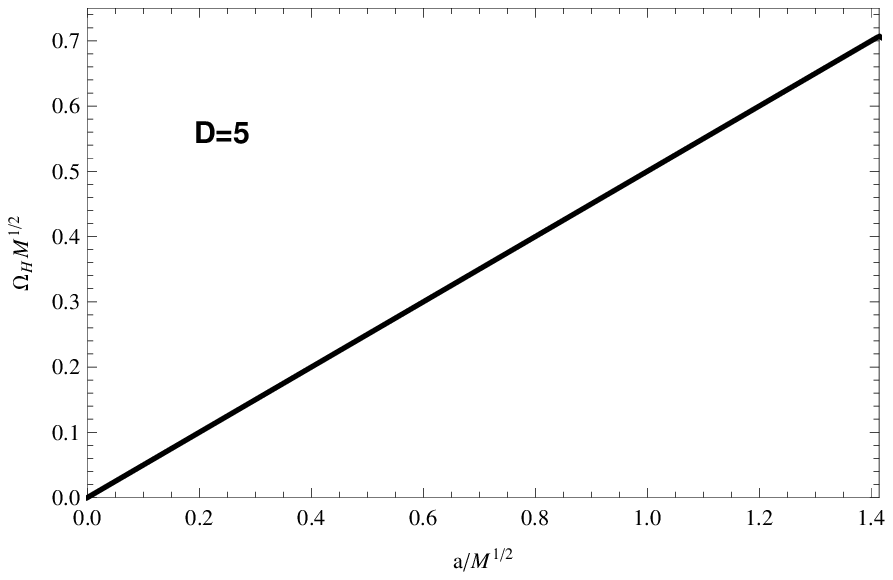,width=7.5cm,angle=0,clip=true}
\\
\epsfig{file=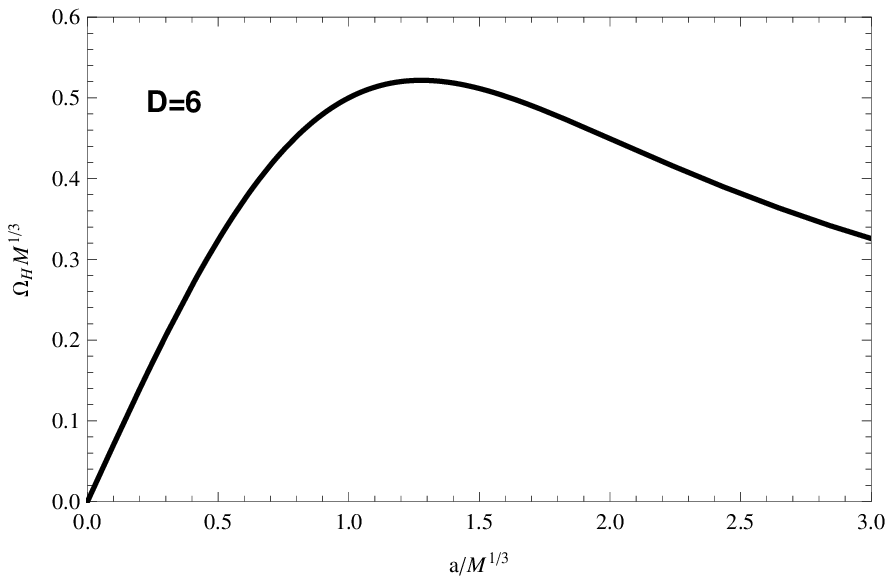,width=7.5cm,angle=0,clip=true}&
\epsfig{file=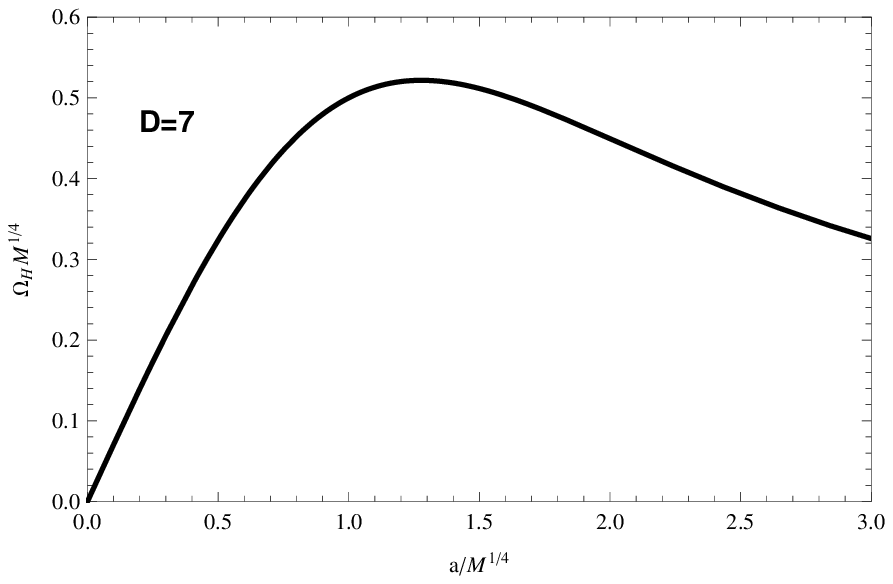,width=7.5cm,angle=0,clip=true}
\end{tabular}
\caption[Superradiance: Angular velocity at the horizon of locally nonrotating observers $\Omega_H={a}/({r_H^2+a^2})$ as a function of the spin parameter $a/M^{1/(1+n)}$]{Angular velocity at the horizon of locally nonrotating observers $\Omega_H={a}/({r_H^2+a^2})$ as a function of the spin parameter $a/M^{1/(1+n)}$.\label{fig:omega}}
\end{center}
\end{figure*}

The behavior of the maximum amplification factor with the black hole spin, can be partially understood computing the proper volume of the ergoregion as a function of the black hole spin. The proper volume can be computed using~\cite{Pani:2010prd},
\be
V=4\pi\int_0^{2\pi} d\theta_n\int_0^{\pi}\prod_{i=1}^{n-1}d\theta_i
\int_0^{\pi/2}d\vartheta\,\int_{r_i}^{r_f}dr\sqrt{g_{rr}g_{\vartheta\vartheta}g_{\phi\phi}
\prod_{i=1}^{n}g_{\theta_i\theta_i}}\,,
\ee
where we have considered a constant time slice, we have used the reflection symmetry of the singly spinning Myers--Perry black hole, and we have already integrated out the $\phi$ dependence. The integration limits, $r_i$ and $r_f$, are set noting that the ergoregion extends between the event horizon, $r_i\equiv r_H$, and the ergosphere radius, $r_f=r_e$; i.e.~, the stationary limit. This occurs where the temporal component of the metric \eqref{Myers-Perry} vanishes. Thus, we can compute $r_e$ solving the equation $g_{tt}(r_e)=0$.
In Fig.~\ref{fig:volume} we show the proper volume of the ergoregion as a function of the spin parameter in different dimensions. In $D=4$ (upper-left panel) the proper volume grows monotonically with $a/M$, diverging for $a=M$. On the other hand, in higher dimensions, the proper volume increases with the black hole spin, for small spins, but eventually reaches a maximum value and then decreases monotonically with $a/M^{1/(1+n)}$. 

Comparing Fig.~\ref{fig:Amax} and Fig.~\ref{fig:volume}, it is evident that there exist a correlation between the proper volume of the ergoregion and the superradiant amplification. This explains why the maximum amplification factor does not grow without limit as we increase the spin. At high spins the ergoregion proper volume goes to zero, constraining the energy extracted from the black hole. In fact, energy extraction from the black hole via superradiance is related to the existence of an ergoregion. Inside the ergoregion negative energy states are possible. If one scatter a wave off the black hole, the wave can excite negative energy modes which will fall into the black hole and extract energy from it. If the proper volume of the ergoregion goes to zero, then the wave will spend less time inside the ergoregion, extracting less energy from the black hole and consequently, the maximum amplification factor will also asymptotically vanish.  

For a matter of completeness, we can also compare the maximum amplification with the angular velocity at the horizon of locally nonrotating observers $\Omega_H={a}/({r_H^2+a^2})$. Comparing Fig.~\ref{fig:Amax} and Fig.~\ref{fig:omega}, there seems to be a better correlation between the superradiant amplification with the proper volume of the ergoregion than with $\Omega_H$. That is evident from the behavior of $\Omega_H$ with the spin parameter in $D=5$. Unlike the maximum amplification it grows linearly with the spin. Moreover, in $D=6$ and $D=7$ we can see that $\Omega_H$ decays slower at large spins than the maximum amplification. 

\begin{figure*}[htb]
\begin{center}
\begin{tabular}{cccc}
\epsfig{file=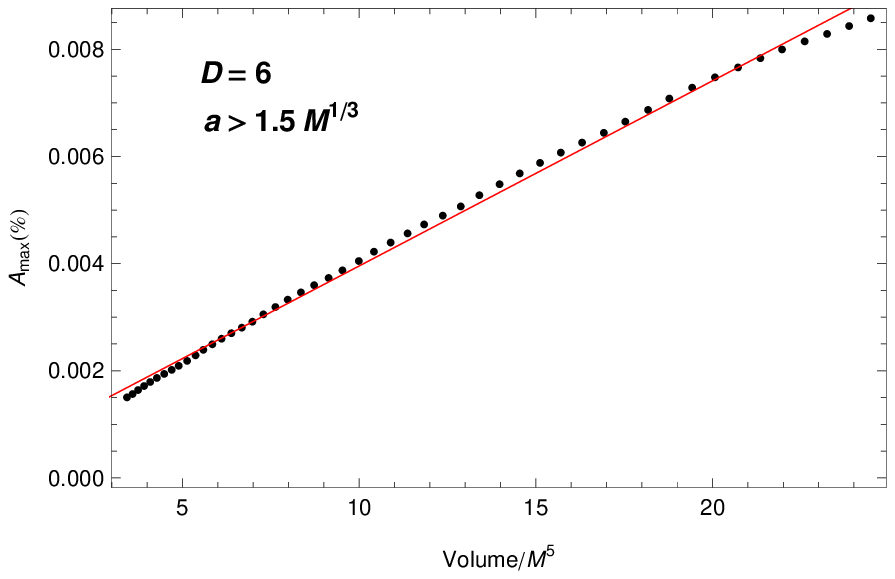,width=7.5cm,angle=0,clip=true}&
\epsfig{file=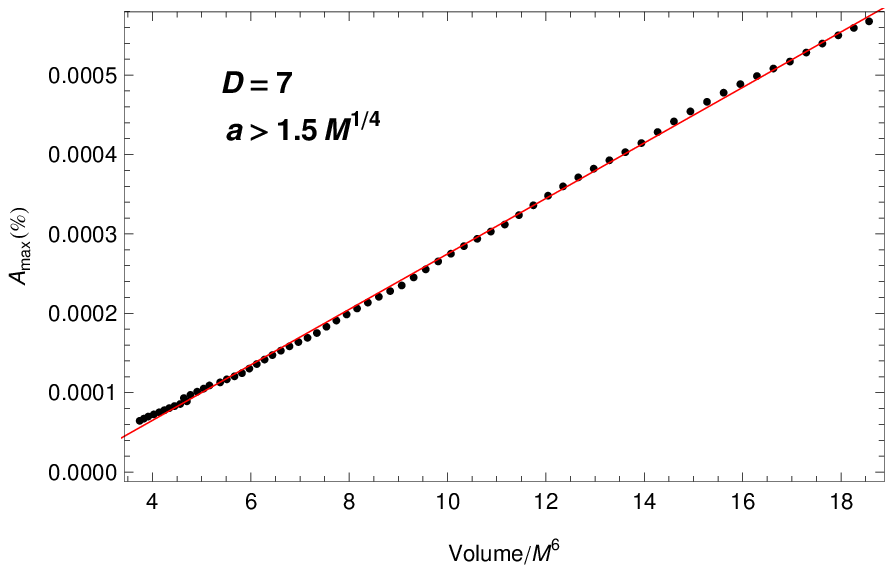,width=7.5cm,angle=0,clip=true}
\end{tabular}
\caption[Superradiance: Maximum amplification factor $A_{\rm max}$ as a function of the ergoregion proper volume  for $a>1.5 M^{1/(1+n)}$]{Maximum amplification factor $A_{\rm max}$ as a function of the ergoregion proper volume  for $a>1.5 M^{1/(1+n)}$. The dots are the numerical results and the straight lines correspond to the fitted curves.\label{fig:Amaxvol2}}
\end{center}
\end{figure*}

An evident functional correlation between the maximum amplification and the ergoregion proper volume is difficult to find, and as we can see comparing Fig.~\ref{fig:Amax} and Fig.~\ref{fig:volume}, the maximum peak of the amplification factor and of the proper volume does not occur at the same value of the spin $a$. However, it is interesting to note that, for $a>1.5M^{1/(1+n)}$, $A_{\rm max}$ grows linearly with the ergoregion proper volume.
In Fig.~\ref{fig:Amaxvol2} we show the maximum amplification factor as a function of the ergoregion proper volume in $D=6$ (left panel) and $D=7$ (right panel), for large spins. Fitting the data to 
\be
A_{\rm max}=a_1+b_1 V\,,
\ee 
we find, for $D=6$,
\begin{equation}
a_1=0.000500012\,,\quad b_1=0.000345673 \,,
\end{equation} 
and, for $D=7$,
\begin{equation}
a_1=-0.000074633\,,\quad b_1=0.000034944 \,.
\end{equation} 
We checked this linear relation between the maximum amplification and the ergoregion proper for large spins up to $D=9$, which led us to say that it should be valid for any dimension.

\cleardoublepage

\chapter{Tides for charged interactions in $4+n$ dimensions}\label{chap:tides}

In the second scenario discussed in this thesis, we consider a particle orbiting a higher-dimensional rotating black hole. We will see that in higher dimensions, strong tidal effects arise and should be crucial to determine the binary evolution.

In fact, gravitational binaries are intrinsically complicated systems that display a wealth of interesting effects. One important effect, which occurs, for example, in several planet-moon systems, are tides generated by differential gravitational forces.
Tidal forces in the Earth--Moon system have long ago locked the Moon in a synchronous rotation with the Earth and have increased the Earth--Moon distance~\cite{darwin,hut}.
These processes, so-called tidal locking and tidal acceleration, respectively, are possible only if there is some dissipation mechanism in the system. Because  of friction, tides extract energy from the binary, and, since angular momentum must be conserved, this provides a mechanism to exchange angular momentum between the Earth and the Moon.
In the Earth--Moon case the dissipation is caused by the friction between the oceans and the Earth surface. In binaries containing rotating black holes the event horizon can play the role of a dissipative membrane, and tidal acceleration is known for many years under a different name: superradiance~\cite{zeldo1,zeldo2,teunature,Cardoso:2012zn}. 

%Consider a Kerr black hole whose angular velocity of the horizon is $\Omega_H$. A wave with frequency 
%$\omega<m\Omega_H$ scattering off the black hole (where m is the azimuthal quantum number) is amplified, and it extracts rotational energy from the black hole.
As stated in the last chapter, superradiance is responsible for many interesting effects, and one of them is the possible existence of ``floating orbits'' around black holes. Generically, orbiting bodies around black holes spiral {\it inward} as a consequence of gravitational-wave emission. 
When the condition for superradiance is met, it is possible to imagine the existence of floating orbits, i.e.~, orbits in which the energy radiated to infinity by the body is entirely compensated by the energy extracted from the black hole~\cite{misner,teunature}. Within general relativity in four dimensions, tidal effects are in general completely washed out by gravitational-wave emission and orbiting bodies always spiral inward \cite{Cardoso:2012zn}. However, when coupling to scalar fields is allowed, an induced dipole moment produces a tidal acceleration (or polarization acceleration \cite{Cardoso:2012zn}), which might be orders of magnitude stronger than tidal quadrupolar effects.
Furthermore, in theories where massive scalar fields are present, the coupling of the scalar field to matter can produce resonances in the scalar energy flux, which can lead to floating orbits outside the innermost stable circular orbit~\cite{Cardoso:2011xi,Yunes:2011aa}.

It was recently argued via a tidal analysis framework that higher-dimensional black holes in general relativity should be prone to strong tidal effects \cite{Cardoso:2012zn}. One of the consequences of those studies was that orbiting bodies around higher-dimensional rotating black holes always spiral \emph{outward}, if the tidal acceleration (or, equivalently, the superradiance) condition is met. 

In this chapter, we use a fully relativistic analysis, albeit in the test-particle limit, to prove this behavior.
For simplicity, we consider the coupling of massless scalar fields to matter around a rotating black hole in higher-dimensional spacetimes. We show that, for spacetime dimensions
$D>5$, tidal effects are so strong, that the energy extracted from the black hole is greater than the energy radiated to infinity. Higher-dimensional spacetimes are of interest in a number of theories and scenarios~\cite{Cardoso:2012qm}; in our case we view them as a proof of principle for strong tidal effects in black hole physics, without the need for resonances.
We do not consider gravitational perturbations; gravitational effects should be subdominant with respect to the dipolar effects discussed here \cite{Cardoso:2012zn}. Nevertheless, the arguments
presented in Ref.~\cite{Cardoso:2012zn} together with the present results show that a purely gravitational interaction also displays this phenomenon, which likely leads to new interesting effects in higher-dimensional black hole physics.

\section{``Polarization'' acceleration}

The flux emitted by a particle orbiting a spinning black hole can be estimated at Newtonian level in terms of black hole tidal acceleration and by applying the membrane paradigm~\cite{Cardoso:2012zn}. In this section, we generalize the computation sketched in Ref.~\cite{Cardoso:2012zn} to higher dimensions and to massless scalar fields.

Let us consider the interaction of a particle with scalar charge $q_p$ and gravitational mass $m_p$ orbiting a neutral central object of mass $M$ and radius $R$. If the object has a dielectric constant $\epsilon=\epsilon_r\epsilon_0$, the particle external field induces a polarization surface charge density on the central object and a dipole moment, which are given, respectively, by~\cite{jackson}
\begin{align}
\sigma_{\rm pol}&=(3+n)\epsilon_0\beta E_0 \cos\vartheta\,,\\
p&=\Omega_{(n+3)}\epsilon_0\beta R^{3+n}E_0\,, \label{dipmom}
\end{align}
where 
\begin{equation}
 E_0= \frac{q_p}{\Omega_{(n+3)}\epsilon_0 r_0^{2+n}}\,,
\end{equation}
and $r_0$ is the orbital distance, $\Omega_{(n+3)}$ is the solid angle of the $(n+3)$-sphere, $\beta$ is some constant that depends on the relative dielectric constant of the object, and $\vartheta$ is the polar angle with respect to the single axis of rotation of the central object. 

Assuming circular orbits, the tangential force on the charge $q_p$ due to the induced electric field is given by
\be 
F_{\vartheta}=\frac{q_p p}{\Omega_{(n+3)} \epsilon_0 r_0^{3+n}}\sin\vartheta\,.
\ee
Without dissipation, the dipole moment would be aligned with the particle's position vector. Here, we consider that dissipation introduces a small time lag $\tau$, such that the dipole moment leads the particle's position vector by a constant angle $\phi$ given by (see~\cite{hut,Cardoso:2012zn} for details)
\be 
\phi=(\Omega-\Omega_H)\tau\,,
\ee
where $\Omega_H$ and $\Omega$ are the rotational angular velocity and the orbital angular velocity, respectively.
At first order in $\phi$, the tangential component of the force reads
\be 
F_{\vartheta}\sim\frac{q_p p}{\Omega_{(n+3)} \epsilon_0 r_0^{3+n}}(\Omega-\Omega_H)\tau\,.
\ee
This exerts a torque $r_0 F_{\vartheta}$ and the change in orbital energy over one orbit reads
\begin{eqnarray}
\dot E_{\rm orbital}&=&\frac{1}{2\pi}\int_0^{2\pi} r_0 F_{\vartheta}{\Omega}\,d\vartheta=\Omega r_0 F_{\vartheta}\nonumber\\
&=&\frac{\beta q^2_p\,R^{3+n}}{\Omega_{(n+3)} \epsilon_0 r_0^{4+2n}}\Omega(\Omega-\Omega_H)\tau\,,
\end{eqnarray}
where, in the last step, we used Eq.~\eqref{dipmom}.

\subsection{``Polarization'' acceleration of black holes}
Remarkably, the equation above qualitatively describes the energy flux across the horizon of a rotating black hole if one identifies $\Omega_H$ with the angular velocity of the black hole and the lag $\tau$ with the light-crossing time, $\tau\sim R/c$, where $R^{1+n}=G_D M/[(n+1)c^2]$, $G_D$ is the $D$-dimensional gravitational constant, and $M$ is the black hole mass~\cite{Cardoso:2012zn}. Accordingly, a particle orbiting a rotating black hole in $4+n$ dimensions dissipates energy at the event horizon at a rate of roughly
\be 
\dot E_{H}=\frac{\beta\,q^2_p\,G_D^{\frac{4+n}{1+n}}}{\Omega_{(n+3)} \epsilon_0 c^{\frac{3(n+3)}{1+n}}(n+1)^{\frac{4+n}{1+n}}}
\frac{M^{\frac{4+n}{1+n}}}{r_0^{4+2n}}\Omega(\Omega-\Omega_H)\,.\label{EdotH}
\ee
On the other hand, charged accelerating particles radiate to infinity according to Larmor's formula, which in $4+n$ dimensions reads~\cite{Cardoso:2007uy} [this will also be derived in the next section, see Eq.~\eqref{anafluxinf}]
\be 
\dot E_{\infty}=\frac{q_p^2\gamma^2}{c^{3+n}}\Omega^{4+n}r_0^2\,,\label{EdotINF}
\ee
where $\gamma$ is some coupling constant.

Tidal acceleration~\cite{Cardoso:2012zn} occurs when the orbit of the particle is pushed outward due to energy dissipation in the central object.
This is possible only if two conditions are satisfied: (i) $\Omega<\Omega_H$, so that $\dot{E}_H<0$ and the energy flows out of the black hole; and (ii) $|\dot E_H|>\dot E_{\infty}$, i.e.~the rate at which energy is dissipated to infinity must be smaller than the rate at which energy is extracted from the black hole. From Eqs.~\eqref{EdotH} and~\eqref{EdotINF}, and using $\Omega\sim r_0^{-(3+n)/2}$~\cite{Cardoso:2008bp}, we find
\begin{align}\label{tideratio}
\frac{|\dot E_H|}{\dot E_{\infty}}&=\frac{\beta G_D^{\frac{4+n}{1+n}}}{\Omega_{(n+3)} \epsilon_0 \gamma^2 c^{\frac{(n+3)(2-n)}{n+1}}(n+1)^{\frac{4+n}{1+n}}}\times\nonumber\\
&\frac{M^{\frac{4+n}{1+n}}}{\Omega^{3+n}r_0^{2(3+n)}}(\Omega_H-\Omega)
\sim \left(\frac{v}{c}\right)^{-\frac{(n-1)(n+3)}{n+1}}\,.
\end{align}
where we have assumed $\Omega_H\gg\Omega$ and we have defined the orbital velocity 
%%%
\begin{equation}
 v=\left[M(n+1)\right]^{\frac{1}{n+3}}\Omega^{\frac{n+1}{n+3}}\,.\label{vel}
\end{equation}

%%%
At large distance, $v\sim r_0\Omega$, and when $n=0$, we recover the standard definition, $v=(M\Omega)^{1/3}$.
Surprisingly, for $n>1$ ($D>5$) tidal acceleration dominates at large distances. This simple argument suggests that test particles orbiting rotating black holes in dimensions greater than five would generically extract energy from the black hole horizon at a larger rate than the energy emitted in gravitational waves to infinity. As a consequence, the orbital separation will increase in time; i.e.~the system will ``outspiral''. Using the framework presented in chapter~\ref{chap:myers}, in the next sections we shall prove this is indeed the case, by computing the linear response of a higher-dimensional spinning black hole to a test particle in circular orbit.

\section{Tidal acceleration: Analytical solution at low frequencies}\label{chap:analytical}
This section contains one of the main results of this thesis. We solve the wave equation, Eq.~\eqref{radial1}, analytically in the low-frequency regime (see e.g.~\cite{poisson1,poisson2,kanti,chen,Creek:2007plb}), obtaining a formula for the scalar flux which will be compared with the results obtained by a direct numerical integration of the wave equation. 

To solve the radial equation in this regime we use a well-known matching technique. We first solve the equation in the \emph{near-horizon} regime, then we find the equivalent in the \emph{far field} limit. Finally, we stretch and match the two solutions in an intermediate zone. This way we can construct an analytic expression for the radial part of the field valid in the entire spacetime. 
 
\subsection{Setup}
Extending the results of the last section, we will restrict to equatorial circular orbits ($\dot\vartheta=0$, $\vartheta=\pi/2$), which is an unrealistic approximation in higher dimensions: generic circular orbits
are unstable, with an instability time scale of order of the orbital period \cite{Cardoso:2008bp}. Nevertheless, our purpose here is to show that tidal effects can dominate, and it is not clear what the overall combined effect of tidal acceleration and circular geodesic motion instability is. Extending the present analysis to generic orbits and relaxing the test-particle approximation are interesting future developments.

For prograde orbits around a singly spinning Myers--Perry black hole~(\ref{Myers-Perry}) the energy, angular momentum, and frequency of the point particle with mass $m_p$ orbiting at $r=r_0$ read~\cite{Cardoso:2008bp}
\be
\frac{E_p}{m_p}=\frac{a \sqrt{(n+1)M}+r_0^{\frac{3+n}{2}}-2Mr_0^{\frac{1-n}{2}}}{r_0^{\frac{3+n}{4}} \sqrt{2 a \sqrt{(n+1)M}+r_0^{\frac{3+n}{2}}-(n+3)Mr_0^{\frac{1-n}{2}}}}\,,\label{Ep}
\ee
\be
\frac{L_p}{m_p}=\frac{\sqrt{(n+1)M} \left(r_0^2-2 a \sqrt{\frac{M}{n+1}}r_0^\frac{1-n}{2}+a^2\right)}{r_0^{\frac{3(n+1)}{4}} \sqrt{2 a \sqrt{(n+1)M}+r_0^{\frac{3+n}{2}}-(n+3)Mr_0^{\frac{1-n}{2}}}}\,,\label{Lp}
\ee
\be
\Omega_p=\frac{\sqrt{(n+1)M}}{a\sqrt{(n+1)M}+r_0^{\frac{3+n}{2}}}\,.\label{Omegap}
\ee
The only nonvanishing components of the $(4+n)$-velocity $U^\nu$ of the particle on a timelike geodesic are given by
%%%
\begin{eqnarray}
 m_p\Delta_{r=r_0} U^t&=& \left(r_0^2+a^2+\frac{2Ma^2}{r_0^{n+1}}\right)E_p-\frac{2Ma L_p}{r_0^{n+1}} \,,\\
 m_p\Delta_{r=r_0}  U^\phi&=& \frac{2Ma E_p}{r_0^{n+1}}+\left(1-\frac{2M}{r_0^{n+1}}\right)L_p
\end{eqnarray}

As seen in chapter~\ref{chap:myers} the nonhomogeneous radial equation for the scalar field reads
\be\label{radial2}
\left[\frac{d^2}{dr_*^2}+V\right]X_{{lmj}}(r^*)=\frac{\Delta}{(r^2+a^2)^{3/2}}r^{n/2}T_{{kmj}}\,,
\ee
where the source term $T_{{kmj}}$ is given, in this case, by
\begin{equation}
T_{{kmj}}=-\frac{q_p\alpha}{U^tr^n}S^*_{lmj}(\pi/2)Y_{j}^{*}(\pi/2,\pi/2,\ldots)
\delta(r-r_0)\delta(m\Omega_p-\omega)\,,
\label{tlmw}
\end{equation}
which has been derived from the stress-energy tensor of the point particle in an equatorial circular geodesic around the black hole (see Appendix~\ref{chapter:appendix2}).
We note that at $\vartheta=\pi/2$ only higher dimensional spheroidal harmonics with $j=0$ are nonvanishing. This can be seen from Eq.~\eqref{zerothorder}. Thus, in order to calculate the fluxes on circular orbits, one only needs to consider terms with $j=0$. In this case, the hyperspherical harmonics $Y_0$ are constant. 

\subsection{Energy Fluxes}
Using the Green's function approach we can derive formulae for the energy fluxes at infinity and at the horizon. 

For very large values of $r$, Eq.~\eqref{greensol} has the following asymptotic form:
\begin{equation}
X_{{kjm}}(r\to\infty)=
\frac{e^{i k_{\infty} r_*}}{2ik_{\infty} A_{{\rm{in}}}} \int_{-\infty}^{\infty} T_{{kjm}}(r^{\prime})X_{{kjm}}^{r_H}\frac{\Delta r^{\prime n/2}}{(r^{\prime 2}+a^2)^{3/2}}dr_*^{\prime}
=Z_{{kjm}}^{\infty}\delta(\omega-m\Omega_p) e^{i k_{\infty} r_*}\,,\label{rlm}
\end{equation} 
where, using Eq.~\eqref{tlmw},  
\be
Z_{{kjm}}^{\infty}=-\alpha\frac{X_{{kjm}}^{r_H}(r_0)}{WU^t}
\frac{S_{kjm}^{*}(\pi/2)Y_{j}^{*}(\pi/2,\pi/2,\ldots)}{\sqrt{r_0^2+a^2}r_0^{n/2}}q_p\,.\label{insol}
\ee
Likewise, at the horizon we get
\be
X_{{kjm}}(r_*\to -\infty)=Z_{{kjm}}^{r_H}\delta(\omega-m\Omega_p)e^{-ik_H r_*}\label{rhor}
\ee
where, 
\be\label{horsol}
Z_{{kjm}}^{r_H}=-\alpha\frac{X_{{kjm}}^{\infty}(r_0)}{W U^t}\frac{S_{kjm}^{*}(\pi/2)Y_{j}^{*}(\pi/2,\pi/2,\ldots)}{\sqrt{r_0^2+a^2}r_0^{n/2}}q_p\,.
\ee
%
%%%%%%%%%%%%%%%%%%%%%%%%%%%%%%%%%%%%%%%%%%%%%%%%%%%%%%%%%%%%%%%%%%%%%%%%%%%
%\subsection*{Energy Fluxes}
%%%%%%%%%%%%%%%%%%%%%%%%%%%%%%%%%%%%%%%%%%%%%%%%%%%%%%%%%%%%%%%%%%%%%%%%%%%

The scalar energy flux at the horizon and at infinity are defined as
\be
\dot{E}_{H,\infty}=\lim_{r\to r_H,\infty} \int d\vartheta d\phi \prod_{i=1}^n d\vartheta_i \sqrt{-g}T^{r}_{t}\,,
\ee
where the stress tensor reads
\be
T_{\mu\nu}=(\nabla_{\mu}\varphi\nabla_{\nu}\varphi^*-\frac{1}{2}g_{\mu\nu}\nabla_{\alpha}\varphi\nabla^{\alpha}\varphi^*)\,.
\ee
Finally, using Eqs.~\eqref{rlm} and \eqref{rhor}, we get
%%%%%
\be\label{flux}
\dot{E}_{H,\infty}=\sum_{kjm}m\Omega_p k_{H,\infty}|Z^{r_H,\infty}_{{kjm}}|^2 \,.
\ee
%%%%%
The equation above shows that, if the superradiant condition $k_H<0$ ($\omega<m\Omega_H$) is met, the energy flux at the horizon can be \emph{negative}; $\dot{E}_{H}<0$, i.e.~energy can be extracted from a spinning black hole~\cite{Teukolsky:1974yv,teunature}. 
Note that this condition is exactly the same as the one obtained from the Wronskian analysis, Eq.~(\ref{eq:consuper}).
In four dimensions, $|\dot{E}_H|\ll\dot{E}_\infty$ and the superradiant extraction is generically negligible. As we show below, in higher dimensions the opposite is true, $|\dot{E}_H|\gg\dot{E}_\infty$ and superradiance dominates over gravitational-wave emission.

\subsection{Solution $X^{r_H}_{{kjm}}$}
\subsubsection*{Near-horizon regime}
Let us first focus on the solution $X^{r_H}_{{kjm}}$, which is regular at the horizon.

We first make the following change of variable: 
\be
h=\frac{\Delta}{r^2+a^2} \Rightarrow \frac{dh}{dr}=(1-h)r\frac{A(r)}{r^2+a^2}\,, 
\ee
where $A(r)=(n+1)+(n-1)a^2/r^2$. Then, near the horizon $r\sim r_H$, the radial Eq.~(\ref{teuradial}) can be written as
\begin{equation}\label{NHrot}
h(1-h)\frac{d^2R}{dh^2}+(1-D_{*}h)\frac{dR}{dh}+
\left[\frac{P^2}{A(r_H)^2 h(1-h)}-
\frac{\Lambda}{r_H^2 A(r_H)^2 (1-h)}\right]R=0\,,
\end{equation}
where 
\begin{align}
&P=\omega(r_H+a^2/r_H)-ma/r_H\,,\\
&\Lambda=[l(l+n+1)+j(j+n-1)a^2/r_H^2](r_H^2+a^2)\,,\\ 
&D_{*}=1-\frac{4a^2 r_H^2}{\left[(n+1)r_H^2+(n-1)a^2\right]^2}\,.
\end{align}
Using the redefinition $R(h)=h^{\alpha}(1-h)^{\beta}F(h)$ the above equation takes the form
\be\label{NH1rot}
h(1-h)\frac{d^2F}{dh^2}+[c-(a+b+1)h]\frac{dF}{dh}-
(ab)F=0\,,
\ee
with 
\be
a=\alpha+\beta+D_{*}-1\,,\hspace{3mm} b=\alpha+\beta\,,\hspace{3mm} c=1+2\alpha\,,
\ee
where $\alpha$ and $\beta$ must satisfy the following algebraic equations:  
\begin{eqnarray}
&& \alpha^2+\frac{P^2}{A(r_H)^2}=0\,,\\
&& \beta^2+\beta(D_{*}-2)+\frac{P^2}{A(r_H)^2}-\frac{\Lambda}{r_H^2 A(r_H)^2}=0\,,
\end{eqnarray}
whose solutions read
\begin{eqnarray}
 \alpha_{\pm}&=&\pm i\frac{P}{A(r_H)}\,,\\
\beta_{\pm}&=&\frac{1}{2}\Bigg[(2-D_{*})\nonumber\\
&&\pm\sqrt{(D_{*}-2)^2-4\frac{P^2}{A(r_H)^2}+4\frac{\Lambda}{r_H^2 A(r_H)^2}}\,\Bigg]\,.
\end{eqnarray}
The two linearly independent solutions of Eq.~\eqref{NH1rot} are $F(a,b;c;h)$ and $h^{1-c}F(a+1-c,b+1-c;2-c;h)$, where $F$ is the hypergeometric function. Convergence requires $\rm{Re}\left[c-a-b\right]>0$, which can be obtained only if the minus sign is chosen in the solutions above. In the following, we shall identify $\beta\equiv\beta_-$ and $\alpha\equiv\alpha_-$.
The general solution of Eq.~(\ref{NHrot}) is then
\begin{equation}
R(h)=A_1 h^{\alpha}(1-h)^{\beta}F(a,b,c;h)
+B_1 h^{-\alpha}(1-h)^{\beta}F(a+1+c,b+1-c,2-c;h)\,.
\end{equation}
Expanding the above result near the horizon, we get
\be
R(h)=A_1 h^{-i\frac{P}{A(r_H)}}+B_1 h^{i\frac{P}{A(r_H)}}=A_1 e^{-ik_H r_{*}}+B_1 e^{ik_H r_{*}}.
\ee
Regularity at the horizon requires $B_1=0$. The near-horizon solution can be written as (see~\cite{handmath})
\begin{align}
&R(h)=A_1 h^{\alpha}(1-h)^{\beta}
\frac{\Gamma[1+2\alpha]\Gamma[2-D_{*}-2\beta]}{\Gamma[2-D_{*}+\alpha-\beta]
\Gamma[1+\alpha-\beta]}F(a,b,a+b-c;1-h)\nonumber\\
&+A_1 h^{\alpha}(1-h)^{2-D_{*}-\beta}
\frac{\Gamma[1+2\alpha]\Gamma[2\beta+D_{*}-2]}{\Gamma[\alpha+\beta+D_{*}-1]
\Gamma[\alpha+\beta]}
F(c-a,c-b,c-a-b+1;1-h)\,.
\end{align}
We can now expand this result in the low-frequency regime and for small values of $a/r_H$, in the region where $1-h\ll 1$ and $r\gg r_H$, 
\begin{equation}\label{NH2rot}
R\sim\frac{X_{kjm}^{r_H}}{r^{1+n/2}}
\sim\frac{r^l}{(2M)^{\frac{2l+n+1}{2(n+1)}}r_H^{\frac{1}{2}}}
\frac{\Gamma[1+2\alpha]\Gamma[2-D_{*}-2\beta]}{\Gamma[2-D_{*}+\alpha-\beta]
\Gamma[1+\alpha-\beta]}\,.
\end{equation}

\subsubsection*{Far-field limit}
We will follow Poisson to solve the wave equation at large distances~\cite{poisson1}. It is useful to rewrite the radial Eq.~(\ref{radial2}) in terms of the dimensionless variable $z=\omega r$. At large distances Eq.~\eqref{radial2} reads
\begin{equation}\label{approxrot}
\left[f\frac{d^2}{dz^2}+\frac{(n+1)\epsilon}{z^{2+n}}\frac{d}{dz}+1
-\frac{l(l+n+1)+\frac{n}{2}(1+\frac{n}{2})}{z^2}
-\frac{\epsilon\left[1+n\left(\frac{n}{4}+1\right)\right]}{z^{3+n}}\right]X_{{kjm}}(z)=0\,,
\end{equation}
where $f=1-\epsilon/z^{n+1}$, $\epsilon=2M \omega^{n+1}$ is a dimensionless parameter, and we used the fact that at large distances the eigenvalues take the form $A_{ljm}=l(l+n+1)$. We can rewrite it in a simpler form if we define the quantum number $J(J+1)=l(l+n+1)+\frac{n}{2}(1+\frac{n}{2})$. Solving for $J$, and assuming $J$ is a non-negative number, we get
\be \label{quantumJrot}
J=l+\frac{n}{2}.
\ee 
In the limit $\epsilon\ll 1$, Eq.~\eqref{approxrot} reads
\be\label{eqauxismallrot}
\left[\frac{d^2}{dz^2}+1-\frac{J(J+1)}{z^2}\right]X_{{kjm}}(z)=0\,.
\ee
The solution can be written in terms of a linear combination of Riccati--Bessel functions, $\sqrt{z}J_{J+1/2}(z)$ and $\sqrt{z}N_{J+1/2}(z)$.
The requirement that $X^{r_H}_{{kjm}}(z)$ be regular at the horizon demands
\be\label{Xhorizonrot}
X^{r_H}_{{kjm}}(z)=B \sqrt{z}\,J_{J+1/2}(z)\,,
\ee
where $B$ is a constant. The asymptotic expansions for the Bessel functions are well known and read 
\be \label{outerrot}
X^{r_H}_{{kjm}}(z \ll 1)\sim \frac{B z^{J+1}}{2^{J+1/2}\Gamma[J+3/2]}\left[1+O(z^2)\right]\,.
\ee
At large distance, the second term within the square brackets is subdominant and we shall ignore it.
Matching~(\ref{outerrot}) to~(\ref{NH2rot}) we get
\begin{equation}
B=\frac{2^{J+1/2}\Gamma[J+3/2]\Gamma[1+2\alpha]\Gamma[2-D_{*}-2\beta]}{\Gamma[2-D_{*}+\alpha-\beta]
\Gamma[1+\alpha-\beta]}\,
\frac{(2M)^{1/(2n+2)}}{\epsilon^{(J+1)/(n+1)}r_H^{1/2}}\left[1+O(\epsilon)\right]\,.
\end{equation}
The parameter $A_{\rm in}$ can be extracted from the behavior of the function near $z=\infty$. Recalling the large argument of the Bessel functions, $X^{r_H}_{{kjm}}(z\rightarrow\infty)\sim B\sqrt{2/\pi}\sin(z-J\pi/2)$, and, using Eq.~(\ref{boundinf}), it follows that 
\begin{equation}
A_{\rm in}=\frac{2^{J}\Gamma[J+3/2]\Gamma[1+2\alpha]\Gamma[2-D_{*}-2\beta]}{\sqrt{\pi}\Gamma[2-D_{*}+\alpha-\beta]
\Gamma[1+\alpha-\beta]}\,
\left(\frac{i}{\epsilon^{1/(n+1)}}\right)^{J+1}\frac{(2M)^{1/(2n+2)}}{r_H^{1/2}}
\left[1+O(\epsilon)\right]\,.
\end{equation}

\subsubsection*{Flux at infinity}
With all of this at hand, we can now compute the flux at infinity in the low-frequency regime. 
From Eqs.~(\ref{insol}) and~(\ref{flux}) we get
\begin{align}
\dot{E}_{\infty}&=m^2\Omega_p^2\left|Z^{\infty}_{{kjm}}\right|^2 \nonumber\\
&=m^{2+2l+n}
\left[\frac{\alpha q_p\sqrt{\pi}}{2^{l+n/2+1}\Gamma[l+n/2+3/2]}\right]^2\,
\left[(n+1)M\right]^{l+n/2+1}
|S_{kjm}(\pi/2)|^2|Y_{j}(\pi/2,\pi/2,\ldots)|^2 \nonumber\\
&\times r_0^{-\frac{2l(n+1)+(n+2)(n+3)}{2}}\,.\label{anafluxinf}
\end{align}
where we used the fact that for small frequencies (large distances) $r^2+a^2\sim r^2$,
$U^t\sim 1$, and $\omega=m\Omega_p\sim m\sqrt{(n+1)M}r_0^{-(3+n)/2}$.

\subsection{Solution $X_{{kjm}}^{\infty}$}
Let us now perform the same calculation for the solution $X_{{kjm}}^{\infty}$, which satisfies outgoing-wave boundary conditions at infinity. The method is analogous to that already described above. However, since for this case the boundary condition is imposed at infinity, we do not require regularity at the horizon. In the limit $\epsilon\ll 1$, $X_{{kjm}}^{\infty}$ can be identified, up to a normalization constant, with
\be
X^{\infty}_{{kjm}}(z)=C\,\sqrt{z}\,H_{J+1/2}^{(1)}(z)\,,
\ee
where $H_{J+1/2}^{(1)}(z)=J_{J+1/2}(z)+iN_{J+1/2}(z)$ is the Hankel function. To determine the constant $C$ we match this solution in the limit $z\rightarrow \infty$ to the required boundary condition
\be 
X^{\infty}_{{kjm}}(z\rightarrow \infty)\sim e^{iz}\,.
\ee
Recalling the asymptotic behavior of the Hankel functions, we have
\be
C \sqrt{z} H_{J+1/2}^{(1)}(z\rightarrow\infty)\sim C \sqrt{\frac{2}{\pi}}e^{iz}[(-i)^{J+1}+O(1/z)]\,,
\ee
from which we get $C=i^{J+1}\sqrt{\pi/2}$.

The small-argument behavior of the Bessel functions reads (when $J>0$)
\begin{align}
J_{J+1/2}(z\ll 1)\sim \frac{z^{J+1/2}}{2^{J+1/2}\Gamma[J+3/2]}\left[1+O(z^2)\right]\,,\nonumber\\
N_{J+1/2}(z\ll 1)\sim -\frac{2^{J+1/2}}{\pi}\frac{\Gamma[J+1/2]}{z^{J+1/2}}\left[1+O(z^2)\right]\,.
\end{align}
At leading order and near $z=0$, the function $N_{J+1/2}$ dominates over $J_{J+1/2}$. Hence, we get
\be 
X^{\infty}_{{kjm}}(z\ll 1)\sim i^J\frac{2^J}{\sqrt{\pi}}\Gamma[J+1/2]z^{-J}\,.
\ee

\subsubsection*{Flux across the horizon} 

We can now compute the flux across the horizon. Using Eqs.~(\ref{horsol}) and~(\ref{flux}) we get 
\begin{align}
\dot{E}_{H}&=m \Omega_p k_H\left|Z^{r_H}_{{kjm}}\right|^2 \nonumber\\
&=m k_H (\alpha\,q_p)^2 \Gamma_1^2 \left(\frac{n+1}{2}\right)^{1/2} r_H(2M)^{\frac{2l+\frac{3n}{2}+\frac{3}{2}}{n+1}}\,
|S_{kjm}(\pi/2)|^2|Y_{j}(\pi/2,\pi/2,\ldots)|^2 \, r_0^{-\frac{4l+5n+7}{2}}\,,
\end{align}
where $\Gamma_1=\frac{\Gamma[l+n/2+1/2]\Gamma[2-D_{*}+\alpha-\beta]
\Gamma[1+\alpha-\beta]}{2\Gamma[l+n/2+3/2]
\Gamma[1+2\alpha]\Gamma[2-D_{*}-2\beta]}$. 

\subsection{Ratio of the fluxes} 

We can now obtain an expression for the ratio of the fluxes on the horizon and at infinity for general $l$, $m$, and $n$. Using the expressions for $\dot{E}_H$ and $\dot{E}_\infty$ calculated above, we find 
\begin{equation}
\frac{\dot E_H}{\dot{E_{\infty}}}=\frac{k_H r_H\left[2^{l+n/2+1}\Gamma[l+n/2+3/2]\right]^2}{\pi m^{2l+n+1}}\Gamma^2_1
\,\left(\frac{2}{n+1}\right)^{\frac{2l+n+1}{2}}
(2M)^{-\frac{(n-1)(2l+n+1)}{2(1+n)}} r_0^{\frac{(n-1)(n+1+2l)}{2}}\,,
\end{equation}
This can be written as a function of the orbital velocity,
\begin{equation}\label{ratioflux}
\frac{\dot E_H}{\dot{E_{\infty}}}=\frac{k_H r_H \left[2^{l+n/2+1}\Gamma[l+n/2+3/2]\right]^2}{\pi m^{2l+n+1}}\Gamma^2_1\,
\left(\frac{2}{n+1}\right)^{\frac{1+2l+n}{1+n}}
\times v^{-\frac{(n-1)(n+1+2l)}{n+1}}\,,
\end{equation}
where we have used Eq.~\eqref{vel} at large distance.
For sufficiently small orbital frequencies, such that the superradiance condition is met, and the flux at the horizon is negative, we then find that the ratio between the fluxes {\it grows} in magnitude with $r_0$ and the particle is tidally accelerated outward.
For the dipolar mode, $l=1$, this expression is in complete agreement with the expected behavior derived from a Newtonian tidal analysis; cf. Equation~(\ref{tideratio}). 

\begin{figure}[htb]
\begin{center}
% \begin{tabular}{c}
\epsfig{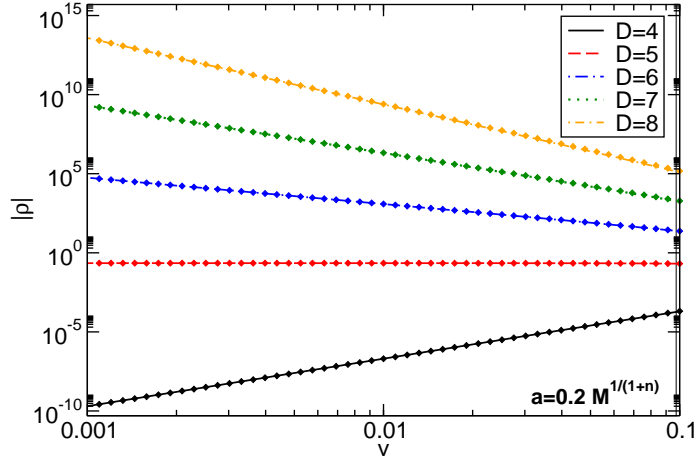}
% \end{tabular}
\caption[Tidal acceleration: Comparison between the flux ratio $\rho=\dot E_H^{\rm Tot}/\dot E_{\infty}^{\rm Tot}$ (in absolute value) calculated analytically and numerically, as a function of the particle velocity $v$ for $n=0,1,2,3,4$ ($D=4,5,6,7,8$) and $a=0.2M^{1/(1+n)}$.]{Comparison between the flux ratio $\rho=\dot E_H^{\rm Tot}/\dot E_{\infty}^{\rm Tot}$ (in absolute value) calculated analytically and numerically, as a function of the particle velocity $v$ for $n=0,1,2,3,4$ ($D=4,5,6,7,8$) and $a=0.2M^{1/(1+n)}$. The straight curves correspond to the analytical formula with $l=1$, and the dots are the numerical results discussed in section~\ref{chap:numerical}. In the slowly rotating regime, numerical results are in very good agreement with the analytical formula. \label{fig:flux_comp}}\end{center}
\end{figure}

Note that these results were derived under the assumption of slow rotation, $a\ll r_H$. This approximation is particularly severe in the near-extremal, five-dimensional case, where $r_H\to0$. Nevertheless, as we discuss in the next section, our method captures the correct scaling of the energy fluxes for \emph{any} spin, and it even gives overall coefficients that are in very good agreement with the numerical ones in the slowly rotating case. This is shown in Fig.~\ref{fig:flux_comp}, where we compare the analytical results of this section with the numerical fluxes computed in the next section.
%

%%%%%%%%%%%%%%%%%%%%%%%%%%%%%%%%%%%%%%%%%%%%%%%%%%%%%%%%%%%%%%%%%%%%%%%%%%%%%
\section{Numerical Results}\label{chap:numerical}
%%%%%%%%%%%%%%%%%%%%%%%%%%%%%%%%%%%%%%%%%%%%%%%%%%%%%%%%%%%%%%%%%%%%%%%%%%%%%
The Green function approach described above can be implemented numerically using standard methods similar to the ones already discussed in the previous chapters and used in Ref.~\cite{Detweiler:1978ge,Cardoso:2011xi,Yunes:2011aa}. For given values of $r_0$, $a$, and $n$, we can compute the fluxes by truncating the sum in Eq.~\eqref{flux} to some $k_{\rm max}$, $m_{\rm max}$, and $j_{\rm max}$. As discussed before, for circular orbits only $j=0$ terms give a nonvanishing contribution. 

For small and moderately large orbital velocities, the sum converges rapidly even for small truncation orders, and we typically set $k_{\rm max}=3$ and $m_{\rm max}=6$. 
However, the convergence is very poor when the orbital velocity approaches the speed of light, i.e.~ when the orbit is close to the prograde null circular geodesic. Recall that circular orbits around Myers--Perry black holes in higher dimensions are unstable~\cite{Cardoso:2008bp} and, in particular, there is no innermost stable circular orbit for $D>4$. Thus, for our purposes we could in principle consider particles in circular orbit up to the light ring, which exists for any dimension~\cite{Cardoso:2008bp}. As the particle approaches the light ring, the flux is dominated by increasingly higher multipoles, thus affecting the convergence properties of the series~\eqref{flux}. For this reason, the plots presented below are extended up to some value of the velocity that guarantees good convergence.

Furthermore, for large orbital velocity and highly spinning black holes, the zeroth order angular eigenfunctions and the corresponding eigenfrequencies~\eqref{Akjm} might not be accurate. Therefore, when $a\omega\gtrsim1$, we have used exact numerical values of $A_{kjm}$ obtained by solving Eq.~\eqref{ang} with the continued fraction
method described in chapter~\ref{chap:myers} and in Ref.~\cite{Berti:2005gp}. We note, however, that Eq.~\eqref{Akjm} reproduces the exact results surprisingly well, even when $a\omega\sim1$.

We checked our method by reproducing the results of Ref.~\cite{Yunes:2011aa} for the massless case in four dimensions. In addition, we can compute the energy flux in any number of dimensions.
The fluxes $\dot E_H$ and $\dot E_\infty$ for $D=5$ and $D=6$ are shown in Tables~\ref{tab:flux1} and~\ref{tab:flux2} for $r_0=10 r_H$. We show the total flux as well as the first multipolar contributions. 

%%%%%%%%%%%%%%%%%%%%%%%
 \begin{table}[th!]
%%%%%%%%%%%%%%%%%%%%%%%
\begin{center}
%%%%%%%%%%%%%%%%%%%%%%
\begin{tabular}{c|c|c|c|c|c|c}
\hline
\hline
$k$ & $m$ & $j$ & $r_0/r_H$ & $\dot E_H(\alpha q_p)^{-2}$ & $\dot E_{\infty}(\alpha q_p)^{-2}$ &	 $|\dot E_H|/\dot E_{\infty}$\\
\hline 
$0$	&	$1$	& $0$ &	$10$	&	$-1.4759 \times 10^{-9}$ & $1.5790 \times 10^{-9}$ & $0.9347$\\
$0$	&	$2$	& $0$ &	$10$	&	$-8.3175 \times 10^{-11}$ & $1.6288 \times 10^{-10}$ & $0.5107$\\
$0$	&	$3$	& $0$ &	$10$	&	$-3.1691 \times 10^{-12}$ & $1.0080 \times 10^{-11}$ & $0.3144$\\
$1$	&	$1$	& $0$ &	$10$	&	$-1.2718 \times 10^{-14}$ & $5.1192 \times 10^{-16}$ & $24.844$\\
\hline
\multicolumn{3}{c|}{$\sum_{kmj}$}   &$10$ & $-3.1248\times 10^{-9}$ & $3.5052\times 10^{-9}$ & $0.8915$\\
\hline
\end{tabular}
\caption[Tidal acceleration: Fluxes across the horizon and to infinity for $n=1$ ($D=5$), $a=M^{1/(1+n)}$, and $r_0=10 r_H$.]{\label{tab:flux1} Fluxes across the horizon and to infinity for $n=1$ ($D=5$), $a=M^{1/(1+n)}$, and $r_0=10 r_H$. In the last row we show the total flux obtained summing up to $k_{\rm max}=3$ and $m_{\rm max}=6$.}
\end{center}
\end{table}
%%%%%%%%%%%%%%%%%%%%%%%%%%%%%%%%%%%
 \begin{table}[th!]
%%%%%%%%%%%%%%%%%%%%%%%%%%%%%%%%%%%
\begin{center}
%%%%%%%%%%%%%%%%%%%%%%%%%%%%%%%%%%
\begin{tabular}{c|c|c|c|c|c|c}
\hline
\hline
$k$ & $m$ & $j$ & $r_0/r_H$ & $\dot E_H(\alpha q_p)^{-2}$ & $\dot E_{\infty}(\alpha q_p)^{-2}$ &	 $|\dot E_H|/\dot E_{\infty}$\\
\hline
$0$	&	$1$	& $0$ &	$10$	&	$-4.1768 \times 10^{-12}$ & $1.3915 \times 10^{-14}$ & $300.155$\\
$0$	&	$2$	& $0$ &	$10$	&	$-3.0159 \times 10^{-13}$ & $3.8070 \times 10^{-16}$ & $792.180$\\
$0$	&	$3$	& $0$ &	$10$	&	$-1.3677 \times 10^{-14}$ & $4.8651 \times 10^{-18}$ & $2811.35$\\
$1$	&	$1$	& $0$ &	$10$	&	$-2.8126 \times 10^{-16}$ & $1.0497 \times 10^{-22}$ & $2.6796\times 10^6$\\
\hline
\multicolumn{3}{c|}{$\sum_{kmj}$}   &$10$ & $-8.9858\times 10^{-12}$ & $2.8601\times 10^{-14}$ & $314.176$\\
\hline
\end{tabular}
\caption[Tidal acceleration: Same as in Table~\ref{tab:flux1} but for $n=2$ ($D=6$).]{\label{tab:flux2} Same as in Table~\ref{tab:flux1} but for $n=2$ ($D=6$).
}
\end{center}
\end{table}
%%%%%%%%%%%%%%%%%%%%%%%%%%%%%%%%%%%%%%%%%%%

\begin{figure}[htb]
\begin{center}
% \begin{tabular}{c}
\epsfig{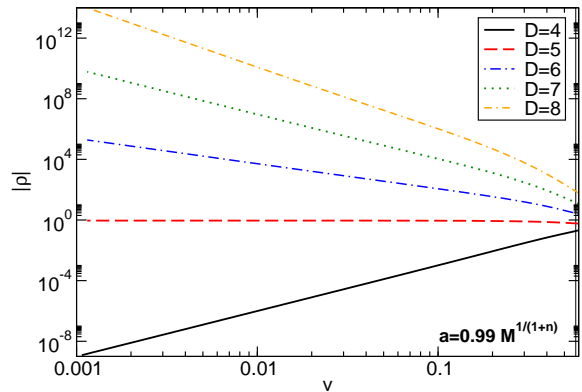}
% \end{tabular}
\caption[Tidal acceleration: The flux ratio $\rho=\dot E_H^{\rm Tot}/\dot E_{\infty}^{\rm Tot}$ (in absolute value) as a function of the particle velocity $v$.]{The flux ratio $\rho=\dot E_H^{\rm Tot}/\dot E_{\infty}^{\rm Tot}$ (in absolute value) as a function of the particle velocity $v$ defined in Eq.~\eqref{vel} for $n=0,1,2,3,4$ ($D=4,5,6,7,8$) and $a=0.99M^{1/(1+n)}$. \label{fig:flux_n}}
\end{center}
\end{figure}

\begin{figure*}[htb]
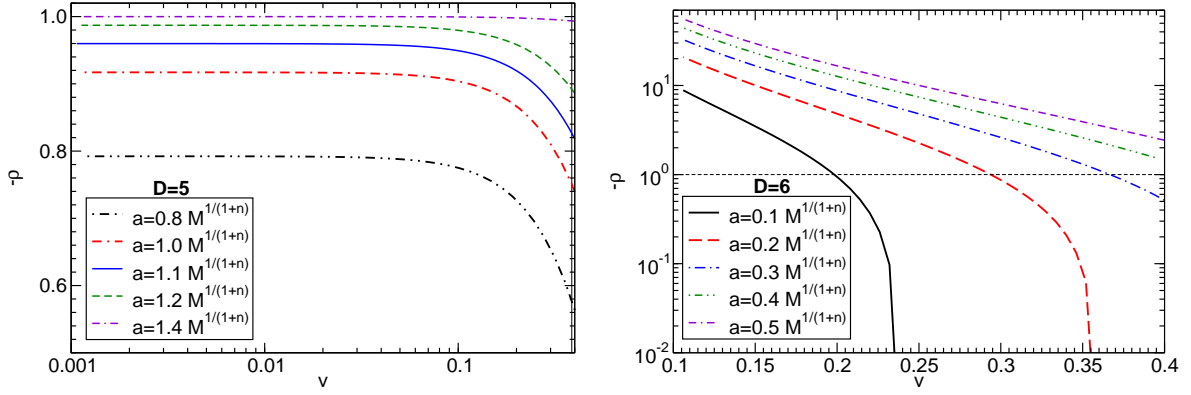

\begin{center}
\begin{tabular}{cc}
\epsfig{file=fluxes_n1.eps,width=7.5cm,angle=0,clip=true}&
\epsfig{file=fluxes_n2.eps,width=7.5cm,angle=0,clip=true}
\end{tabular}
\caption[Tidal acceleration: The ratio $\rho=\dot E_H^{\rm Tot}/\dot E_{\infty}^{\rm Tot}$ as a function of the orbital velocity for several values of $a$.]{The ratio $\rho=\dot E_H^{\rm Tot}/\dot E_{\infty}^{\rm Tot}$ as a function of the orbital velocity defined in Eq.~\eqref{vel} for several values of $a$. Left panel: when $D=5$, the ratio is constant in the small $v$ region, and it approaches unity in the extremal limit, $a\to\sqrt{2M}$. Right panel: when $D=6$, the flux at the horizon can exceed the flux at infinity. For each curve, the intersection with the horizontal line corresponds to a floating orbit, $-\rho=1$. Note that, at large orbital velocity, the superradiant condition is not met and $\dot{E}_H>0$. \label{fig:flux5d6d}}
\end{center}
\end{figure*}

Tables~\ref{tab:flux1} and~\ref{tab:flux2} confirm our analytical expectations that the behavior for $n>1$ ($D>5$) is qualitatively different: the energy flux across the horizon is larger (in modulus) than the flux at infinity. This is shown in Fig.~\ref{fig:flux_n}, where we compare the flux ratio $\rho=\dot E^{\rm Tot}_H/\dot E^{\rm Tot}_{\infty}$ as a function of the orbital velocity $v$ for $a=0.99M^{1/(1+n)}$ in various dimensions. 
This figure is analogous to Fig.~\ref{fig:flux_comp} but for $a=0.99M^{1/(1+n)}$, i.e.~a regime that is not well described by the analytical formula~\eqref{ratioflux}.
For $D=4$, we find the usual behavior; i.e.~the flux at the horizon is usually negligible with respect to that at infinity, and the ratio decreases rapidly at large distance. The case $D=5$ marks a transition, because $\rho$ is constant at large distance. This is better shown in the left panel of Fig.~\ref{fig:flux5d6d}. On the other hand, for any $D>5$ the flux across the horizon generically dominates over the flux at infinity.
%

% %
% \begin{figure}[htb]
% \begin{center}
% % \begin{tabular}{c}
% \epsfig{file=Plots/circ_null.eps,width=7.5cm,angle=0,clip=true}
% % \end{tabular}
% \caption{Radius of the prograde null circular orbit as a function of the rotation parameter in various dimensions. For $D=4,5$ curves terminate at extremality.\label{fig:circ_null}}
% \end{center}
% \end{figure}
% %

In Fig.~\ref{fig:flux5d6d} we show the flux ratio $\rho$ for some selected value of the spin parameter $a$ in five dimensions (left panel) and in six dimensions (right panel). When $D=5$, the ratio is constant in the small $v$ region, and it approaches unity in the extremal limit, $a\to\sqrt{2M}$. As shown in the right panel of Fig.~\ref{fig:flux5d6d}, when $D=6$ there exist some orbital velocity for which $-\rho=1$, corresponding to a total vanishing flux, $\dot{E}_H+\dot{E}_\infty=0$. These orbital frequencies correspond to ``floating'' orbits~\cite{teunature,Cardoso:2011xi}.
Although in the right panel of Fig.~\ref{fig:flux5d6d} this is shown only for $a/M^{1/3}=0.1,0.2,0.3$, we expect this to be a generic feature also for larger values of the spin. The poor convergence properties of the series~\eqref{flux} prevent us from extending the curves to larger values of $v$, where floating orbits for $a>0.3M^{1/3}$ are expected to occur. 
At smaller velocity, the energy flux contribution dominates, and the motion of the test particle is generically dominated by tidal acceleration. Similar results can be obtained for any $D\geq6$.

\cleardoublepage

%%%%%%%%%%%%%%%%%%%%%%%%%%%%%%%%%%%%%%%%%%%%%%%%%%%%%%%%%%%%%%%%%%%%%%%%%%%%%%
\chapter{Conclusions}\label{chap:conclusion}
%%%%%%%%%%%%%%%%%%%%%%%%%%%%%%%%%%%%%%%%%%%%%%%%%%%%%%%%%%%%%%%%%%%%%%%%%%%%%%
The main motivation of this thesis was to study what kind of effects and signatures can arise in black hole spacetimes due to the perturbation of a test particle coupled to a scalar field. The interaction of black holes with fields is an interesting topic by itself and also fundamental for the understanding of astrophysical processes, and scenarios such as the $AdS$/CFT duality and TeV-scale gravity.  

Through the use of black hole perturbation theory, we computed the massive scalar radiation emitted by a test-particle falling radially into a Schwarzschild black hole. The energy spectra suffers a cutoff at the massive scalar field fundamental quasinormal frequency, thus showing the relevance of the quasinormal modes of a black hole. The signal for massless and massive scalar radiation was shown to be quite different. As for the massless case, at intermediate late times, the lowest multipoles of the massive radiation were expected to be dominated by a quasinormal ringing. Instead, the signal was shown to be dominated by an oscillatory tail, which decays slower than any power law. Due to the power-law decay of massless fields at very late-times, this behavior implies that massive perturbations decay slower than any massless perturbation. Furthermore, no information about the black hole parameters can be extracted from the lowest multipoles signal of massive scalar radiation in this kind of processes, since these oscillatory tails are quite universal and do not depend on the black hole geometry.

In the second part of this thesis, we computed the amplification of massless scalar waves scattering off a singly spinning Myers--Perry black hole due to superradiance. This phenomenon also arise in higher dimensions and in dimensions greater than five, where there is no upper limit on the black hole spin, the amplification factor does not increase without limit when we increase the spin parameter. It was then conjectured that this behavior is due to the asymptotic behaviour of the ergoregion proper volume which, as for the amplification factor, goes to zero for very large spins.  

Finally, we concluded with the main result of this thesis, by computing the rate at which the energy is extracted from a singly spinning, higher-dimensional black hole when a massless scalar field is coupled to a test particle in circular orbit. We showed that, for dimensions greater than five and small orbital velocities, the energy flux radiated to infinity becomes negligible compared to the energy extracted from the black hole via superradiance.

Although we considered scalar-wave emission, we expect our results to be generic in higher dimensions. In particular, superradiance should be a dominant effect also for gravitational radiation. At leading order, the ratio $|\dot E_H|/\dot E_{\infty}$ for gravitational radiation should scale with the velocity as described by Eq.~(\ref{ratioflux}). The dominant quadrupole term ($l=2$) reads~\cite{Cardoso:2012zn}
%%%
\begin{equation}
 \frac{|\dot E_H|}{\dot E_{\infty}}\sim v^{-\frac{(n-1)(n+5)}{n+1}}\,.\nonumber
\end{equation}
%%%
By comparing the formula above to Eq.~\eqref{ratioflux} with $l=1$, we note that dipolar effects are dominant over their quadrupolar counterpart. Nevertheless, even in the purely gravitational case, tidal acceleration and floating orbits around spinning black holes are generic and distinctive effects of higher dimensions. 

In principle, gravitational waveforms would carry a clear signature of floating orbits \cite{Cardoso:2011xi,Yunes:2011aa}. Does floating or these strong tidal effects 
have any significance in higher-dimensional black hole physics? We should start by stressing that circular geodesics in higher dimensions are unstable,
on a time scale comparable to the one discussed here~\cite{Cardoso:2008bp}; however, our analysis suggests that, while more pronounced for circular orbits, tidal
acceleration is generic and in no way dependent on the stability of the orbit under consideration. We are thus led to conjecture that tidal effects are crucial to determine binary evolution in higher dimensions. It is possible that tidal effects already play a role in the numerical simulations of the kind recently reported in Refs.~\cite{Witek:2010xi,Okawa:2011fv,Cardoso:2012qm}, but further study is necessary. One of the consequences of our results for those types of simulations is, for instance, that in higher-dimensional black hole collisions the amount of gravitational radiation accretion might play an important role. 
It would certainly be an interesting topic for further study to understand tidal effects for generic orbits, and to include finite-size effects in the calculations.
 
Such effects in higher dimensional spacetimes were discussed in this thesis for the first time. In the formation of micro black holes at the LHC, if ever recorded, strong tidal effects should certainly play a relevant role. This could also be the case in the $AdS$/CFT correspondence. Does this kind of tidal effects exist in $AdS$ backgrounds in $D+1$ dimensions? What other phenomenon could arise do to these strong tidal effects and the presence of massive fields? The understanding of the physics involved when dealing with black holes in four and higher dimensions is far from over and further work in this direction is mandatory.      

\cleardoublepage

\appendix
\chapter{Regularity of the boundary condition at the horizon}
\label{chapter:appendix1}

Physically motivated boundary conditions are imposed by requiring that using well-behaved coordinates at the horizon, i.~e.~, coordinates that remove the singularity at the horizon, we have neither singular nor identically zero fields.

First we note that the Boyer--Lindquist coordinates are singular at the horizon. It takes an infinite coordinate time for any particle to fall into the black hole, $t \to \infty$ as $r\to r_H$, and the world lines are infinitely twisted around the horizon, $\phi\to\infty$ as $r\to r_H$. We can remove the coordinate singularity transforming the Boyer--Lindquist coordinates to the so-called Kerr "ingoing" coordinates~\cite{gravitation}. In place of $t$ and $\phi$ we introduce the new variables,
\begin{align}\label{eq:kerrvar}
dv&=dt+\frac{r^2+a^2}{\Delta}dr,\nonumber \\
d\tilde{\phi}&=d\phi+\frac{a}{\Delta}dr.
\end{align}
By a direct substitution of the relations (\ref{eq:kerrvar}) into the line element (\ref{Myers-Perry}), the Myers--Perry metric in these new coordinates can be brought to the form
\begin{align}
ds^2=&-\frac{\Delta-a^2\sin^2\vartheta}{\Sigma}dv^2+2dvdr
-\frac{2a(r^2+a^2-\Delta)\sin^2\vartheta}{\Sigma}dvd\tilde{\phi} + \frac{(r^2+a^2)^2-\Delta a^2\sin^2\vartheta}{\Sigma}\sin^2\vartheta d\tilde{\phi}^2\nonumber\\
&-2a\sin^2\vartheta drd\tilde{\phi}
+\Sigma d\vartheta^2+r^2\cos^2\vartheta d\Omega_n^2\,.
\end{align} 
This metric is clearly non-singular at the horizon, but it is singular at $\Sigma=0$, or equivalently at $r=0$ and $\phi=\pi/2$, which is the only true singularity of the Kerr--Myers--Perry space-time.

Under these new coordinates the scalar field equation is separable using the ansatz
\begin{equation}\label{eq:ansatz3}
\varphi(v,r,\vartheta,\tilde{\phi})=\sum_{l,m,j}\int d\omega e^{im\tilde{\phi}-i\omega v}\tilde{R}(r)S_{lmj}(\vartheta)Y_j\,.
\end{equation}
Inserting the Kerr coordinates in the form,
\begin{align}
v&=t+\int\frac{r^2+a^2}{\Delta}dr\,,\nonumber \\
\tilde{\phi}&=\phi+\int\frac{a}{\Delta}dr\,,
\end{align}
into (\ref{eq:ansatz3}), the relation between $R(r)$ and $\tilde{R}(r)$ is easily obtained,
\begin{align}
&\sum_{l,m,j}\int d\omega\, e^{im\phi-i\omega t}R(r)S_{lmj}(\vartheta)Y_j=
\sum_{l,m,j}\int d\omega\, 
e^{im\left(\tilde{\phi}-\int\frac{a}{\Delta}dr\right)-i\omega\left(v-\int\frac{r^2+a^2}{\Delta}dr\right)}
R(r)S_{lmj}(\vartheta)Y_j=\nonumber\\
=&\sum_{l,m,j}\int d\omega\, e^{im\tilde{\phi}-i\omega v}R(r)e^{i\int\frac{(r^2+a^2)\omega-am}{\Delta}dr}S_{lmj}(\vartheta)Y_j
\Rightarrow \nonumber \\
\Rightarrow &\tilde{R}(r)=R(r)e^{i\int\frac{(r^2+a^2)\omega-am}{\Delta}dr}\,.
\end{align}

At the horizon we can have either ingoing our outgoing waves: 
\begin{equation}
R\sim e^{\pm ik_H r_*},
\end{equation}
where $k_H=\omega-m\Omega_H$. Requiring the regularity of $\tilde{R}(r)$ around $r=r_H$ lead us to the correct boundary condition:
\begin{equation}
\tilde{R}(r\to r_H)\sim e^{\pm ik_H r_*}e^{i\int\frac{(r_H^2+a^2)\omega-am}{\Delta}dr}\approx e^{i\int\frac{\pm(r_H^2+a^2)k_H+(r_H^2+a^2)k_H}{\Delta}dr},
\end{equation}
where we have used the definition of the tortoise coordinate, $dr/dr_*=\Delta/(r^2+a^2)$.
Since $\Delta\to 0$ as $r\to r_H$, we must have $\pm(r^2+a^2)k_H+(r^2+a^2)k_H\to 0$ as $r\to r_H$, to assure the regularity of the solution. The regular boundary condition is thus,
\begin{equation}\label{eq:correctwave}
R\sim e^{-ik_H r_*}.
\end{equation}

Another way to see that (\ref{eq:horizonwave}) represent ingoing waves at the horizon, is checking the radial group velocity of the wave. For a well-behaved observer at the horizon the group velocity must be negative; i.e. he must see the wave progressing toward the black hole. 

The group and phase velocity of the regular solution (\ref{eq:correctwave}), close to the horizon, are given, respectively, by
\begin{align}\label{eq:group}
v_{group}&=-\frac{d\omega}{dk_H}=-1\,,\\
\label{eq:phase}
v_{phase}&=-\frac{\omega}{k_H}=-\left(1-\frac{m\Omega_H}{\omega}\right)^{-1}\,.
\end{align}
We see that the group velocity (\ref{eq:group}) is negative, which means that for any local observer the waves are progressing towards the horizon. However, for an observer at infinity, $v_{phase}$ is the relevant velocity. We see by (\ref{eq:phase}) that if $\frac{m\Omega_H}{\omega}>1$ (the superradiant frequencies), the wave will appear as emerging from the black hole. For the observer at infinity the wave is amplified after scattering off the black hole.
\cleardoublepage

\chapter{Derivation of the source terms $T_{lmj}$ and $S$}
\label{chapter:appendix2}

\section*{Source $T_{lmj}$}
The trace of the stress-energy tensor for a point particle in an equatorial circular orbit can be written in the form 
\be
T=-\frac{q_p}{\sqrt{-g} U^t}\delta(r-r_0)\delta(\vartheta-\pi/2)\delta(\phi-\Omega_p t)\prod_{i=1}^n \delta(\theta_i-\pi/2)\,,\label{te}
\ee
where, using the line element \eqref{Myers-Perry}, the square root of the metric determinant takes the form
\be\label{metricdet1}
\sqrt{-g}=\Sigma\sin\vartheta r^n\cos^n\vartheta\prod_{i=1}^{n-1}\sin\theta_i^{n-i}.
\ee

To separate the wave equation, Eq.~\eqref{fieldeq}, we define   
\be\label{eq:ansatz2}
\varphi(t,r,\vartheta,\phi)=\sum_{l,m,j}\int d\omega e^{im\phi-i\omega t}R(r)S_{lmj}(\vartheta)Y_j\,,
\ee
and
\be
\alpha \Sigma T=\sum_{l,m,j}\int d\omega e^{im\phi-i\omega t}T_{lmj}S_{lmj}(\vartheta)Y_j\,. \label{defTlm2}
\ee

Upon direct substitution of Eq.~\eqref{eq:ansatz2}, and Eq.~\eqref{defTlm2} into the wave equation, we get the separate angular and radial equation, Eq.~\eqref{teuradial} and Eq.~\eqref{ang}. The source term $T_{lmj}$ can then be obtained inverting Eq.~\eqref{defTlm2}.

Using \eqref{te} and \eqref{metricdet1}, we can write the function $\Sigma T$ in the following form 
\be
\Sigma T=-\frac{q_p}{U^t\,r^n\sin\vartheta\cos^n\vartheta\prod_{i=1}^{n-1}\sin\theta_i^{n-i}}
\delta(r-r_0)\delta(\vartheta-\pi/2)\delta(\phi-\Omega_p t)\prod_{i=1}^n \delta(\theta_i-\pi/2)\,.
\ee
Then, using (\ref{defTlm2}), we get 
\begin{align}
&-\frac{q_p\alpha}{U^t\,r^n\sin\vartheta\cos^n\vartheta\prod_{i=1}^{n-1}\sin\theta_i^{n-i}}
\delta(r-r_0)\delta(\vartheta-\pi/2)\delta(\phi-\Omega_p t)\prod_{i=1}^n \delta(\theta_i-\pi/2)=\nonumber\\
&=\sum_{l,m,j}\int d\omega e^{im\phi-i\omega t}T_{lmj}S_{lmj}(\vartheta)Y_j\,.
\end{align}
Multiplying both sides by $S_{l'm'j'}^{*}Y_{j'}^{*}e^{-im'\phi}$ and integrating on the ($n+2$)-sphere, we get 
\be
-\frac{q_p\alpha S_{l'm'j'}^{*}(\pi/2)Y_{j'}^{*}(\pi/2,\pi/2,\ldots)e^{-im'\Omega_p t}}{U^t\,r^n}\delta(r-r_0)=\int d\omega e^{-i\omega t}T_{l'm'j'}\,.
\ee
Note that, for real frequencies, the spin-weighted harmonics are orthogonal on the sphere. We chose the normalization, $\int d\Omega e^{i\phi(m-m')}S_{lmj}S^*_{l'm'j'}Y_{j}Y^*_{j'}=\delta_{l'l}\delta_{m'm}\delta_{j'j}$, which means that
\be
\int d\vartheta\sin\vartheta\cos^n\vartheta\prod_{i=1}^n d\theta_i\sin\theta_i^{n-i} S_{lmj}S^*_{lmj}Y_{j}Y^*_{j}=1\,.
\ee
Finally, multiplying both sides by $e^{i\omega' t}$ and integrating in the time variable, we get the expression for $T_{lmj}$, 
\begin{equation}
T_{{lmj}}=-\frac{q_p\alpha}{U^tr^n}S^*_{lmj}(\pi/2)Y_{j}^{*}(\pi/2,\pi/2,\ldots)
\delta(r-r_0)\delta(m\Omega_p-\omega)\,.
\end{equation}

\section*{Source $S$}

For the case of a radial geodesic in a Schwarzschild geometry, the trace of the stress-energy tensor for a point particle can be written in the form 
\be
T=-\frac{q_p}{\sqrt{-g} U^t}\delta(r-R(\tau))\delta(\theta-0)\delta(\phi-0)\,,\label{te1}
\ee
where the square root of the metric determinant takes the form
\be\label{metricdet3}
\sqrt{-g}=r^2\sin\theta.
\ee

The nonhomogeneous equation for the scalar field can written in the form,
\be\label{radial3}
\left[\frac{d^2}{dr_*^2}+\omega^2-V\right]X_{{lm}}(r^*)=\frac{f(r)}{r}T_{{lm}}=f(r)S\,,
\ee
where $f(r)=1-2M/r$ and $S=T_{lm}/r$.

Once more, to separate the wave equation we define   
\be
r^2 T=\frac{1}{\sqrt{2\pi}}\sum_{l,m}\int d\omega e^{-i\omega t}T_{lm}Y_{lm}(\theta,\phi)\,. \label{defTlm3}
\ee

The source term $T_{lm}$ can then be obtained inverting Eq.~\eqref{defTlm3}.

Using \eqref{te1} and \eqref{metricdet3}, we can write the function $r^2 T$ in the following form 
\be
r^2 T=-\frac{q_p}{U^t\sin\theta}
\delta(r-R(\tau))\delta(\theta-0)\delta(\phi-0)\,.
\ee
Then, using (\ref{defTlm3}), we get 
\begin{align}
&-\frac{\sqrt{2\pi}q_p}{U^t\sin\theta}
\delta(r-R(\tau))\delta(\theta-0)\delta(\phi-0)=\nonumber\\
&=\sum_{l,m}\int d\omega e^{-i\omega t}T_{lm}Y_{lm}(\theta,\phi)\,.
\end{align}
Multiplying both sides by $Y_{l'm'}^{*}$ and integrating on the sphere, we get 
\be
-\frac{\sqrt{2\pi}q_p Y_{l'm'}^{*}(0,0)}{U^t}\delta(r-R(\tau))=\int d\omega e^{-i\omega t}T_{l'm'}\,.
\ee

Finally, multiplying both sides by $e^{i\omega' t}$ and integrating in the time variable, we get 
\begin{align}
T_{{lm}}&=-\frac{q_p}{\sqrt{2\pi}}Y^*_{lm}(0,0)\int dt (\frac{dt}{d\tau})^{-1} e^{i\omega t}
\delta(r-R(\tau))=\nonumber \\
&=-\frac{q_p}{\sqrt{2\pi}}Y^*_{lm}(0,0)\int dr (\frac{dt}{dr})(\frac{dt}{d\tau})^{-1} e^{i\omega t}\delta(r-R(\tau))=\nonumber \\
&=-\frac{q_p}{\sqrt{2\pi}}Y^*_{lm}(0,0)e^{i\omega T(r)}(\frac{dr}{d\tau})^{-1}\,.
\end{align}

So, the source term $S$ can be written,
\be
S=-\frac{q_p}{\sqrt{2\pi}r}Y^*_{lm}(0,0)e^{i\omega T(r)}(\frac{dr}{d\tau})^{-1}\,.
\ee

\cleardoublepage

%\cleardoublepage
\phantomsection
\bibliographystyle{apsrev4-1}
\bibliography{ref}

\begin{thebibliography}{94}%
\makeatletter
\providecommand \@ifxundefined [1]{%
 \@ifx{#1\undefined}
}%
\providecommand \@ifnum [1]{%
 \ifnum #1\expandafter \@firstoftwo
 \else \expandafter \@secondoftwo
 \fi
}%
\providecommand \@ifx [1]{%
 \ifx #1\expandafter \@firstoftwo
 \else \expandafter \@secondoftwo
 \fi
}%
\providecommand \natexlab [1]{#1}%
\providecommand \enquote  [1]{``#1''}%
\providecommand \bibnamefont  [1]{#1}%
\providecommand \bibfnamefont [1]{#1}%
\providecommand \citenamefont [1]{#1}%
\providecommand \href@noop [0]{\@secondoftwo}%
\providecommand \href [0]{\begingroup \@sanitize@url \@href}%
\providecommand \@href[1]{\@@startlink{#1}\@@href}%
\providecommand \@@href[1]{\endgroup#1\@@endlink}%
\providecommand \@sanitize@url [0]{\catcode `\\12\catcode `\$12\catcode
  `\&12\catcode `\#12\catcode `\^12\catcode `\_12\catcode `\%12\relax}%
\providecommand \@@startlink[1]{}%
\providecommand \@@endlink[0]{}%
\providecommand \url  [0]{\begingroup\@sanitize@url \@url }%
\providecommand \@url [1]{\endgroup\@href {#1}{\urlprefix }}%
\providecommand \urlprefix  [0]{URL }%
\providecommand \Eprint [0]{\href }%
\providecommand \doibase [0]{http://dx.doi.org/}%
\providecommand \selectlanguage [0]{\@gobble}%
\providecommand \bibinfo  [0]{\@secondoftwo}%
\providecommand \bibfield  [0]{\@secondoftwo}%
\providecommand \translation [1]{[#1]}%
\providecommand \BibitemOpen [0]{}%
\providecommand \bibitemStop [0]{}%
\providecommand \bibitemNoStop [0]{.\EOS\space}%
\providecommand \EOS [0]{\spacefactor3000\relax}%
\providecommand \BibitemShut  [1]{\csname bibitem#1\endcsname}%
\let\auto@bib@innerbib\@empty
%</preamble>
\bibitem [{\citenamefont {Brito}\ \emph {et~al.}(2012)\citenamefont {Brito},
  \citenamefont {Cardoso},\ and\ \citenamefont {Pani}}]{Brito:2012prd}%
  \BibitemOpen
  \bibfield  {author} {\bibinfo {author} {\bibfnamefont {R.}~\bibnamefont
  {Brito}}, \bibinfo {author} {\bibfnamefont {V.}~\bibnamefont {Cardoso}}, \
  and\ \bibinfo {author} {\bibfnamefont {P.}~\bibnamefont {Pani}},\ }\href
  {\doibase 10.1103/PhysRevD.86.024032} {\bibfield  {journal} {\bibinfo
  {journal} {Phys.Rev.}\ }\textbf {\bibinfo {volume} {D86}},\ \bibinfo {pages}
  {024032} (\bibinfo {year} {2012})},\ \Eprint
  {http://arxiv.org/abs/1207.0504v2} {1207.0504v2 [gr-qc]} \BibitemShut
  {NoStop}%
%%CITATION = ARXIV:1207.0504v2;%%
\bibitem [{\citenamefont {Misner}\ \emph {et~al.}(1973)\citenamefont {Misner},
  \citenamefont {Thorne},\ and\ \citenamefont {Wheeler}}]{gravitation}%
  \BibitemOpen
  \bibfield  {author} {\bibinfo {author} {\bibfnamefont {C.}~\bibnamefont
  {Misner}}, \bibinfo {author} {\bibfnamefont {K.}~\bibnamefont {Thorne}}, \
  and\ \bibinfo {author} {\bibfnamefont {J.}~\bibnamefont {Wheeler}},\
  }\href@noop {} {\emph {\bibinfo {title} {Gravitation}}}\ (\bibinfo
  {publisher} {W.H.Freeman and Company},\ \bibinfo {year} {1973})\BibitemShut
  {NoStop}%
\bibitem [{\citenamefont {Chandrasekhar}(1983)}]{chandra}%
  \BibitemOpen
  \bibfield  {author} {\bibinfo {author} {\bibfnamefont {S.}~\bibnamefont
  {Chandrasekhar}},\ }\href@noop {} {\emph {\bibinfo {title} {The Mathematical
  Theory of Black Holes}}}\ (\bibinfo  {publisher} {Oxford University Press},\
  \bibinfo {year} {1983})\BibitemShut {NoStop}%
\bibitem [{\citenamefont {Melia}(2007)}]{Melia:2007vt}%
  \BibitemOpen
  \bibfield  {author} {\bibinfo {author} {\bibfnamefont {F.}~\bibnamefont
  {Melia}},\ }\href@noop {} {\  (\bibinfo {year} {2007})},\ \Eprint
  {http://arxiv.org/abs/0705.1537} {0705.1537 [astro-ph]} \BibitemShut
  {NoStop}%
%%CITATION = ARXIV:0705.1537;%%
\bibitem [{\citenamefont {Hawking}(1975)}]{Hawking:1974sw}%
  \BibitemOpen
  \bibfield  {author} {\bibinfo {author} {\bibfnamefont {S.~W.}\ \bibnamefont
  {Hawking}},\ }\href {\doibase 10.1007/BF02345020} {\bibfield  {journal}
  {\bibinfo  {journal} {Commun.Math.Phys.}\ }\textbf {\bibinfo {volume} {43}},\
  \bibinfo {pages} {199} (\bibinfo {year} {1975})}\BibitemShut {NoStop}%
%%CITATION = CMPHA,43,199;%%
\bibitem [{\citenamefont {Cardoso}\ \emph {et~al.}(2012)\citenamefont
  {Cardoso}, \citenamefont {Gualtieri}, \citenamefont {Herdeiro}, \citenamefont
  {Sperhake} \emph {et~al.}}]{Cardoso:2012qm}%
  \BibitemOpen
  \bibfield  {author} {\bibinfo {author} {\bibfnamefont {V.}~\bibnamefont
  {Cardoso}}, \bibinfo {author} {\bibfnamefont {L.}~\bibnamefont {Gualtieri}},
  \bibinfo {author} {\bibfnamefont {C.}~\bibnamefont {Herdeiro}}, \bibinfo
  {author} {\bibfnamefont {U.}~\bibnamefont {Sperhake}},  \emph {et~al.},\
  }\href@noop {} {\  (\bibinfo {year} {2012})},\ \Eprint
  {http://arxiv.org/abs/1201.5118} {1201.5118 [hep-th]} \BibitemShut {NoStop}%
%%CITATION = ARXIV:1201.5118;%%
\bibitem [{\citenamefont {Maldacena}(1999)}]{Maldacena:1997zz}%
  \BibitemOpen
  \bibfield  {author} {\bibinfo {author} {\bibfnamefont {J.~M.}\ \bibnamefont
  {Maldacena}},\ }\href {\doibase 10.1063/1.59653, 10.1023/A:1026654312961}
  {\bibfield  {journal} {\bibinfo  {journal} {AIP Conf.Proc.}\ }\textbf
  {\bibinfo {volume} {484}},\ \bibinfo {pages} {51} (\bibinfo {year}
  {1999})}\BibitemShut {NoStop}%
%%CITATION = APCPC,484,51;%%
\bibitem [{\citenamefont {Witten}(1998)}]{Witten:1998qj}%
  \BibitemOpen
  \bibfield  {author} {\bibinfo {author} {\bibfnamefont {E.}~\bibnamefont
  {Witten}},\ }\href@noop {} {\bibfield  {journal} {\bibinfo  {journal}
  {Adv.Theor.Math.Phys.}\ }\textbf {\bibinfo {volume} {2}},\ \bibinfo {pages}
  {253} (\bibinfo {year} {1998})},\ \Eprint
  {http://arxiv.org/abs/hep-th/9802150} {hep-th/9802150 [hep-th]} \BibitemShut
  {NoStop}%
%%CITATION = HEP-TH/9802150;%%
\bibitem [{\citenamefont {Polchinski}\ and\ \citenamefont
  {Strassler}(2002)}]{Polchinski:2001tt}%
  \BibitemOpen
  \bibfield  {author} {\bibinfo {author} {\bibfnamefont {J.}~\bibnamefont
  {Polchinski}}\ and\ \bibinfo {author} {\bibfnamefont {M.~J.}\ \bibnamefont
  {Strassler}},\ }\href {\doibase 10.1103/PhysRevLett.88.031601} {\bibfield
  {journal} {\bibinfo  {journal} {Phys.Rev.Lett.}\ }\textbf {\bibinfo {volume}
  {88}},\ \bibinfo {pages} {031601} (\bibinfo {year} {2002})},\ \Eprint
  {http://arxiv.org/abs/hep-th/0109174} {hep-th/0109174 [hep-th]} \BibitemShut
  {NoStop}%
%%CITATION = HEP-TH/0109174;%%
\bibitem [{\citenamefont {Polchinski}\ and\ \citenamefont
  {Strassler}(2003)}]{Polchinski:2002jw}%
  \BibitemOpen
  \bibfield  {author} {\bibinfo {author} {\bibfnamefont {J.}~\bibnamefont
  {Polchinski}}\ and\ \bibinfo {author} {\bibfnamefont {M.}~\bibnamefont
  {Strassler}},\ }\href@noop {} {\bibfield  {journal} {\bibinfo  {journal}
  {JHEP}\ }\textbf {\bibinfo {volume} {0305}},\ \bibinfo {pages} {012}
  (\bibinfo {year} {2003})},\ \Eprint {http://arxiv.org/abs/hep-th/0209211}
  {hep-th/0209211 [hep-th]} \BibitemShut {NoStop}%
%%CITATION = HEP-TH/0209211;%%
\bibitem [{\citenamefont {Giddings}(2003)}]{Giddings:2002cd}%
  \BibitemOpen
  \bibfield  {author} {\bibinfo {author} {\bibfnamefont {S.~B.}\ \bibnamefont
  {Giddings}},\ }\href {\doibase 10.1103/PhysRevD.67.126001} {\bibfield
  {journal} {\bibinfo  {journal} {Phys.Rev.}\ }\textbf {\bibinfo {volume}
  {D67}},\ \bibinfo {pages} {126001} (\bibinfo {year} {2003})},\ \Eprint
  {http://arxiv.org/abs/hep-th/0203004} {hep-th/0203004 [hep-th]} \BibitemShut
  {NoStop}%
%%CITATION = HEP-TH/0203004;%%
\bibitem [{\citenamefont {Kang}\ and\ \citenamefont
  {Nastase}(2005)}]{Kang:2004jd}%
  \BibitemOpen
  \bibfield  {author} {\bibinfo {author} {\bibfnamefont {K.}~\bibnamefont
  {Kang}}\ and\ \bibinfo {author} {\bibfnamefont {H.}~\bibnamefont {Nastase}},\
  }\href {\doibase 10.1103/PhysRevD.72.106003} {\bibfield  {journal} {\bibinfo
  {journal} {Phys.Rev.}\ }\textbf {\bibinfo {volume} {D72}},\ \bibinfo {pages}
  {106003} (\bibinfo {year} {2005})},\ \Eprint
  {http://arxiv.org/abs/hep-th/0410173} {hep-th/0410173 [hep-th]} \BibitemShut
  {NoStop}%
%%CITATION = HEP-TH/0410173;%%
\bibitem [{\citenamefont {Luzum}\ and\ \citenamefont
  {Romatschke}(2008)}]{Luzum:2008cw}%
  \BibitemOpen
  \bibfield  {author} {\bibinfo {author} {\bibfnamefont {M.}~\bibnamefont
  {Luzum}}\ and\ \bibinfo {author} {\bibfnamefont {P.}~\bibnamefont
  {Romatschke}},\ }\href {\doibase 10.1103/PhysRevC.78.034915,
  10.1103/PhysRevC.79.039903} {\bibfield  {journal} {\bibinfo  {journal}
  {Phys.Rev.}\ }\textbf {\bibinfo {volume} {C78}},\ \bibinfo {pages} {034915}
  (\bibinfo {year} {2008})},\ \Eprint {http://arxiv.org/abs/0804.4015}
  {0804.4015 [nucl-th]} \BibitemShut {NoStop}%
%%CITATION = ARXIV:0804.4015;%%
\bibitem [{\citenamefont {Arkani-Hamed}\ \emph {et~al.}(1998)\citenamefont
  {Arkani-Hamed}, \citenamefont {Dimopoulos},\ and\ \citenamefont
  {Dvali}}]{ArkaniHamed:1998rs}%
  \BibitemOpen
  \bibfield  {author} {\bibinfo {author} {\bibfnamefont {N.}~\bibnamefont
  {Arkani-Hamed}}, \bibinfo {author} {\bibfnamefont {S.}~\bibnamefont
  {Dimopoulos}}, \ and\ \bibinfo {author} {\bibfnamefont {G.~R.}\ \bibnamefont
  {Dvali}},\ }\href {\doibase 10.1016/S0370-2693(98)00466-3} {\bibfield
  {journal} {\bibinfo  {journal} {Phys.Lett.}\ }\textbf {\bibinfo {volume}
  {B429}},\ \bibinfo {pages} {263} (\bibinfo {year} {1998})},\ \Eprint
  {http://arxiv.org/abs/hep-ph/9803315} {hep-ph/9803315 [hep-ph]} \BibitemShut
  {NoStop}%
%%CITATION = HEP-PH/9803315;%%
\bibitem [{\citenamefont {Arkani-Hamed}\ \emph {et~al.}(1999)\citenamefont
  {Arkani-Hamed}, \citenamefont {Dimopoulos},\ and\ \citenamefont
  {Dvali}}]{ArkaniHamed:1998nn}%
  \BibitemOpen
  \bibfield  {author} {\bibinfo {author} {\bibfnamefont {N.}~\bibnamefont
  {Arkani-Hamed}}, \bibinfo {author} {\bibfnamefont {S.}~\bibnamefont
  {Dimopoulos}}, \ and\ \bibinfo {author} {\bibfnamefont {G.~R.}\ \bibnamefont
  {Dvali}},\ }\href {\doibase 10.1103/PhysRevD.59.086004} {\bibfield  {journal}
  {\bibinfo  {journal} {Phys.Rev.}\ }\textbf {\bibinfo {volume} {D59}},\
  \bibinfo {pages} {086004} (\bibinfo {year} {1999})},\ \Eprint
  {http://arxiv.org/abs/hep-ph/9807344} {hep-ph/9807344 [hep-ph]} \BibitemShut
  {NoStop}%
%%CITATION = HEP-PH/9807344;%%
\bibitem [{\citenamefont {Antoniadis}\ \emph {et~al.}(1998)\citenamefont
  {Antoniadis}, \citenamefont {Arkani-Hamed}, \citenamefont {Dimopoulos},\ and\
  \citenamefont {Dvali}}]{Antoniadis:1998ig}%
  \BibitemOpen
  \bibfield  {author} {\bibinfo {author} {\bibfnamefont {I.}~\bibnamefont
  {Antoniadis}}, \bibinfo {author} {\bibfnamefont {N.}~\bibnamefont
  {Arkani-Hamed}}, \bibinfo {author} {\bibfnamefont {S.}~\bibnamefont
  {Dimopoulos}}, \ and\ \bibinfo {author} {\bibfnamefont {G.~R.}\ \bibnamefont
  {Dvali}},\ }\href {\doibase 10.1016/S0370-2693(98)00860-0} {\bibfield
  {journal} {\bibinfo  {journal} {Phys.Lett.}\ }\textbf {\bibinfo {volume}
  {B436}},\ \bibinfo {pages} {257} (\bibinfo {year} {1998})},\ \Eprint
  {http://arxiv.org/abs/hep-ph/9804398} {hep-ph/9804398 [hep-ph]} \BibitemShut
  {NoStop}%
%%CITATION = HEP-PH/9804398;%%
\bibitem [{\citenamefont {Randall}\ and\ \citenamefont
  {Sundrum}(1999)}]{Randall:1999vf}%
  \BibitemOpen
  \bibfield  {author} {\bibinfo {author} {\bibfnamefont {L.}~\bibnamefont
  {Randall}}\ and\ \bibinfo {author} {\bibfnamefont {R.}~\bibnamefont
  {Sundrum}},\ }\href {\doibase 10.1103/PhysRevLett.83.4690} {\bibfield
  {journal} {\bibinfo  {journal} {Phys.Rev.Lett.}\ }\textbf {\bibinfo {volume}
  {83}},\ \bibinfo {pages} {4690} (\bibinfo {year} {1999})},\ \Eprint
  {http://arxiv.org/abs/hep-th/9906064} {hep-th/9906064 [hep-th]} \BibitemShut
  {NoStop}%
%%CITATION = HEP-TH/9906064;%%
\bibitem [{\citenamefont {DeWolfe}\ and\ \citenamefont
  {Giddings}(2003)}]{DeWolfe:2002nn}%
  \BibitemOpen
  \bibfield  {author} {\bibinfo {author} {\bibfnamefont {O.}~\bibnamefont
  {DeWolfe}}\ and\ \bibinfo {author} {\bibfnamefont {S.~B.}\ \bibnamefont
  {Giddings}},\ }\href {\doibase 10.1103/PhysRevD.67.066008} {\bibfield
  {journal} {\bibinfo  {journal} {Phys.Rev.}\ }\textbf {\bibinfo {volume}
  {D67}},\ \bibinfo {pages} {066008} (\bibinfo {year} {2003})},\ \Eprint
  {http://arxiv.org/abs/hep-th/0208123} {hep-th/0208123 [hep-th]} \BibitemShut
  {NoStop}%
%%CITATION = HEP-TH/0208123;%%
\bibitem [{\citenamefont {Banks}\ and\ \citenamefont
  {Fischler}(1999)}]{Banks:1999gd}%
  \BibitemOpen
  \bibfield  {author} {\bibinfo {author} {\bibfnamefont {T.}~\bibnamefont
  {Banks}}\ and\ \bibinfo {author} {\bibfnamefont {W.}~\bibnamefont
  {Fischler}},\ }\href@noop {} {\  (\bibinfo {year} {1999})},\ \Eprint
  {http://arxiv.org/abs/hep-th/9906038} {hep-th/9906038 [hep-th]} \BibitemShut
  {NoStop}%
%%CITATION = HEP-TH/9906038;%%
\bibitem [{\citenamefont {Giddings}\ and\ \citenamefont
  {Thomas}(2002)}]{Giddings:2001bu}%
  \BibitemOpen
  \bibfield  {author} {\bibinfo {author} {\bibfnamefont {S.~B.}\ \bibnamefont
  {Giddings}}\ and\ \bibinfo {author} {\bibfnamefont {S.~D.}\ \bibnamefont
  {Thomas}},\ }\href {\doibase 10.1103/PhysRevD.65.056010} {\bibfield
  {journal} {\bibinfo  {journal} {Phys.Rev.}\ }\textbf {\bibinfo {volume}
  {D65}},\ \bibinfo {pages} {056010} (\bibinfo {year} {2002})},\ \Eprint
  {http://arxiv.org/abs/hep-ph/0106219} {hep-ph/0106219 [hep-ph]} \BibitemShut
  {NoStop}%
%%CITATION = HEP-PH/0106219;%%
\bibitem [{\citenamefont {Dimopoulos}\ and\ \citenamefont
  {Landsberg}(2001)}]{Dimopoulos:2001hw}%
  \BibitemOpen
  \bibfield  {author} {\bibinfo {author} {\bibfnamefont {S.}~\bibnamefont
  {Dimopoulos}}\ and\ \bibinfo {author} {\bibfnamefont {G.~L.}\ \bibnamefont
  {Landsberg}},\ }\href {\doibase 10.1103/PhysRevLett.87.161602} {\bibfield
  {journal} {\bibinfo  {journal} {Phys.Rev.Lett.}\ }\textbf {\bibinfo {volume}
  {87}},\ \bibinfo {pages} {161602} (\bibinfo {year} {2001})},\ \Eprint
  {http://arxiv.org/abs/hep-ph/0106295} {hep-ph/0106295 [hep-ph]} \BibitemShut
  {NoStop}%
%%CITATION = HEP-PH/0106295;%%
\bibitem [{\citenamefont {Giddings}\ and\ \citenamefont
  {Katz}(2001)}]{Giddings:2000ay}%
  \BibitemOpen
  \bibfield  {author} {\bibinfo {author} {\bibfnamefont {S.~B.}\ \bibnamefont
  {Giddings}}\ and\ \bibinfo {author} {\bibfnamefont {E.}~\bibnamefont
  {Katz}},\ }\href {\doibase 10.1063/1.1377036} {\bibfield  {journal} {\bibinfo
   {journal} {J.Math.Phys.}\ }\textbf {\bibinfo {volume} {42}},\ \bibinfo
  {pages} {3082} (\bibinfo {year} {2001})},\ \Eprint
  {http://arxiv.org/abs/hep-th/0009176} {hep-th/0009176 [hep-th]} \BibitemShut
  {NoStop}%
%%CITATION = HEP-TH/0009176;%%
\bibitem [{\citenamefont {Khachatryan}\ \emph {et~al.}(2011)\citenamefont
  {Khachatryan} \emph {et~al.}}]{Khachatryan:2010wx}%
  \BibitemOpen
  \bibfield  {author} {\bibinfo {author} {\bibfnamefont {V.}~\bibnamefont
  {Khachatryan}} \emph {et~al.} (\bibinfo {collaboration} {CMS
  Collaboration}),\ }\href {\doibase 10.1016/j.physletb.2011.02.032} {\bibfield
   {journal} {\bibinfo  {journal} {Phys.Lett.}\ }\textbf {\bibinfo {volume}
  {B697}},\ \bibinfo {pages} {434} (\bibinfo {year} {2011})},\ \Eprint
  {http://arxiv.org/abs/1012.3375} {1012.3375 [hep-ex]} \BibitemShut {NoStop}%
%%CITATION = ARXIV:1012.3375;%%
\bibitem [{\citenamefont {Chatrchyan}\ \emph {et~al.}(2012)\citenamefont
  {Chatrchyan} \emph {et~al.}}]{Chatrchyan:2012taa}%
  \BibitemOpen
  \bibfield  {author} {\bibinfo {author} {\bibfnamefont {S.}~\bibnamefont
  {Chatrchyan}} \emph {et~al.} (\bibinfo {collaboration} {CMS Collaboration}),\
  }\href {\doibase 10.1007/JHEP04(2012)061} {\bibfield  {journal} {\bibinfo
  {journal} {JHEP}\ }\textbf {\bibinfo {volume} {1204}},\ \bibinfo {pages}
  {061} (\bibinfo {year} {2012})},\ \Eprint {http://arxiv.org/abs/1202.6396}
  {1202.6396 [hep-ex]} \BibitemShut {NoStop}%
%%CITATION = ARXIV:1202.6396;%%
\bibitem [{\citenamefont {Cardoso}\ and\ \citenamefont
  {Pani}(2012)}]{Cardoso:2012zn}%
  \BibitemOpen
  \bibfield  {author} {\bibinfo {author} {\bibfnamefont {V.}~\bibnamefont
  {Cardoso}}\ and\ \bibinfo {author} {\bibfnamefont {P.}~\bibnamefont {Pani}},\
  }\href@noop {} {\  (\bibinfo {year} {2012})},\ \Eprint
  {http://arxiv.org/abs/1205.3184} {1205.3184 [gr-qc]} \BibitemShut {NoStop}%
%%CITATION = ARXIV:1205.3184;%%
\bibitem [{\citenamefont {Thorne}\ \emph {et~al.}(1986)\citenamefont {Thorne},
  \citenamefont {Price},\ and\ \citenamefont {Macdonald}}]{Thorne:1986iy}%
  \BibitemOpen
  \bibfield  {author} {\bibinfo {author} {\bibfnamefont {K.~S.}\ \bibnamefont
  {Thorne}}, \bibinfo {author} {\bibfnamefont {R.~H.}\ \bibnamefont {Price}}, \
  and\ \bibinfo {author} {\bibfnamefont {D.~A.}\ \bibnamefont {Macdonald}},\
  }\href@noop {} {\emph {\bibinfo {title} {{Black holes: the membrane
  paradigm}}}}\ (\bibinfo {year} {1986})\BibitemShut {NoStop}%
\bibitem [{\citenamefont {Arvanitaki}\ \emph {et~al.}(2010)\citenamefont
  {Arvanitaki}, \citenamefont {Dimopoulos}, \citenamefont {Dubovsky},
  \citenamefont {Kaloper},\ and\ \citenamefont
  {March-Russell}}]{Arvanitaki:2010prd}%
  \BibitemOpen
  \bibfield  {author} {\bibinfo {author} {\bibfnamefont {A.}~\bibnamefont
  {Arvanitaki}}, \bibinfo {author} {\bibfnamefont {S.}~\bibnamefont
  {Dimopoulos}}, \bibinfo {author} {\bibfnamefont {S.}~\bibnamefont
  {Dubovsky}}, \bibinfo {author} {\bibfnamefont {N.}~\bibnamefont {Kaloper}}, \
  and\ \bibinfo {author} {\bibfnamefont {J.}~\bibnamefont {March-Russell}},\
  }\href {\doibase 10.1103/PhysRevD.81.123530} {\bibfield  {journal} {\bibinfo
  {journal} {Phys.Rev.}\ }\textbf {\bibinfo {volume} {D81}},\ \bibinfo {pages}
  {123530} (\bibinfo {year} {2010})},\ \Eprint {http://arxiv.org/abs/0905.4720}
  {0905.4720 [hep-th]} \BibitemShut {NoStop}%
%%CITATION = ARXIV:0905.4720;%%
\bibitem [{\citenamefont {Arvanitaki}\ and\ \citenamefont
  {Dubovsky}(2011)}]{Arvanitaki:2011prd}%
  \BibitemOpen
  \bibfield  {author} {\bibinfo {author} {\bibfnamefont {A.}~\bibnamefont
  {Arvanitaki}}\ and\ \bibinfo {author} {\bibfnamefont {S.}~\bibnamefont
  {Dubovsky}},\ }\href {\doibase 10.1103/PhysRevD.83.044026} {\bibfield
  {journal} {\bibinfo  {journal} {Phys.Rev.}\ }\textbf {\bibinfo {volume}
  {D83}},\ \bibinfo {pages} {044026} (\bibinfo {year} {2011})},\ \Eprint
  {http://arxiv.org/abs/1004.3558} {1004.3558 [hep-th]} \BibitemShut {NoStop}%
%%CITATION = ARXIV:1004.3558;%%
\bibitem [{\citenamefont {Fujii}\ and\ \citenamefont {Maeda}(2003)}]{Fujii}%
  \BibitemOpen
  \bibfield  {author} {\bibinfo {author} {\bibfnamefont {Y.}~\bibnamefont
  {Fujii}}\ and\ \bibinfo {author} {\bibfnamefont {K.-I.}\ \bibnamefont
  {Maeda}},\ }\href@noop {} {\emph {\bibinfo {title} {The Scalar-Tensor Theory
  of Gravitation}}}\ (\bibinfo  {publisher} {Cambridge University Press},\
  \bibinfo {year} {2003})\BibitemShut {NoStop}%
\bibitem [{\citenamefont {Sotiriou}\ and\ \citenamefont
  {Faraoni}(2010)}]{Sotiriou:2008rp}%
  \BibitemOpen
  \bibfield  {author} {\bibinfo {author} {\bibfnamefont {T.~P.}\ \bibnamefont
  {Sotiriou}}\ and\ \bibinfo {author} {\bibfnamefont {V.}~\bibnamefont
  {Faraoni}},\ }\href {\doibase 10.1103/RevModPhys.82.451} {\bibfield
  {journal} {\bibinfo  {journal} {Rev.Mod.Phys.}\ }\textbf {\bibinfo {volume}
  {82}},\ \bibinfo {pages} {451} (\bibinfo {year} {2010})},\ \Eprint
  {http://arxiv.org/abs/0805.1726} {0805.1726 [gr-qc]} \BibitemShut {NoStop}%
%%CITATION = ARXIV:0805.1726;%%
\bibitem [{\citenamefont {Regge}\ and\ \citenamefont
  {Wheeler}(1957)}]{Regge:1957td}%
  \BibitemOpen
  \bibfield  {author} {\bibinfo {author} {\bibfnamefont {T.}~\bibnamefont
  {Regge}}\ and\ \bibinfo {author} {\bibfnamefont {J.~A.}\ \bibnamefont
  {Wheeler}},\ }\href {\doibase 10.1103/PhysRev.108.1063} {\bibfield  {journal}
  {\bibinfo  {journal} {Phys.Rev.}\ }\textbf {\bibinfo {volume} {108}},\
  \bibinfo {pages} {1063} (\bibinfo {year} {1957})}\BibitemShut {NoStop}%
%%CITATION = PHRVA,108,1063;%%
\bibitem [{\citenamefont {Zerilli}(1970)}]{Zerilli:1971wd}%
  \BibitemOpen
  \bibfield  {author} {\bibinfo {author} {\bibfnamefont {F.~J.}\ \bibnamefont
  {Zerilli}},\ }\href {\doibase 10.1103/PhysRevD.2.2141} {\bibfield  {journal}
  {\bibinfo  {journal} {Phys.Rev.}\ }\textbf {\bibinfo {volume} {D2}},\
  \bibinfo {pages} {2141} (\bibinfo {year} {1970})}\BibitemShut {NoStop}%
%%CITATION = PHRVA,D2,2141;%%
\bibitem [{\citenamefont {{Press}}\ and\ \citenamefont
  {{Teukolsky}}(1972)}]{teunature}%
  \BibitemOpen
  \bibfield  {author} {\bibinfo {author} {\bibfnamefont {W.~H.}\ \bibnamefont
  {{Press}}}\ and\ \bibinfo {author} {\bibfnamefont {S.~A.}\ \bibnamefont
  {{Teukolsky}}},\ }\href {\doibase 10.1038/238211a0} {\bibfield  {journal}
  {\bibinfo  {journal} {Nature}\ }\textbf {\bibinfo {volume} {238}},\ \bibinfo
  {pages} {211} (\bibinfo {year} {1972})}\BibitemShut {NoStop}%
\bibitem [{\citenamefont {Brill}\ \emph {et~al.}(1972)\citenamefont {Brill},
  \citenamefont {Chrzanowski}, \citenamefont {Pereira}, \citenamefont
  {Fackerell},\ and\ \citenamefont {Ipser}}]{Brill:1972prd}%
  \BibitemOpen
  \bibfield  {author} {\bibinfo {author} {\bibfnamefont {D.~R.}\ \bibnamefont
  {Brill}}, \bibinfo {author} {\bibfnamefont {P.~L.}\ \bibnamefont
  {Chrzanowski}}, \bibinfo {author} {\bibfnamefont {C.~M.}\ \bibnamefont
  {Pereira}}, \bibinfo {author} {\bibfnamefont {E.~D.}\ \bibnamefont
  {Fackerell}}, \ and\ \bibinfo {author} {\bibfnamefont {J.~R.}\ \bibnamefont
  {Ipser}},\ }\href {\doibase 10.1103/PhysRevD.5.1913} {\bibfield  {journal}
  {\bibinfo  {journal} {Phys. Rev.}\ }\textbf {\bibinfo {volume} {D5}},\
  \bibinfo {pages} {1913} (\bibinfo {year} {1972})}\BibitemShut {NoStop}%
%%CITATION = PRD/v5/i8/p1913_1;%%
\bibitem [{\citenamefont {Carter}(1968)}]{Carter:1968pr}%
  \BibitemOpen
  \bibfield  {author} {\bibinfo {author} {\bibfnamefont {B.}~\bibnamefont
  {Carter}},\ }\href {\doibase 10.1103/PhysRev.174.1559} {\bibfield  {journal}
  {\bibinfo  {journal} {Phys. Rev.}\ }\textbf {\bibinfo {volume} {174}},\
  \bibinfo {pages} {1559} (\bibinfo {year} {1968})}\BibitemShut {NoStop}%
%%CITATION = PR/v174/i5/p1559_1;%%
\bibitem [{\citenamefont {Teukolsky}(1973)}]{teu}%
  \BibitemOpen
  \bibfield  {author} {\bibinfo {author} {\bibfnamefont {S.~A.}\ \bibnamefont
  {Teukolsky}},\ }\href {\doibase 10.1086/152444} {\bibfield  {journal}
  {\bibinfo  {journal} {Astrophys.J.}\ }\textbf {\bibinfo {volume} {185}},\
  \bibinfo {pages} {635} (\bibinfo {year} {1973})},\ \bibinfo {note} {ph.D.
  Thesis (Advisor: Kip S. Thorne)}\BibitemShut {NoStop}%
%%CITATION = ASJOA,185,635;%%
\bibitem [{\citenamefont {Myers}\ and\ \citenamefont {Perry}(1986)}]{myers}%
  \BibitemOpen
  \bibfield  {author} {\bibinfo {author} {\bibfnamefont {R.~C.}\ \bibnamefont
  {Myers}}\ and\ \bibinfo {author} {\bibfnamefont {M.~J.}\ \bibnamefont
  {Perry}},\ }\href {\doibase 10.1016/0003-4916(86)90186-7} {\bibfield
  {journal} {\bibinfo  {journal} {Annals Phys.}\ }\textbf {\bibinfo {volume}
  {172}},\ \bibinfo {pages} {304} (\bibinfo {year} {1986})}\BibitemShut
  {NoStop}%
%%CITATION = APNYA,172,304;%%
\bibitem [{\citenamefont {Ida}\ \emph {et~al.}(2003)\citenamefont {Ida},
  \citenamefont {Uchida},\ and\ \citenamefont {Morisawa}}]{Ida}%
  \BibitemOpen
  \bibfield  {author} {\bibinfo {author} {\bibfnamefont {D.}~\bibnamefont
  {Ida}}, \bibinfo {author} {\bibfnamefont {Y.}~\bibnamefont {Uchida}}, \ and\
  \bibinfo {author} {\bibfnamefont {Y.}~\bibnamefont {Morisawa}},\ }\href
  {\doibase 10.1103/PhysRevD.67.084019} {\bibfield  {journal} {\bibinfo
  {journal} {Phys.Rev.}\ }\textbf {\bibinfo {volume} {D67}},\ \bibinfo {pages}
  {084019} (\bibinfo {year} {2003})},\ \Eprint
  {http://arxiv.org/abs/gr-qc/0212035} {gr-qc/0212035 [gr-qc]} \BibitemShut
  {NoStop}%
%%CITATION = GR-QC/0212035;%%
\bibitem [{\citenamefont {Cardoso}\ \emph {et~al.}(2005)\citenamefont
  {Cardoso}, \citenamefont {Siopsis},\ and\ \citenamefont
  {Yoshida}}]{Cardoso:2004cj}%
  \BibitemOpen
  \bibfield  {author} {\bibinfo {author} {\bibfnamefont {V.}~\bibnamefont
  {Cardoso}}, \bibinfo {author} {\bibfnamefont {G.}~\bibnamefont {Siopsis}}, \
  and\ \bibinfo {author} {\bibfnamefont {S.}~\bibnamefont {Yoshida}},\ }\href
  {\doibase 10.1103/PhysRevD.71.024019} {\bibfield  {journal} {\bibinfo
  {journal} {Phys.Rev.}\ }\textbf {\bibinfo {volume} {D71}},\ \bibinfo {pages}
  {024019} (\bibinfo {year} {2005})},\ \Eprint
  {http://arxiv.org/abs/hep-th/0412138} {hep-th/0412138 [hep-th]} \BibitemShut
  {NoStop}%
%%CITATION = HEP-TH/0412138;%%
\bibitem [{\citenamefont {Cardoso}\ and\ \citenamefont
  {Yoshida}(2005)}]{Cardoso:2005vk}%
  \BibitemOpen
  \bibfield  {author} {\bibinfo {author} {\bibfnamefont {V.}~\bibnamefont
  {Cardoso}}\ and\ \bibinfo {author} {\bibfnamefont {S.}~\bibnamefont
  {Yoshida}},\ }\href {\doibase 10.1088/1126-6708/2005/07/009} {\bibfield
  {journal} {\bibinfo  {journal} {JHEP}\ }\textbf {\bibinfo {volume} {0507}},\
  \bibinfo {pages} {009} (\bibinfo {year} {2005})},\ \Eprint
  {http://arxiv.org/abs/hep-th/0502206} {hep-th/0502206 [hep-th]} \BibitemShut
  {NoStop}%
%%CITATION = HEP-TH/0502206;%%
\bibitem [{\citenamefont {Berti}\ \emph {et~al.}(2006)\citenamefont {Berti},
  \citenamefont {Cardoso},\ and\ \citenamefont {Casals}}]{Berti:2005gp}%
  \BibitemOpen
  \bibfield  {author} {\bibinfo {author} {\bibfnamefont {E.}~\bibnamefont
  {Berti}}, \bibinfo {author} {\bibfnamefont {V.}~\bibnamefont {Cardoso}}, \
  and\ \bibinfo {author} {\bibfnamefont {M.}~\bibnamefont {Casals}},\ }\href
  {\doibase 10.1103/PhysRevD.73.024013, 10.1103/PhysRevD.73.109902} {\bibfield
  {journal} {\bibinfo  {journal} {Phys.Rev.}\ }\textbf {\bibinfo {volume}
  {D73}},\ \bibinfo {pages} {024013} (\bibinfo {year} {2006})},\ \Eprint
  {http://arxiv.org/abs/gr-qc/0511111} {gr-qc/0511111 [gr-qc]} \BibitemShut
  {NoStop}%
%%CITATION = GR-QC/0511111;%%
\bibitem [{\citenamefont {Leaver}(1985)}]{Leaver:1985qnm}%
  \BibitemOpen
  \bibfield  {author} {\bibinfo {author} {\bibfnamefont {E.~W.}\ \bibnamefont
  {Leaver}},\ }\href {\doibase 10.1098/rspa.1985.0119} {\bibfield  {journal}
  {\bibinfo  {journal} {Proc. R. Soc. London}\ }\textbf {\bibinfo {volume}
  {A402}},\ \bibinfo {pages} {285} (\bibinfo {year} {1985})}\BibitemShut
  {NoStop}%
\bibitem [{\citenamefont {Detweiler}(1978)}]{Detweiler:1978ge}%
  \BibitemOpen
  \bibfield  {author} {\bibinfo {author} {\bibfnamefont {S.~L.}\ \bibnamefont
  {Detweiler}},\ }\href {\doibase 10.1086/156529} {\bibfield  {journal}
  {\bibinfo  {journal} {Astrophys. J.}\ }\textbf {\bibinfo {volume} {225}},\
  \bibinfo {pages} {687} (\bibinfo {year} {1978})}\BibitemShut {NoStop}%
%%CITATION = ASJOA,225,687;%%
\bibitem [{\citenamefont {Davis}\ \emph {et~al.}(1971)\citenamefont {Davis},
  \citenamefont {Ruffini}, \citenamefont {Press},\ and\ \citenamefont
  {Price}}]{Davis:1971gg}%
  \BibitemOpen
  \bibfield  {author} {\bibinfo {author} {\bibfnamefont {M.}~\bibnamefont
  {Davis}}, \bibinfo {author} {\bibfnamefont {R.}~\bibnamefont {Ruffini}},
  \bibinfo {author} {\bibfnamefont {W.~H.}\ \bibnamefont {Press}}, \ and\
  \bibinfo {author} {\bibfnamefont {R.}~\bibnamefont {Price}},\ }\href
  {\doibase 10.1103/PhysRevLett.27.1466} {\bibfield  {journal} {\bibinfo
  {journal} {Phys.Rev.Lett.}\ }\textbf {\bibinfo {volume} {27}},\ \bibinfo
  {pages} {1466} (\bibinfo {year} {1971})}\BibitemShut {NoStop}%
%%CITATION = PRLTA,27,1466;%%
\bibitem [{\citenamefont {Mitsou}(2011)}]{Mitsou:2010jv}%
  \BibitemOpen
  \bibfield  {author} {\bibinfo {author} {\bibfnamefont {E.}~\bibnamefont
  {Mitsou}},\ }\href {\doibase 10.1103/PhysRevD.83.044039} {\bibfield
  {journal} {\bibinfo  {journal} {Phys.Rev.}\ }\textbf {\bibinfo {volume}
  {D83}},\ \bibinfo {pages} {044039} (\bibinfo {year} {2011})},\ \Eprint
  {http://arxiv.org/abs/1012.2028} {1012.2028 [gr-qc]} \BibitemShut {NoStop}%
%%CITATION = ARXIV:1012.2028;%%
\bibitem [{\citenamefont {Ruffini}(1973)}]{Ruffini:1973ky}%
  \BibitemOpen
  \bibfield  {author} {\bibinfo {author} {\bibfnamefont {R.}~\bibnamefont
  {Ruffini}},\ }\href {\doibase 10.1103/PhysRevD.7.972} {\bibfield  {journal}
  {\bibinfo  {journal} {Phys.Rev.}\ }\textbf {\bibinfo {volume} {D7}},\
  \bibinfo {pages} {972} (\bibinfo {year} {1973})}\BibitemShut {NoStop}%
%%CITATION = PHRVA,D7,972;%%
\bibitem [{\citenamefont {Berti}\ \emph {et~al.}(2010)\citenamefont {Berti},
  \citenamefont {Cardoso}, \citenamefont {Hinderer}, \citenamefont {Lemos},
  \citenamefont {Pretorius} \emph {et~al.}}]{Berti:2010ce}%
  \BibitemOpen
  \bibfield  {author} {\bibinfo {author} {\bibfnamefont {E.}~\bibnamefont
  {Berti}}, \bibinfo {author} {\bibfnamefont {V.}~\bibnamefont {Cardoso}},
  \bibinfo {author} {\bibfnamefont {T.}~\bibnamefont {Hinderer}}, \bibinfo
  {author} {\bibfnamefont {M.}~\bibnamefont {Lemos}}, \bibinfo {author}
  {\bibfnamefont {F.}~\bibnamefont {Pretorius}},  \emph {et~al.},\ }\href
  {\doibase 10.1103/PhysRevD.81.104048} {\bibfield  {journal} {\bibinfo
  {journal} {Phys.Rev.}\ }\textbf {\bibinfo {volume} {D81}},\ \bibinfo {pages}
  {104048} (\bibinfo {year} {2010})},\ \Eprint {http://arxiv.org/abs/1003.0812}
  {1003.0812 [gr-qc]} \BibitemShut {NoStop}%
%%CITATION = ARXIV:1003.0812;%%
\bibitem [{\citenamefont {Anninos}\ \emph {et~al.}(1995)\citenamefont
  {Anninos}, \citenamefont {Hobill}, \citenamefont {Seidel}, \citenamefont
  {Smarr},\ and\ \citenamefont {Suen}}]{Anninos:1994gp}%
  \BibitemOpen
  \bibfield  {author} {\bibinfo {author} {\bibfnamefont {P.}~\bibnamefont
  {Anninos}}, \bibinfo {author} {\bibfnamefont {D.}~\bibnamefont {Hobill}},
  \bibinfo {author} {\bibfnamefont {E.}~\bibnamefont {Seidel}}, \bibinfo
  {author} {\bibfnamefont {L.}~\bibnamefont {Smarr}}, \ and\ \bibinfo {author}
  {\bibfnamefont {W.-M.}\ \bibnamefont {Suen}},\ }\href {\doibase
  10.1103/PhysRevD.52.2044} {\bibfield  {journal} {\bibinfo  {journal}
  {Phys.Rev.}\ }\textbf {\bibinfo {volume} {D52}},\ \bibinfo {pages} {2044}
  (\bibinfo {year} {1995})},\ \Eprint {http://arxiv.org/abs/gr-qc/9408041}
  {gr-qc/9408041 [gr-qc]} \BibitemShut {NoStop}%
%%CITATION = GR-QC/9408041;%%
\bibitem [{\citenamefont {Cardoso}\ and\ \citenamefont
  {Lemos}(2002)}]{Cardoso:2002ay}%
  \BibitemOpen
  \bibfield  {author} {\bibinfo {author} {\bibfnamefont {V.}~\bibnamefont
  {Cardoso}}\ and\ \bibinfo {author} {\bibfnamefont {J.~P.~S.}\ \bibnamefont
  {Lemos}},\ }\href {\doibase 10.1016/S0370-2693(02)01961-5} {\bibfield
  {journal} {\bibinfo  {journal} {Phys.Lett.}\ }\textbf {\bibinfo {volume}
  {B538}},\ \bibinfo {pages} {1} (\bibinfo {year} {2002})},\ \Eprint
  {http://arxiv.org/abs/gr-qc/0202019} {gr-qc/0202019 [gr-qc]} \BibitemShut
  {NoStop}%
%%CITATION = GR-QC/0202019;%%
\bibitem [{\citenamefont {Kokkotas}\ and\ \citenamefont
  {Schmidt}(1999)}]{Kokkotas:1999bd}%
  \BibitemOpen
  \bibfield  {author} {\bibinfo {author} {\bibfnamefont {K.~D.}\ \bibnamefont
  {Kokkotas}}\ and\ \bibinfo {author} {\bibfnamefont {B.~G.}\ \bibnamefont
  {Schmidt}},\ }\href@noop {} {\bibfield  {journal} {\bibinfo  {journal}
  {Living Rev.Rel.}\ }\textbf {\bibinfo {volume} {2}},\ \bibinfo {pages} {2}
  (\bibinfo {year} {1999})},\ \Eprint {http://arxiv.org/abs/gr-qc/9909058}
  {gr-qc/9909058 [gr-qc]} \BibitemShut {NoStop}%
%%CITATION = GR-QC/9909058;%%
\bibitem [{\citenamefont {Berti}\ \emph {et~al.}(2009)\citenamefont {Berti},
  \citenamefont {Cardoso},\ and\ \citenamefont {Starinets}}]{Berti:2009kk}%
  \BibitemOpen
  \bibfield  {author} {\bibinfo {author} {\bibfnamefont {E.}~\bibnamefont
  {Berti}}, \bibinfo {author} {\bibfnamefont {V.}~\bibnamefont {Cardoso}}, \
  and\ \bibinfo {author} {\bibfnamefont {A.~O.}\ \bibnamefont {Starinets}},\
  }\href {\doibase 10.1088/0264-9381/26/16/163001} {\bibfield  {journal}
  {\bibinfo  {journal} {Class.Quant.Grav.}\ }\textbf {\bibinfo {volume} {26}},\
  \bibinfo {pages} {163001} (\bibinfo {year} {2009})},\ \Eprint
  {http://arxiv.org/abs/0905.2975} {0905.2975 [gr-qc]} \BibitemShut {NoStop}%
%%CITATION = ARXIV:0905.2975;%%
\bibitem [{\citenamefont {Konoplya}\ and\ \citenamefont
  {Zhidenko}(2011)}]{Konoplya:2011qq}%
  \BibitemOpen
  \bibfield  {author} {\bibinfo {author} {\bibfnamefont {R.~A.}\ \bibnamefont
  {Konoplya}}\ and\ \bibinfo {author} {\bibfnamefont {A.}~\bibnamefont
  {Zhidenko}},\ }\href {\doibase 10.1103/RevModPhys.83.793} {\bibfield
  {journal} {\bibinfo  {journal} {Rev.Mod.Phys.}\ }\textbf {\bibinfo {volume}
  {83}},\ \bibinfo {pages} {793} (\bibinfo {year} {2011})},\ \Eprint
  {http://arxiv.org/abs/1102.4014} {1102.4014 [gr-qc]} \BibitemShut {NoStop}%
%%CITATION = ARXIV:1102.4014;%%
\bibitem [{\citenamefont {Konoplya}\ and\ \citenamefont
  {Zhidenko}(2005)}]{Konoplya:2004wg}%
  \BibitemOpen
  \bibfield  {author} {\bibinfo {author} {\bibfnamefont {R.~A.}\ \bibnamefont
  {Konoplya}}\ and\ \bibinfo {author} {\bibfnamefont {A.}~\bibnamefont
  {Zhidenko}},\ }\href {\doibase 10.1016/j.physletb.2005.01.078} {\bibfield
  {journal} {\bibinfo  {journal} {Phys.Lett.}\ }\textbf {\bibinfo {volume}
  {B609}},\ \bibinfo {pages} {377} (\bibinfo {year} {2005})},\ \Eprint
  {http://arxiv.org/abs/gr-qc/0411059} {gr-qc/0411059 [gr-qc]} \BibitemShut
  {NoStop}%
%%CITATION = GR-QC/0411059;%%
\bibitem [{\citenamefont {Konoplya}\ and\ \citenamefont
  {Zhidenko}(2006)}]{Konoplya:2006br}%
  \BibitemOpen
  \bibfield  {author} {\bibinfo {author} {\bibfnamefont {R.~A.}\ \bibnamefont
  {Konoplya}}\ and\ \bibinfo {author} {\bibfnamefont {A.}~\bibnamefont
  {Zhidenko}},\ }\href {\doibase 10.1103/PhysRevD.73.124040} {\bibfield
  {journal} {\bibinfo  {journal} {Phys.Rev.}\ }\textbf {\bibinfo {volume}
  {D73}},\ \bibinfo {pages} {124040} (\bibinfo {year} {2006})},\ \Eprint
  {http://arxiv.org/abs/gr-qc/0605013} {gr-qc/0605013 [gr-qc]} \BibitemShut
  {NoStop}%
%%CITATION = GR-QC/0605013;%%
\bibitem [{\citenamefont {Ohashi}\ and\ \citenamefont
  {Sakagami}(2004)}]{Ohashi:2004wr}%
  \BibitemOpen
  \bibfield  {author} {\bibinfo {author} {\bibfnamefont {A.}~\bibnamefont
  {Ohashi}}\ and\ \bibinfo {author} {\bibfnamefont {M.-a.}\ \bibnamefont
  {Sakagami}},\ }\href {\doibase 10.1088/0256-307X/21/8/011} {\bibfield
  {journal} {\bibinfo  {journal} {Class.Quant.Grav.}\ }\textbf {\bibinfo
  {volume} {21}},\ \bibinfo {pages} {3973} (\bibinfo {year} {2004})},\ \Eprint
  {http://arxiv.org/abs/gr-qc/0407009} {gr-qc/0407009 [gr-qc]} \BibitemShut
  {NoStop}%
%%CITATION = GR-QC/0407009;%%
\bibitem [{\citenamefont {Price}(1972{\natexlab{a}})}]{Price:1971fb}%
  \BibitemOpen
  \bibfield  {author} {\bibinfo {author} {\bibfnamefont {R.~H.}\ \bibnamefont
  {Price}},\ }\href {\doibase 10.1103/PhysRevD.5.2419} {\bibfield  {journal}
  {\bibinfo  {journal} {Phys.Rev.}\ }\textbf {\bibinfo {volume} {D5}},\
  \bibinfo {pages} {2419} (\bibinfo {year} {1972}{\natexlab{a}})}\BibitemShut
  {NoStop}%
%%CITATION = PHRVA,D5,2419;%%
\bibitem [{\citenamefont {Price}(1972{\natexlab{b}})}]{Price:1972pw}%
  \BibitemOpen
  \bibfield  {author} {\bibinfo {author} {\bibfnamefont {R.~H.}\ \bibnamefont
  {Price}},\ }\href {\doibase 10.1103/PhysRevD.5.2439} {\bibfield  {journal}
  {\bibinfo  {journal} {Phys.Rev.}\ }\textbf {\bibinfo {volume} {D5}},\
  \bibinfo {pages} {2439} (\bibinfo {year} {1972}{\natexlab{b}})}\BibitemShut
  {NoStop}%
%%CITATION = PHRVA,D5,2439;%%
\bibitem [{\citenamefont {Hod}\ and\ \citenamefont {Piran}(1998)}]{Hod:1998ra}%
  \BibitemOpen
  \bibfield  {author} {\bibinfo {author} {\bibfnamefont {S.}~\bibnamefont
  {Hod}}\ and\ \bibinfo {author} {\bibfnamefont {T.}~\bibnamefont {Piran}},\
  }\href {\doibase 10.1103/PhysRevD.58.044018} {\bibfield  {journal} {\bibinfo
  {journal} {Phys.Rev.}\ }\textbf {\bibinfo {volume} {D58}},\ \bibinfo {pages}
  {044018} (\bibinfo {year} {1998})},\ \Eprint
  {http://arxiv.org/abs/gr-qc/9801059} {gr-qc/9801059 [gr-qc]} \BibitemShut
  {NoStop}%
%%CITATION = GR-QC/9801059;%%
\bibitem [{\citenamefont {Koyama}\ and\ \citenamefont
  {Tomimatsu}(2001)}]{Koyama:2001ee}%
  \BibitemOpen
  \bibfield  {author} {\bibinfo {author} {\bibfnamefont {H.}~\bibnamefont
  {Koyama}}\ and\ \bibinfo {author} {\bibfnamefont {A.}~\bibnamefont
  {Tomimatsu}},\ }\href {\doibase 10.1103/PhysRevD.64.044014} {\bibfield
  {journal} {\bibinfo  {journal} {Phys.Rev.}\ }\textbf {\bibinfo {volume}
  {D64}},\ \bibinfo {pages} {044014} (\bibinfo {year} {2001})},\ \Eprint
  {http://arxiv.org/abs/gr-qc/0103086} {gr-qc/0103086 [gr-qc]} \BibitemShut
  {NoStop}%
%%CITATION = GR-QC/0103086;%%
\bibitem [{\citenamefont {Koyama}\ and\ \citenamefont
  {Tomimatsu}(2002)}]{Koyama:2001qw}%
  \BibitemOpen
  \bibfield  {author} {\bibinfo {author} {\bibfnamefont {H.}~\bibnamefont
  {Koyama}}\ and\ \bibinfo {author} {\bibfnamefont {A.}~\bibnamefont
  {Tomimatsu}},\ }\href {\doibase 10.1103/PhysRevD.65.084031} {\bibfield
  {journal} {\bibinfo  {journal} {Phys.Rev.}\ }\textbf {\bibinfo {volume}
  {D65}},\ \bibinfo {pages} {084031} (\bibinfo {year} {2002})},\ \Eprint
  {http://arxiv.org/abs/gr-qc/0112075} {gr-qc/0112075 [gr-qc]} \BibitemShut
  {NoStop}%
%%CITATION = GR-QC/0112075;%%
\bibitem [{\citenamefont {Adler}\ and\ \citenamefont
  {Zeks}(1975)}]{Adler:1975dj}%
  \BibitemOpen
  \bibfield  {author} {\bibinfo {author} {\bibfnamefont {R.~J.}\ \bibnamefont
  {Adler}}\ and\ \bibinfo {author} {\bibfnamefont {B.}~\bibnamefont {Zeks}},\
  }\href {\doibase 10.1103/PhysRevD.12.3007} {\bibfield  {journal} {\bibinfo
  {journal} {Phys.Rev.}\ }\textbf {\bibinfo {volume} {D12}},\ \bibinfo {pages}
  {3007} (\bibinfo {year} {1975})}\BibitemShut {NoStop}%
%%CITATION = PHRVA,D12,3007;%%
\bibitem [{\citenamefont {Smarr}(1977)}]{Smarr:1977fy}%
  \BibitemOpen
  \bibfield  {author} {\bibinfo {author} {\bibfnamefont {L.}~\bibnamefont
  {Smarr}},\ }\href {\doibase 10.1103/PhysRevD.15.2069} {\bibfield  {journal}
  {\bibinfo  {journal} {Phys.Rev.}\ }\textbf {\bibinfo {volume} {D15}},\
  \bibinfo {pages} {2069} (\bibinfo {year} {1977})}\BibitemShut {NoStop}%
%%CITATION = PHRVA,D15,2069;%%
\bibitem [{\citenamefont {He}\ and\ \citenamefont {Jing}(2006)}]{He:2006jv}%
  \BibitemOpen
  \bibfield  {author} {\bibinfo {author} {\bibfnamefont {X.}~\bibnamefont
  {He}}\ and\ \bibinfo {author} {\bibfnamefont {J.}~\bibnamefont {Jing}},\
  }\href {\doibase 10.1016/j.nuclphysb.2006.08.015} {\bibfield  {journal}
  {\bibinfo  {journal} {Nucl.Phys.}\ }\textbf {\bibinfo {volume} {B755}},\
  \bibinfo {pages} {313} (\bibinfo {year} {2006})},\ \Eprint
  {http://arxiv.org/abs/gr-qc/0611003} {gr-qc/0611003 [gr-qc]} \BibitemShut
  {NoStop}%
%%CITATION = GR-QC/0611003;%%
\bibitem [{\citenamefont {Jing}(2005)}]{Jing:2004zb}%
  \BibitemOpen
  \bibfield  {author} {\bibinfo {author} {\bibfnamefont {J.}~\bibnamefont
  {Jing}},\ }\href {\doibase 10.1103/PhysRevD.72.027501} {\bibfield  {journal}
  {\bibinfo  {journal} {Phys.Rev.}\ }\textbf {\bibinfo {volume} {D72}},\
  \bibinfo {pages} {027501} (\bibinfo {year} {2005})},\ \Eprint
  {http://arxiv.org/abs/gr-qc/0408090} {gr-qc/0408090 [gr-qc]} \BibitemShut
  {NoStop}%
%%CITATION = GR-QC/0408090;%%
\bibitem [{\citenamefont {Konoplya}\ and\ \citenamefont
  {Molina}(2007)}]{Konoplya:2006gq}%
  \BibitemOpen
  \bibfield  {author} {\bibinfo {author} {\bibfnamefont {R.~A.}\ \bibnamefont
  {Konoplya}}\ and\ \bibinfo {author} {\bibfnamefont {C.}~\bibnamefont
  {Molina}},\ }\href {\doibase 10.1103/PhysRevD.75.084004} {\bibfield
  {journal} {\bibinfo  {journal} {Phys.Rev.}\ }\textbf {\bibinfo {volume}
  {D75}},\ \bibinfo {pages} {084004} (\bibinfo {year} {2007})},\ \Eprint
  {http://arxiv.org/abs/gr-qc/0602047} {gr-qc/0602047 [gr-qc]} \BibitemShut
  {NoStop}%
%%CITATION = GR-QC/0602047;%%
\bibitem [{\citenamefont {Zel'dovich}(1971)}]{zeldo1}%
  \BibitemOpen
  \bibfield  {author} {\bibinfo {author} {\bibfnamefont {Y.~B.}\ \bibnamefont
  {Zel'dovich}},\ }\href@noop {} {\bibfield  {journal} {\bibinfo  {journal}
  {JETP Lett.}\ }\textbf {\bibinfo {volume} {14}},\ \bibinfo {pages} {180}
  (\bibinfo {year} {1971})}\BibitemShut {NoStop}%
\bibitem [{\citenamefont {Zel'dovich}(1972)}]{zeldo2}%
  \BibitemOpen
  \bibfield  {author} {\bibinfo {author} {\bibfnamefont {Y.~B.}\ \bibnamefont
  {Zel'dovich}},\ }\href@noop {} {\bibfield  {journal} {\bibinfo  {journal}
  {Sov. Phys. JETP}\ }\textbf {\bibinfo {volume} {35}},\ \bibinfo {pages}
  {1085} (\bibinfo {year} {1972})}\BibitemShut {NoStop}%
\bibitem [{\citenamefont {Teukolsky}\ and\ \citenamefont
  {Press}(1974)}]{Teukolsky:1974yv}%
  \BibitemOpen
  \bibfield  {author} {\bibinfo {author} {\bibfnamefont {S.~A.}\ \bibnamefont
  {Teukolsky}}\ and\ \bibinfo {author} {\bibfnamefont {W.~H.}\ \bibnamefont
  {Press}},\ }\href {\doibase 10.1086/153180} {\bibfield  {journal} {\bibinfo
  {journal} {Astrophys.J.}\ }\textbf {\bibinfo {volume} {193}},\ \bibinfo
  {pages} {443} (\bibinfo {year} {1974})}\BibitemShut {NoStop}%
\bibitem [{\citenamefont {Penrose}\ and\ \citenamefont
  {Floyd}(1971)}]{Penrose:1971nature}%
  \BibitemOpen
  \bibfield  {author} {\bibinfo {author} {\bibfnamefont {R.}~\bibnamefont
  {Penrose}}\ and\ \bibinfo {author} {\bibfnamefont {R.}~\bibnamefont
  {Floyd}},\ }\href@noop {} {\bibfield  {journal} {\bibinfo  {journal}
  {Nature}\ }\textbf {\bibinfo {volume} {\textbf{229}}},\ \bibinfo {pages}
  {177} (\bibinfo {year} {1971})}\BibitemShut {NoStop}%
\bibitem [{\citenamefont {R.Penrose}(1969)}]{Penrose:1969rnc}%
  \BibitemOpen
  \bibfield  {author} {\bibinfo {author} {\bibnamefont {R.Penrose}},\
  }\href@noop {} {\bibfield  {journal} {\bibinfo  {journal} {Rivista del Nuovo
  Cimento}\ }\textbf {\bibinfo {volume} {\textbf{1}}},\ \bibinfo {pages} {252}
  (\bibinfo {year} {1969})}\BibitemShut {NoStop}%
\bibitem [{\citenamefont {Misner}(1972)}]{misner}%
  \BibitemOpen
  \bibfield  {author} {\bibinfo {author} {\bibfnamefont {C.~W.}\ \bibnamefont
  {Misner}},\ }\href {\doibase 10.1103/PhysRevLett.28.994} {\bibfield
  {journal} {\bibinfo  {journal} {Phys.Rev.Lett.}\ }\textbf {\bibinfo {volume}
  {28}},\ \bibinfo {pages} {994} (\bibinfo {year} {1972})}\BibitemShut
  {NoStop}%
%%CITATION = PRLTA,28,994;%%
\bibitem [{\citenamefont {Cardoso}\ \emph {et~al.}(2004)\citenamefont
  {Cardoso}, \citenamefont {Dias}, \citenamefont {Lemos},\ and\ \citenamefont
  {Yoshida}}]{Cardoso:2004nk}%
  \BibitemOpen
  \bibfield  {author} {\bibinfo {author} {\bibfnamefont {V.}~\bibnamefont
  {Cardoso}}, \bibinfo {author} {\bibfnamefont {O.~J.~C.}\ \bibnamefont
  {Dias}}, \bibinfo {author} {\bibfnamefont {J.~P.~S.}\ \bibnamefont {Lemos}},
  \ and\ \bibinfo {author} {\bibfnamefont {S.}~\bibnamefont {Yoshida}},\ }\href
  {\doibase 10.1103/PhysRevD.70.044039, 10.1103/PhysRevD.70.049903} {\bibfield
  {journal} {\bibinfo  {journal} {Phys.Rev.}\ }\textbf {\bibinfo {volume}
  {D70}},\ \bibinfo {pages} {044039} (\bibinfo {year} {2004})},\ \Eprint
  {http://arxiv.org/abs/hep-th/0404096} {hep-th/0404096 [hep-th]} \BibitemShut
  {NoStop}%
%%CITATION = HEP-TH/0404096;%%
\bibitem [{\citenamefont {Cardoso}\ \emph {et~al.}(2008)\citenamefont
  {Cardoso}, \citenamefont {Pani}, \citenamefont {Cadoni},\ and\ \citenamefont
  {Cavaglia}}]{Cardoso:2007az}%
  \BibitemOpen
  \bibfield  {author} {\bibinfo {author} {\bibfnamefont {V.}~\bibnamefont
  {Cardoso}}, \bibinfo {author} {\bibfnamefont {P.}~\bibnamefont {Pani}},
  \bibinfo {author} {\bibfnamefont {M.}~\bibnamefont {Cadoni}}, \ and\ \bibinfo
  {author} {\bibfnamefont {M.}~\bibnamefont {Cavaglia}},\ }\href {\doibase
  10.1103/PhysRevD.77.124044} {\bibfield  {journal} {\bibinfo  {journal}
  {Phys.Rev.}\ }\textbf {\bibinfo {volume} {D77}},\ \bibinfo {pages} {124044}
  (\bibinfo {year} {2008})},\ \Eprint {http://arxiv.org/abs/0709.0532}
  {0709.0532} \BibitemShut {NoStop}%
%%CITATION = ARXIV:0709.0532;%%
\bibitem [{\citenamefont {Cardoso}\ \emph {et~al.}(2011)\citenamefont
  {Cardoso}, \citenamefont {Chakrabarti}, \citenamefont {Pani}, \citenamefont
  {Berti},\ and\ \citenamefont {Gualtieri}}]{Cardoso:2011xi}%
  \BibitemOpen
  \bibfield  {author} {\bibinfo {author} {\bibfnamefont {V.}~\bibnamefont
  {Cardoso}}, \bibinfo {author} {\bibfnamefont {S.}~\bibnamefont
  {Chakrabarti}}, \bibinfo {author} {\bibfnamefont {P.}~\bibnamefont {Pani}},
  \bibinfo {author} {\bibfnamefont {E.}~\bibnamefont {Berti}}, \ and\ \bibinfo
  {author} {\bibfnamefont {L.}~\bibnamefont {Gualtieri}},\ }\href {\doibase
  10.1103/PhysRevLett.107.241101} {\bibfield  {journal} {\bibinfo  {journal}
  {Phys.Rev.Lett.}\ }\textbf {\bibinfo {volume} {107}},\ \bibinfo {pages}
  {241101} (\bibinfo {year} {2011})},\ \Eprint {http://arxiv.org/abs/1109.6021}
  {1109.6021 [gr-qc]} \BibitemShut {NoStop}%
%%CITATION = ARXIV:1109.6021;%%
\bibitem [{\citenamefont {Yoshino}\ and\ \citenamefont
  {Kodama}(2012)}]{Yoshino:2012kn}%
  \BibitemOpen
  \bibfield  {author} {\bibinfo {author} {\bibfnamefont {H.}~\bibnamefont
  {Yoshino}}\ and\ \bibinfo {author} {\bibfnamefont {H.}~\bibnamefont
  {Kodama}},\ }\href@noop {} {\  (\bibinfo {year} {2012})},\ \bibinfo {note}
  {38 pages, 18 figures},\ \Eprint {http://arxiv.org/abs/1203.5070} {1203.5070
  [gr-qc]} \BibitemShut {NoStop}%
%%CITATION = ARXIV:1203.5070;%%
\bibitem [{\citenamefont {Yunes}\ \emph {et~al.}(2012)\citenamefont {Yunes},
  \citenamefont {Pani},\ and\ \citenamefont {Cardoso}}]{Yunes:2011aa}%
  \BibitemOpen
  \bibfield  {author} {\bibinfo {author} {\bibfnamefont {N.}~\bibnamefont
  {Yunes}}, \bibinfo {author} {\bibfnamefont {P.}~\bibnamefont {Pani}}, \ and\
  \bibinfo {author} {\bibfnamefont {V.}~\bibnamefont {Cardoso}},\ }\href
  {\doibase 10.1103/PhysRevD.85.102003} {\bibfield  {journal} {\bibinfo
  {journal} {Phys.Rev.}\ }\textbf {\bibinfo {volume} {D85}},\ \bibinfo {pages}
  {102003} (\bibinfo {year} {2012})},\ \Eprint {http://arxiv.org/abs/1112.3351}
  {1112.3351 [gr-qc]} \BibitemShut {NoStop}%
%%CITATION = ARXIV:1112.3351;%%
\bibitem [{\citenamefont {Dolan}(2007)}]{dolan}%
  \BibitemOpen
  \bibfield  {author} {\bibinfo {author} {\bibfnamefont {S.~R.}\ \bibnamefont
  {Dolan}},\ }\href {\doibase 10.1103/PhysRevD.76.084001} {\bibfield  {journal}
  {\bibinfo  {journal} {Phys.Rev.}\ }\textbf {\bibinfo {volume} {D76}},\
  \bibinfo {pages} {084001} (\bibinfo {year} {2007})},\ \Eprint
  {http://arxiv.org/abs/0705.2880} {0705.2880 [gr-qc]} \BibitemShut {NoStop}%
%%CITATION = ARXIV:0705.2880;%%
\bibitem [{\citenamefont {Cardoso}\ and\ \citenamefont
  {Dias}(2004)}]{Cardoso:2004hs}%
  \BibitemOpen
  \bibfield  {author} {\bibinfo {author} {\bibfnamefont {V.}~\bibnamefont
  {Cardoso}}\ and\ \bibinfo {author} {\bibfnamefont {O.~J.~C.}\ \bibnamefont
  {Dias}},\ }\href {\doibase 10.1103/PhysRevD.70.084011} {\bibfield  {journal}
  {\bibinfo  {journal} {Phys.Rev.}\ }\textbf {\bibinfo {volume} {D70}},\
  \bibinfo {pages} {084011} (\bibinfo {year} {2004})},\ \Eprint
  {http://arxiv.org/abs/hep-th/0405006} {hep-th/0405006 [hep-th]} \BibitemShut
  {NoStop}%
%%CITATION = HEP-TH/0405006;%%
\bibitem [{\citenamefont {Hawking}\ and\ \citenamefont
  {Reall}(2000)}]{Hawking:1999dp}%
  \BibitemOpen
  \bibfield  {author} {\bibinfo {author} {\bibfnamefont {S.~W.}\ \bibnamefont
  {Hawking}}\ and\ \bibinfo {author} {\bibfnamefont {H.~S.}\ \bibnamefont
  {Reall}},\ }\href {\doibase 10.1103/PhysRevD.61.024014} {\bibfield  {journal}
  {\bibinfo  {journal} {Phys.Rev.}\ }\textbf {\bibinfo {volume} {D61}},\
  \bibinfo {pages} {024014} (\bibinfo {year} {2000})},\ \Eprint
  {http://arxiv.org/abs/hep-th/9908109} {hep-th/9908109 [hep-th]} \BibitemShut
  {NoStop}%
%%CITATION = HEP-TH/9908109;%%
\bibitem [{\citenamefont {Casals}\ \emph {et~al.}(2008)\citenamefont {Casals},
  \citenamefont {Dolan}, \citenamefont {Kanti},\ and\ \citenamefont
  {Winstanley}}]{Casals:2008jhep}%
  \BibitemOpen
  \bibfield  {author} {\bibinfo {author} {\bibfnamefont {M.}~\bibnamefont
  {Casals}}, \bibinfo {author} {\bibfnamefont {S.}~\bibnamefont {Dolan}},
  \bibinfo {author} {\bibfnamefont {P.}~\bibnamefont {Kanti}}, \ and\ \bibinfo
  {author} {\bibfnamefont {E.}~\bibnamefont {Winstanley}},\ }\href {\doibase
  10.1088/1126-6708/2008/06/071} {\bibfield  {journal} {\bibinfo  {journal}
  {JHEP}\ }\textbf {\bibinfo {volume} {0806}},\ \bibinfo {pages} {071}
  (\bibinfo {year} {2008})},\ \Eprint {http://arxiv.org/abs/0801.4910v2}
  {0801.4910v2 [hep-th]} \BibitemShut {NoStop}%
%%CITATION = ARXIV:0801.4910v2;%%
\bibitem [{\citenamefont {Pani}\ \emph {et~al.}(2010)\citenamefont {Pani},
  \citenamefont {Barausse}, \citenamefont {Berti},\ and\ \citenamefont
  {Cardoso}}]{Pani:2010prd}%
  \BibitemOpen
  \bibfield  {author} {\bibinfo {author} {\bibfnamefont {P.}~\bibnamefont
  {Pani}}, \bibinfo {author} {\bibfnamefont {E.}~\bibnamefont {Barausse}},
  \bibinfo {author} {\bibfnamefont {E.}~\bibnamefont {Berti}}, \ and\ \bibinfo
  {author} {\bibfnamefont {V.}~\bibnamefont {Cardoso}},\ }\href {\doibase
  10.1103/PhysRevD.82.044009} {\bibfield  {journal} {\bibinfo  {journal}
  {Phys.Rev.}\ }\textbf {\bibinfo {volume} {D82}},\ \bibinfo {pages} {044009}
  (\bibinfo {year} {2010})},\ \Eprint {http://arxiv.org/abs/1006.1863v3}
  {1006.1863v3} \BibitemShut {NoStop}%
%%CITATION = ARXIV:1006.1863v3;%%
\bibitem [{\citenamefont {Darwin}(1880)}]{darwin}%
  \BibitemOpen
  \bibfield  {author} {\bibinfo {author} {\bibfnamefont {G.~H.}\ \bibnamefont
  {Darwin}},\ }\href@noop {} {\bibfield  {journal} {\bibinfo  {journal}
  {Philos. Trans. R. Soc. London}\ }\textbf {\bibinfo {volume} {171}},\
  \bibinfo {pages} {713} (\bibinfo {year} {1880})}\BibitemShut {NoStop}%
\bibitem [{\citenamefont {Hut}(1981)}]{hut}%
  \BibitemOpen
  \bibfield  {author} {\bibinfo {author} {\bibfnamefont {P.}~\bibnamefont
  {Hut}},\ }\href@noop {} {\bibfield  {journal} {\bibinfo  {journal} {Astrono.
  Astrophys.}\ }\textbf {\bibinfo {volume} {99}},\ \bibinfo {pages} {126}
  (\bibinfo {year} {1981})}\BibitemShut {NoStop}%
\bibitem [{\citenamefont {Jackson}(1975)}]{jackson}%
  \BibitemOpen
  \bibfield  {author} {\bibinfo {author} {\bibfnamefont {J.~D.}\ \bibnamefont
  {Jackson}},\ }\href@noop {} {\emph {\bibinfo {title} {Classical
  Electrodynamics}}}\ (\bibinfo  {publisher} {John Wiley},\ \bibinfo {address}
  {New York},\ \bibinfo {year} {1975})\BibitemShut {NoStop}%
\bibitem [{\citenamefont {Cardoso}\ \emph {et~al.}(2007)\citenamefont
  {Cardoso}, \citenamefont {Cavaglia},\ and\ \citenamefont
  {Guo}}]{Cardoso:2007uy}%
  \BibitemOpen
  \bibfield  {author} {\bibinfo {author} {\bibfnamefont {V.}~\bibnamefont
  {Cardoso}}, \bibinfo {author} {\bibfnamefont {M.}~\bibnamefont {Cavaglia}}, \
  and\ \bibinfo {author} {\bibfnamefont {J.-Q.}\ \bibnamefont {Guo}},\ }\href
  {\doibase 10.1103/PhysRevD.75.084020} {\bibfield  {journal} {\bibinfo
  {journal} {Phys.Rev.}\ }\textbf {\bibinfo {volume} {D75}},\ \bibinfo {pages}
  {084020} (\bibinfo {year} {2007})},\ \Eprint
  {http://arxiv.org/abs/hep-th/0702138} {hep-th/0702138 [hep-th]} \BibitemShut
  {NoStop}%
%%CITATION = HEP-TH/0702138;%%
\bibitem [{\citenamefont {Cardoso}\ \emph {et~al.}(2009)\citenamefont
  {Cardoso}, \citenamefont {Miranda}, \citenamefont {Berti}, \citenamefont
  {Witek},\ and\ \citenamefont {Zanchin}}]{Cardoso:2008bp}%
  \BibitemOpen
  \bibfield  {author} {\bibinfo {author} {\bibfnamefont {V.}~\bibnamefont
  {Cardoso}}, \bibinfo {author} {\bibfnamefont {A.~S.}\ \bibnamefont
  {Miranda}}, \bibinfo {author} {\bibfnamefont {E.}~\bibnamefont {Berti}},
  \bibinfo {author} {\bibfnamefont {H.}~\bibnamefont {Witek}}, \ and\ \bibinfo
  {author} {\bibfnamefont {V.~T.}\ \bibnamefont {Zanchin}},\ }\href {\doibase
  10.1103/PhysRevD.79.064016} {\bibfield  {journal} {\bibinfo  {journal}
  {Phys.Rev.}\ }\textbf {\bibinfo {volume} {D79}},\ \bibinfo {pages} {064016}
  (\bibinfo {year} {2009})},\ \Eprint {http://arxiv.org/abs/0812.1806}
  {0812.1806 [hep-th]} \BibitemShut {NoStop}%
%%CITATION = ARXIV:0812.1806;%%
\bibitem [{\citenamefont {Poisson}(1993)}]{poisson1}%
  \BibitemOpen
  \bibfield  {author} {\bibinfo {author} {\bibfnamefont {E.}~\bibnamefont
  {Poisson}},\ }\href {\doibase 10.1103/PhysRevD.47.1497} {\bibfield  {journal}
  {\bibinfo  {journal} {Phys.Rev.}\ }\textbf {\bibinfo {volume} {D47}},\
  \bibinfo {pages} {1497} (\bibinfo {year} {1993})}\BibitemShut {NoStop}%
%%CITATION = PHRVA,D47,1497;%%
\bibitem [{\citenamefont {Poisson}\ and\ \citenamefont
  {Sasaki}(1995)}]{poisson2}%
  \BibitemOpen
  \bibfield  {author} {\bibinfo {author} {\bibfnamefont {E.}~\bibnamefont
  {Poisson}}\ and\ \bibinfo {author} {\bibfnamefont {M.}~\bibnamefont
  {Sasaki}},\ }\href {\doibase 10.1103/PhysRevD.51.5753} {\bibfield  {journal}
  {\bibinfo  {journal} {Phys.Rev.}\ }\textbf {\bibinfo {volume} {D51}},\
  \bibinfo {pages} {5753} (\bibinfo {year} {1995})},\ \Eprint
  {http://arxiv.org/abs/gr-qc/9412027} {gr-qc/9412027 [gr-qc]} \BibitemShut
  {NoStop}%
%%CITATION = GR-QC/9412027;%%
\bibitem [{\citenamefont {Kanti}\ and\ \citenamefont
  {March-Russell}(2002)}]{kanti}%
  \BibitemOpen
  \bibfield  {author} {\bibinfo {author} {\bibfnamefont {P.}~\bibnamefont
  {Kanti}}\ and\ \bibinfo {author} {\bibfnamefont {J.}~\bibnamefont
  {March-Russell}},\ }\href {\doibase 10.1103/PhysRevD.66.024023} {\bibfield
  {journal} {\bibinfo  {journal} {Phys.Rev.}\ }\textbf {\bibinfo {volume}
  {D66}},\ \bibinfo {pages} {024023} (\bibinfo {year} {2002})},\ \Eprint
  {http://arxiv.org/abs/hep-ph/0203223} {hep-ph/0203223 [hep-ph]} \BibitemShut
  {NoStop}%
%%CITATION = HEP-PH/0203223;%%
\bibitem [{\citenamefont {Chen}\ \emph {et~al.}(2008)\citenamefont {Chen},
  \citenamefont {Wang}, \citenamefont {Su},\ and\ \citenamefont
  {Hwang}}]{chen}%
  \BibitemOpen
  \bibfield  {author} {\bibinfo {author} {\bibfnamefont {S.}~\bibnamefont
  {Chen}}, \bibinfo {author} {\bibfnamefont {B.}~\bibnamefont {Wang}}, \bibinfo
  {author} {\bibfnamefont {R.-K.}\ \bibnamefont {Su}}, \ and\ \bibinfo {author}
  {\bibfnamefont {W.-Y.~P.}\ \bibnamefont {Hwang}},\ }\href {\doibase
  10.1088/1126-6708/2008/03/019} {\bibfield  {journal} {\bibinfo  {journal}
  {JHEP}\ }\textbf {\bibinfo {volume} {0803}},\ \bibinfo {pages} {019}
  (\bibinfo {year} {2008})},\ \Eprint {http://arxiv.org/abs/0711.3599}
  {0711.3599 [hep-th]} \BibitemShut {NoStop}%
%%CITATION = ARXIV:0711.3599;%%
\bibitem [{\citenamefont {Creek}\ \emph {et~al.}(2007)\citenamefont {Creek},
  \citenamefont {Efthimiou}, \citenamefont {Kanti},\ and\ \citenamefont
  {Tamvakis}}]{Creek:2007plb}%
  \BibitemOpen
  \bibfield  {author} {\bibinfo {author} {\bibfnamefont {S.}~\bibnamefont
  {Creek}}, \bibinfo {author} {\bibfnamefont {O.}~\bibnamefont {Efthimiou}},
  \bibinfo {author} {\bibfnamefont {P.}~\bibnamefont {Kanti}}, \ and\ \bibinfo
  {author} {\bibfnamefont {K.}~\bibnamefont {Tamvakis}},\ }\href {\doibase
  10.1016/j.physletb.2007.09.050} {\bibfield  {journal} {\bibinfo  {journal}
  {Phys.Lett.}\ }\textbf {\bibinfo {volume} {B656}},\ \bibinfo {pages} {102}
  (\bibinfo {year} {2007})},\ \Eprint {http://arxiv.org/abs/0709.0241v2}
  {0709.0241v2 [hep-th]} \BibitemShut {NoStop}%
%%CITATION = ARXIV:0709.0241v2;%%
\bibitem [{\citenamefont {Abramowitz}\ and\ \citenamefont
  {Stegun}(1964)}]{handmath}%
  \BibitemOpen
  \bibfield  {author} {\bibinfo {author} {\bibfnamefont {M.}~\bibnamefont
  {Abramowitz}}\ and\ \bibinfo {author} {\bibfnamefont {I.~A.}\ \bibnamefont
  {Stegun}},\ }\href@noop {} {\emph {\bibinfo {title} {Handbook of Mathematical
  Functions}}}\ (\bibinfo  {publisher} {GPO},\ \bibinfo {year}
  {1964})\BibitemShut {NoStop}%
\bibitem [{\citenamefont {Witek}\ \emph {et~al.}(2010)\citenamefont {Witek},
  \citenamefont {Zilhao}, \citenamefont {Gualtieri}, \citenamefont {Cardoso},
  \citenamefont {Herdeiro} \emph {et~al.}}]{Witek:2010xi}%
  \BibitemOpen
  \bibfield  {author} {\bibinfo {author} {\bibfnamefont {H.}~\bibnamefont
  {Witek}}, \bibinfo {author} {\bibfnamefont {M.}~\bibnamefont {Zilhao}},
  \bibinfo {author} {\bibfnamefont {L.}~\bibnamefont {Gualtieri}}, \bibinfo
  {author} {\bibfnamefont {V.}~\bibnamefont {Cardoso}}, \bibinfo {author}
  {\bibfnamefont {C.}~\bibnamefont {Herdeiro}},  \emph {et~al.},\ }\href
  {\doibase 10.1103/PhysRevD.82.104014} {\bibfield  {journal} {\bibinfo
  {journal} {Phys.Rev.}\ }\textbf {\bibinfo {volume} {D82}},\ \bibinfo {pages}
  {104014} (\bibinfo {year} {2010})},\ \Eprint {http://arxiv.org/abs/1006.3081}
  {1006.3081 [gr-qc]} \BibitemShut {NoStop}%
%%CITATION = ARXIV:1006.3081;%%
\bibitem [{\citenamefont {Okawa}\ \emph {et~al.}(2011)\citenamefont {Okawa},
  \citenamefont {Nakao},\ and\ \citenamefont {Shibata}}]{Okawa:2011fv}%
  \BibitemOpen
  \bibfield  {author} {\bibinfo {author} {\bibfnamefont {H.}~\bibnamefont
  {Okawa}}, \bibinfo {author} {\bibfnamefont {K.-i.}\ \bibnamefont {Nakao}}, \
  and\ \bibinfo {author} {\bibfnamefont {M.}~\bibnamefont {Shibata}},\ }\href
  {\doibase 10.1103/PhysRevD.83.121501} {\bibfield  {journal} {\bibinfo
  {journal} {Phys.Rev.}\ }\textbf {\bibinfo {volume} {D83}},\ \bibinfo {pages}
  {121501} (\bibinfo {year} {2011})},\ \Eprint {http://arxiv.org/abs/1105.3331}
  {1105.3331 [gr-qc]} \BibitemShut {NoStop}%
%%CITATION = ARXIV:1105.3331;%%
\end{thebibliography}%

\addcontentsline{toc}{chapter}{\bibname}
\cleardoublepage

\end{document}